\documentclass[notitlepage,aps,prd,twocolumn,superscriptaddress,showpacs]{revtex4-2}

\usepackage{natbib}
\usepackage[english]{babel}
\usepackage{graphicx}

\usepackage{amsmath, amssymb}
\usepackage{picinpar}
\usepackage{mathtools}
\usepackage[pdfborderstyle={/S/U/W 1}]{hyperref}
\usepackage{bbold}
\usepackage{empheq}
\usepackage{booktabs}
\usepackage[most]{tcolorbox}
\usepackage{siunitx}
\usepackage{tensor}
\usepackage{latexsym}
\usepackage{placeins}
\usepackage{csquotes}
\usepackage[acronym]{glossaries}
\glsdisablehyper
\usepackage{xfrac}
\usepackage{multirow}
\usepackage{titlesec}
\usepackage{csquotes}
\usepackage[normalem]{ulem}
\usepackage{colortbl}
\usepackage[caption=false,justification=justified]{subfig}  
\usepackage{todonotes}
\usepackage{hhline}
\usepackage{physics}
\usepackage{xargs}
\usepackage{ifmtarg}
\usepackage[h]{esvect}
\usepackage{bm}
\usepackage{svg}
\usepackage{adjustbox}

\DeclareMathSizes{10}{10}{8}{6}
\DeclareSIUnit{\persqrthz}{\ensuremath{\text{Hz}^{-1/2}}}
\renewcommand{\arraystretch}{1.5}
\allowdisplaybreaks

\definecolor{Gray}{gray}{0.9}

\everymath{\displaystyle}

\DeclareUnicodeCharacter{2009}{\,} 

\begin{document}

\title{LISA Dynamics \& Control: Closed-loop Simulation and Numerical Demonstration of Time Delay Interferometry}



\author{Lavinia Heisenberg}
\affiliation{Institut f\"ur Theoretische Physik, Universit\"at Heidelberg, Philosophenweg 16, 69120 Heidelberg, Germany}
\affiliation{Institute for Theoretical Physics, ETH Z\"urich, Wolfgang-Pauli-Strasse 27, 8093, Z\"urich, Switzerland}
\affiliation{Perimeter Institute for Theoretical Physics, 31 Caroline St N, Waterloo, Ontario, N2L 6B9, Canada}

\author{Henri Inchausp\'e}
\email[Corresponding author: ]{inchauspe@tphys.uni-heidelberg.de}
\affiliation{Institut f\"ur Theoretische Physik, Universit\"at Heidelberg, Philosophenweg 16, 69120 Heidelberg, Germany}
\affiliation{Universit\'e Paris Cit\'e, CNRS, CNES, Astroparticule et Cosmologie, F-75013 Paris, France}

\author{Dam Quang Nam}
\affiliation{Universit\'e Paris Cit\'e, CNRS, Astroparticule et Cosmologie, F-75013 Paris, France}
\affiliation{Laboratoire de Physique Corpusculaire de Caen, Universit\'e de Caen/CNRS, 14000 Caen, France}

\author{Orion Sauter}
\affiliation{Department of Mechanical and Aerospace Engineering, MAE-A, P.O. Box 116250, University of Florida, Gainesville, Florida 32611, USA}

\author{Ricardo Waibel}
\affiliation{Institut f\"ur Theoretische Physik, Universit\"at Heidelberg, Philosophenweg 16, 69120 Heidelberg, Germany}

\author{Peter Wass}
\affiliation{Department of Mechanical and Aerospace Engineering, MAE-A, P.O. Box 116250, University of Florida, Gainesville, Florida 32611, USA}


\begin{abstract}

    The Laser Interferometer Space Antenna (LISA), space-based gravitational wave observatory involves a complex multidimensional closed-loop dynamical system. Its instrument performance is expected to be less efficiently isolated from platform motion than was its simpler technological demonstrator, LISA Pathfinder. It is of crucial importance to understand and model LISA dynamical behavior accurately to understand the propagation of dynamical excitations through the response of the instrument down to the interferometer data streams. More generally, simulation of the system allows for the preparation of the processing and interpretation of in-flight metrology data. In this work, we present a comprehensive mathematical modeling of the closed-loop system dynamics and its numerical implementation within the LISA Consortium simulation suite. We provide, for the first time, a full time-domain numerical demonstration of post-processing Time Delay Interferometer techniques combining multiple position measurements with realistic control loops to create a synthetic Michelson interferometer. We show that in the absence of physical coupling to spacecraft and telescope motion (through tilt-to-length, stiffness and actuation cross-talk) the effect of noisy spacecraft motion is efficiently suppressed to a level below the noise originating in the rest of the instrument.
    
\end{abstract}

\maketitle

\newacronym{e2e}{E2E}{End-To-End}
\newacronym{inrep}{INREP}{Initial Noise REduction Pipeline}
\newacronym{tdi}{TDI}{Time Delay Interferometry}
\newacronym{ttl}{TTL}{Tilt-To-Length}
\newacronym{dfacs}{DFACS}{Drag-Free and Attitude Control System}
\newacronym{ldc}{LDC}{LISA Data Challenge}
\newacronym{ldpg}{LDPG}{LISA Data Processing Group}
\newacronym{lsg}{LSG}{LISA Science Group}
\newacronym{lig}{LIG}{LISA Instrument Group}
\newacronym{simwg}{LDPG-SimWG}{Simulation Working Group}
\newacronym{aei}{AEI}{Albert Einstein Institute}
\newacronym{pssl}{PSSL}{Precision Space Systems Laboratory}
\newacronym{apc}{APC}{AstroParticule et Cosmologie}
\newacronym{syrte}{SYRTE}{Systèmes de Référence Temps-Espace}
\newacronym{l2it}{L2IT}{Laboratoire des deux Infinis de Toulouse}
\newacronym{irap}{IRAP}{Institut de Recherche en Astrophysique et Planétologie}
\newacronym{lisa}{LISA}{Laser Interferometer Space Antenna}
\newacronym{siso}{SISO}{Single-Input Single-Ouput}
\newacronym{mimo}{MIMO}{Multiple-Input Multiple-Output}
\newacronym{emri}{EMRI}{Extreme Mass Ratio Inspiral}
\newacronym{ifo}{IFO}{Interferometry System}
\newacronym{grs}{GRS}{Gravitational Reference System}
\newacronym{dws}{DWS}{Differential Wavefront Sensing}
\newacronym{mosa}{MOSA}{Moving Optical Sub-Assembly}
\newacronym{eom}{EOM}{Equations Of Motion}
\newacronym{com}{CoM}{center of mass}
\newacronym{gw}{GW}{gravitational wave}
\newacronym{isi}{ISI}{inter-spacecraft interferometer}
\newacronym{tmi}{TMI}{test-mass interferometer}
\newacronym{ldws}{LDWS}{Long-arm Differential Wavefront Sensing}
\newacronym{dof}{DoF}{Degree of Freedom}
\newacronym{icrs}{ICRS}{International Celestial Reference System}
\newacronym{rk4}{RK4}{Runge-Kutta 4}
\newacronym{rkf45}{RKF45}{Runge-Kutta-Fehlberg 4 (5)}
\newacronym{lte}{LTE}{Local Truncation Error}
\newacronym{gte}{GTE}{Global Truncation Error}
\newacronym{tm}{TM}{Test Mass}
\newacronym{lti}{LTI}{Linear Time-Invariant}

\newcommand{\indice}[1]{{\scriptscriptstyle #1}}
\newcommand{\exposant}[1]{{\scriptscriptstyle #1}}
\newcommand{\myvec}[2]{\va*{#1}_\indice{#2}}
\newcommandx{\mylongvec}[4][2=,4=]{\vv{\mathbf{#1}_\indice{#2}\mathbf{#3}_\indice{#4}}}
\newcommand{\myexpr}[3]{#1_\indice{#2}^\exposant{#3}}
\newcommandx{\ddt}[3][3=]{ \dv{t}\Bigg|_\indice{\mathcal{#2}_{#3}} \mspace{-10mu} \left[ #1 \right]}
\newcommandx{\ddtddt}[3][3=]{ \dv[2]{t}\Bigg|_\indice{\mathcal{#2}_{#3}} \mspace{-10mu} \left[ #1 \right]}
\newcommand{\myhyperref}[1]{\hyperref[#1]{\ref{#1}}}
\newcommand{\identite}[1]{{\displaystyle \mathbb{1}_{\indice{#1}}}}
\newcommand{\myat}[2][]{#1|_{#2}}
\newcommand{\timestentothe}[1]{\times 10^{#1}}
\newcommand{\msout}[1]{\text{\sout{\ensuremath{#1}}}}
\newcommandx{\cardanvec}[4][2=,4=]{\myvec{\alpha}{\mathcal{#1}_\indice{#2}/\mathcal{#3}_\indice{#4}}}
\newcommandx{\thetacardan}[4][2=,4=]{\theta_\indice{{\mathcal{#1}_\indice{#2}/\mathcal{#3}_\indice{#4}}}}
\newcommandx{\etacardan}[4][2=,4=]{\eta_\indice{{\mathcal{#1}_\indice{#2}/\mathcal{#3}_\indice{#4}}}}
\newcommandx{\phicardan}[4][2=,4=]{\phi_\indice{{\mathcal{#1}_\indice{#2}/\mathcal{#3}_\indice{#4}}}}
\newcommandx{\cardanvecdot}[4][2=,4=]{\dv{\myvec{\alpha}{\mathcal{#1}_\indice{#2}/\mathcal{#3}_\indice{#4}}}{t}}
\newcommandx{\thetacardandot}[4][2=,4=]{\dot{\theta}_\indice{{\mathcal{#1}_\indice{#2}/\mathcal{#3}_\indice{#4}}}}
\newcommandx{\etacardandot}[4][2=,4=]{\dot{\eta}_\indice{{\mathcal{#1}_\indice{#2}/\mathcal{#3}_\indice{#4}}}}
\newcommandx{\phicardandot}[4][2=,4=]{\dot{\phi}_\indice{{\mathcal{#1}_\indice{#2}/\mathcal{#3}_\indice{#4}}}}
\newcommandx{\vangvel}[6][2=,3=,5=,6=]{\myvec{\omega}{\mathcal{#1}_\indice{#2}^\exposant{#3}/\mathcal{#4}_\indice{#5}^\exposant{#6}}}
\newcommandx{\framevu}[3][3=]{\vu{e}_{#2,\mathcal{#1}_\indice{#3}}}
\newcommand{\framevus}[2]{\vu{e}_{#1_\indice{#2}}}
\newcommandx{\rotmat}[4][2=,4=]{\myexpr{T}{\mathcal{#3}_\indice{#4}}{\mathcal{#1}_\indice{#2}}}

\newcommand{\sint}[1]{s_{#1}}
\newcommand{\cost}[1]{c_{#1}}

\newcommandx{\posvec}[4][2=,4=]{\myvec{r}{#1_\indice{#2} / #3_\indice{#4}}}
\newcommandx{\velvec}[4][2=,4=]{\myvec{v}{#1_\indice{#2} / #3_\indice{#4}}}
\newcommandx{\posvecexp}[6][2=,4=,6=]{\myexpr{r}{#1_\indice{#2} / #3_\indice{#4}}{\mathcal{#5}_\indice{#6}}}
\newcommandx{\velvecexp}[6][2=,4=,6=]{\myexpr{\dot{r}}{#1_\indice{#2} / #3_\indice{#4}}{\mathcal{#5}_\indice{#6}}}
\newcommandx{\accvecexp}[6][2=,4=,6=]{\myexpr{\ddot{r}}{#1_\indice{#2} / #3_\indice{#4}}{\mathcal{#5}_\indice{#6}}}

\newcommandx{\vangvelexp}[6][2=,4=,6=]{\myexpr{\omega}{\mathcal{#1}_\indice{#2} / \mathcal{#3}_\indice{#4}}{\mathcal{#5}_\indice{#6}}}
\newcommandx{\vangaccexp}[6][2=,4=,6=]{\myexpr{\dot{\omega}}{\mathcal{#1}_\indice{#2} / \mathcal{#3}_\indice{#4}}{\mathcal{#5}_\indice{#6}}}

\newcommandx{\force}[4][2=,4=]{\myexpr{f}{#1_\indice{#2}}{\mathcal{#3}_\indice{#4}}}
\newcommandx{\torque}[4][2=,4=]{\myexpr{t}{#1_\indice{#2}}{\mathcal{#3}_\indice{#4}}}

\newcommand{\mytensor}[2]{\vb{#1}_\indice{#2}}
\newcommandx{\itensor}[3][3=]{\mytensor{I}{\text{#1}/#2_\indice{#3}}}
\newcommandx{\itensorexp}[5][3=,5=]{\myexpr{I}{\text{#1}/#2_\indice{#3}}{\mathcal{#4}_\indice{#5}}}

\newcommand{\myskew}[1]{\left[ #1 \right]^{\times}}

\renewcommand{\arraystretch}{1.5}

\newcommand{\sensor}[1]{\exposant{\text{{#1}}}}

\newcommand{\euler}{\operatorname{e}}
\newcommand{\unitmat}{\mathbb{1}}

\section{Space-based gravitational waves astronomy and LISA detector}

The \gls{lisa} is a space-based, \gls{gw} observatory planned to launch in 2035 \cite{amaro-seoane_laser_2017}. It aims to open a new window on the Universe in the milliHertz bandwidth of the \gls{gw} sky, which is expected to harbor a rich and diverse collection of astrophysical and cosmological sources, including the merger events of supermassive black hole binaries \cite{klein_science_2016, lisa_sirens_2022}, believed as among the most energetic events in the observable Universe.

The mission, led by the European Space Agency, consists of three spacecraft arranged in a nearly equilateral triangular constellation, whose barycenter follows the Earth in a heliocentric orbit. Interferometric measurements of the spacecraft separation will be used to measure \gls{gw} as they pass through the constellation. The scale of the detector and its space environment allow for operation in the millihertz bandwidth. The 2.5 million km arm-lengths result in picometer optical path length variations due to \gls{gw}, and antenna nulls in the $40\,\si{\milli\hertz}$ regime.
Operation in space eliminates acoustic, seismic and Newtonian noise disturbances which limit the sensitivity of ground-based detectors below 1~Hz \cite{buikema_sensitivity_2020}.

\subsection{Picometer laser interferometry in space}
\label{subsection: space interferometer}

To observe \gls{gw} with strain amplitudes of order 10$^{-20}$, the LISA instrument must overcome two major challenges.
Firstly, the satellites are poor references of inertia, being subjected to force noise, from solar wind, solar radiation pressure, and their own micro-propulsion system \cite{lisa_pathfinder_collaboration_lisa_2019}. To overcome this obstacle, cubic Au-Pt alloy $1.92\,\si{\kilo\gram}$ test-masses within the spacecraft are used as the end-mirrors of the interferometer. 
Protected from environmental force noise, they constitute a very good approximation to local inertial reference frames on-board the spacecraft. They are housed in vacuum chambers and surrounded by electrodes which allow for position sensing and actuation \cite{lisa_pathfinder_collaboration_capacitive_2017}. Interferometric sensing is preferred to probe the test-mass displacement relative to their housings along the directions of the arm lengths \cite{armano_sensor_2021, armano_sensor_2022} (cf. constellation and  geometry spacecraft in Figure \myhyperref{figure: constellation} and \myhyperref{figure: spacecraft}). These \glspl{tmi} are exploited to monitor and suppress the spacecraft acceleration in the long-range science interferometers---called \glspl{isi} \cite{bayle_unified_2023}. The ability to fly $1.92\,\si{\kilo\gram}$ cubic references of inertia with stability performance compatible with \gls{gw} astronomy has been successfully demonstrated by the LISA Pathfinder mission \cite{armano_sub-femto-g_2016, armano_beyond_2018, lpf_platform_stability}.
    
Secondly, a Michelson-type interferometer with  million-km arm-lengths cannot be realized in space without post-processing techniques. Reflected laser roundtrips along the arms are prohibited by the available laser power on-board, and frequency noise suppression by optical path length matching is made impossible as the spacecraft orbital mechanics result in arm-lengths that are mismatched and changing over time. Instead, a collection of heterodyne interferometric measurements between local laser beams and propagated distant laser beams is performed across the constellation \cite{bayle_unified_2023}, and the resulting beat notes are time synchronized on-ground accounting for light propagation time delays, and linearly combined so that the laser noise is suppressed: this post-processing technique is called \gls{tdi} \cite{tinto_time-delay_2020, vallisneri_geometric_2005}.

Tackling these two technological challenges renders the measurement complex and composite, relying strongly on multiple post-processing steps to produce data exploitable for astronomy. Preparing for the successful operation of the mission, it is therefore crucial to understand, model, and simulate the instrument response and data generation in order to test with a high degree of representativity data interpretation and analysis strategies and methods.

\subsection{Metrology on-board}
\label{subsection: metrology}
To guarantee the best stability of the apparatus, and precise centering and alignment of the test-masses in their housings, the spacecraft embed high-precision sensor and actuation systems. 
Local optical \glspl{ifo} provide picometer-stable test-mass displacement sensing along the telescope axes (see Figure \myhyperref{figure: spacecraft} for the spacecraft geometry). These optical readouts are used as \gls{tmi} in the overall \gls{tdi} and contribute to the final interferometer data streams. 
An alternative combination of the interferometer  quadrant photodiode readout signals allows for a high-precision sensing of test-mass angular motion in two degrees of freedom using a method known as \gls{dws}~\cite{armano_sensor_2022}. 
The \gls{grs} provides $\sim$nm and $\sim$100\,nRad electrostatic sensing of all test-mass longitudinal and angular displacements, through the measurement of differential capacitance changes on a set of electrodes surrounding the test-mass~\cite{lisa_pathfinder_collaboration_capacitive_2017}. This set of capacitors is also used to exert correction forces and torques on the test-masses through audio frequency voltages applied to the electrodes~\cite{dolesi_gravitational_2003}. Control forces on the test-masses are triggered only by their differential motion~\cite{lisa_dfacs_scheme}. 
The \gls{dws} technique is exploited again to sense the angle of the long-range inter-spacecraft laser beam w.r.t. the telescope axes. The magnification of the telescope allows for sub-nanoradian sensing of spacecraft attitude and telescopes: this measurement is referred to as the \gls{ldws}. 
The assembly of the telescope, the optical bench, and the \gls{grs} housing the test-masses is named the \gls{mosa}. It will rotate about the test-mass axis to allow for the variation in the opening angle of the constellation due to orbital variations known as {\it breathing}. 
Spacecraft longitudinal and attitude control are performed using a system of micronewton thrusters \cite{lisa_pathfinder_collaboration_lisa_2019}, which allows the satellite to track the test-masses in their free-fall ({\it drag-free} control), and to lock its attitude on the laser beam received from the distant spacecraft ({\it attitude} control). A rotation mechanism is used for slow actuation of the \gls{mosa} ensuring the pointing direction tracks the incoming laser beam direction and nulls the \gls{ldws} output. The sensing, actuation, and control systems are summarized in Table \myhyperref{table: DFACS}. Noise performance values, which will be considered in the simulation, will be discussed in section \myhyperref{section: dfacs} and summarized in Table \myhyperref{table: NoiseLevels}. Performance estimates are based on the current LISA design and measurements from the LISA Pathfinder mission~\cite{armano_sub-femto-g_2016, lpf_platform_stability, armano_sensor_2022}.

\begin{table}
\scriptsize
\caption{For each control coordinate, the table lists the respective control type and actuator used, as well as the subsystem they are sensed with. Capital letters are used for spacecraft coordinates, while lower case and indices are used for test-mass coordinates. The table is adapted from Table I. of \cite{lisa_dfacs_scheme}, to which the control of the telescopes' opening angle $\hat{\phi}_{m}$ has been added in row 7.
}
\label{table: DFACS}
\begin{ruledtabular}
\begin{tabular}{|c|c|c|c|c|c|}
\# & Coord. & Sensor & Control Mode & Actuator & Command \\ \hline\hline
1 & $\hat{x}_\indice{1}$ & \gls{ifo} & Drag-Free & $\mu$-thrust & $f_{X}^{\text{drag-free}}$ \\
2 & $\hat{x}_\indice{2}$ & \gls{ifo} & Drag-Free & $\mu$-thrust & $f_{Y}^{\text{drag-free}}$ \\
3 & $\hat{z}_\indice{1}$ & \gls{grs} & Drag-Free & $\mu$-thrust & $f_{Z}^{\text{drag-free}}$ \\
4 & $\hat{\Theta}$ & \gls{ldws} & Attitude & $\mu$-thrust & $t_{X}^{\text{att}}$ \\
5 & $\hat{H}$ & \gls{ldws} & Attitude & $\mu$-thrust & $t_{Y}^{\text{att}}$ \\
6 & $\hat{\Phi}$ & \gls{ldws} & Attitude & $\mu$-thrust & $t_{Z}^{\text{att}}$ \\
7 & $\hat{\phi}_{m}$ & \gls{ldws} & Tel. Pointing & \acrshort{mosa} mechanism & $t_{z_\indice{1}}^{\text{tel}}$ / $t_{z_\indice{2}}^{\text{tel}}$ \\
8 & $\hat{y}_\indice{1}$ & \gls{grs} & Suspension & \gls{grs} & $f_{y_1}^{\text{sus}}$ \\
9 & $\hat{y}_\indice{2}$ & \gls{grs} & Suspension & \gls{grs} & $f_{y_2}^{\text{sus}}$ \\
10 & $\hat{z}_\indice{2}$ & \gls{grs} & Suspension & \gls{grs} & $f_{z_1}^{\text{sus}}$ / $f_{z_2}^{\text{sus}}$ \\
11 & $\hat{\theta}_\indice{1}$ & \gls{grs} & Suspension & \gls{grs} & $t_{x_1}^{\text{sus}}$ \\
12 & $\hat{\eta}_\indice{1}$ & \gls{ifo} & Suspension & \gls{grs} & $t_{y_1}^{\text{sus}}$ \\
13 & $\hat{\phi}_\indice{1}$ & \gls{ifo} & Suspension & \gls{grs} & $t_{z_1}^{\text{sus}}$ \\
14 & $\hat{\theta}_\indice{2}$ & \gls{grs} & Suspension & \gls{grs} & $t_{x_2}^{\text{sus}}$ \\
15 & $\hat{\eta}_\indice{2}$ & \gls{ifo} & Suspension & \gls{grs} & $t_{y_2}^{\text{sus}}$ \\
16 & $\hat{\phi}_\indice{2}$ & \gls{ifo} & Suspension & \gls{grs} & $t_{z_2}^{\text{sus}}$ \\
\end{tabular}
\end{ruledtabular}
\end{table}

\subsection{Operating without a direct optical differential channel}
\label{subsection: diff channel}

\begin{figure}[h!]
\frame{\centerline{\includegraphics[width=\linewidth, trim={0.0cm 0.0cm 0.0cm 0.0cm}, clip]{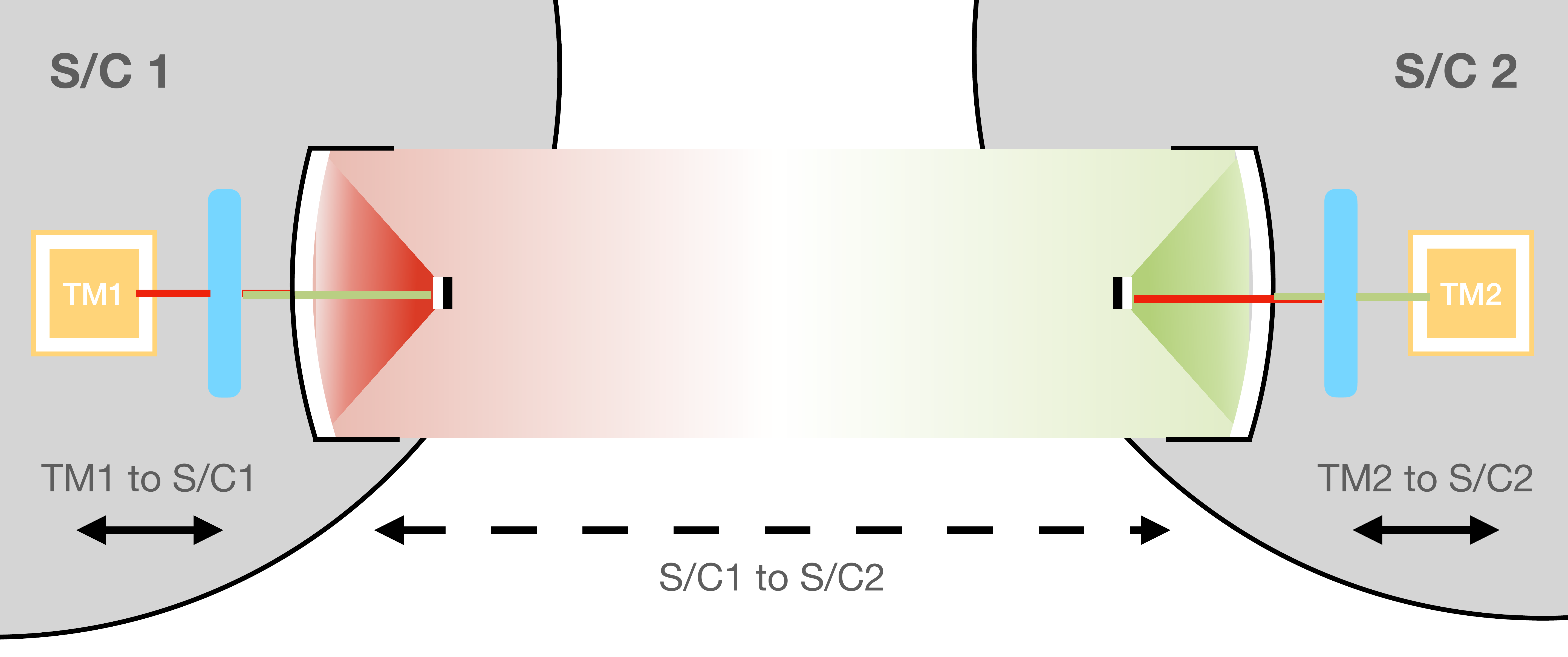}}}
\caption{The overall test-mass-to-test-mass optical measurement is split into three pieces: one \gls{isi} and two \gls{tmi}s per arm. Nominally, local spacecraft (1) accelerations towards the distant spacecraft (2) shorten the \gls{isi} optical path, but increase the local \gls{tmi} optical path length by the same amount, thus \gls{tdi} outputs remain unaffected. In this nominal case, the geometrical point of the spacecraft probed by the \gls{isi} is matching the one probed by the \gls{tmi}. We denoted such a geometrical point by the letter $C$. Illustration from \cite{lisa_dfacs_scheme}.}
\label{figure: Images/Images_longrange}
\end{figure}

A key difference LISA Pathfinder and LISA is that no direct, differential optical measurements between test-masses will be available for LISA, neither locally within the spacecraft, nor at the constellation scale between distant test-masses. Indeed, as will be discussed in more detail in the final section of this work (cf. Figure \myhyperref{figure: Images/Images_longrange}), the long-range optical measurement is split into three pieces: one local test-mass to spacecraft \gls{tmi} measurement, one long-range \gls{isi} measurement, and a second \gls{tmi} measurement at the other end of the interferometer arm. 
The necessary decomposition of the long-range interferometers makes the detector performance more liable to spacecraft and telescope dynamical stability than in the LISA Pathfinder case, where test-mass-to-test-mass measurements were by construction efficiently isolated from noisy spacecraft motion.
\cite{lpf_platform_stability, armano_sub-femto-g_2016}. In particular, optical geometrical misalignment and off-centering will introduce \gls{ttl} coupling to the spacecraft and telescope noisy angular motion \cite{chwalla_optical_2020, paczkowski_postprocessing_2022}. This effect is expected to be one of the leading noise contributors to the overall noise budget of the instrument. In addition, electrostatic and gravitational fields create a force gradient at the test-mass---referred to as stiffnesses---that couples to spacecraft motion and generates acceleration noise.

Dynamical stability plays a driving role in both of these disturbances and is a key property for LISA performance. It is of crucial importance to model the closed-loop dynamics of the spacecraft-telescopes-test-masses system to assess accurately such dynamical noise, and demonstrate our ability to mitigate the impact of dynamical artifacts on LISA data analysis. LISA dynamics modeling has been already addressed in recent literature \cite{vidano_lisa_2020}, in the scope of \gls{dfacs} design, and implemented in a proprietary software \texttt{Matlab/Simscape}. Here we present an implementation of a full closed-loop dynamics simulation integrated into the \texttt{python} based consortium \gls{e2e} simulation suite and dedicated to LISA data processing and analysis. We provide the modeling framework and the \gls{eom} at play in comprehensive detail, as we believe it deserves an elaborated and formalized reference in the scope of future LISA in-flight data diagnostics and interpretation.

\subsection{Simulating LISA dynamics}
\label{subsection: simulating dynamics}

Among the simulation tools developed by the LISA consortium, the \texttt{LISANode} software~\cite{bayle_lisanode_2022} is particularly well-suited for spacecraft dynamics and control simulation. Its graph-based, modular framework naturally lends itself to control system implementation (in a similar way to commonly-used \texttt{Matlab-Simulink} tools) and its generation and management of time series data, operating on quantities as they flow between graphs, enables long simulations with efficient use of memory~\cite{bayle_simulation_2019}. 

In this paper, the full derivation of the \gls{eom} of the LISA dynamical system is developed and its implementation in the \texttt{LISANode} simulation tool is described. This includes implementation of the \gls{eom}, simulations of the sensing systems, and implementation of the feedback loop, interface with the interferometer measurements, and post-processing the resulting beat notes with \gls{tdi}. For the first time, the impact of spacecraft and test-mass dynamics, and control on LISA data at the level of heterodyne beat notes fluctuations and the \gls{tdi} channels can be studied. 

This article is organized as follows: We introduce the LISA dynamical system, describing the relevant reference frames in Section \myhyperref{section: dynamics}. Simulation of the \gls{eom} for the test-masses and spacecraft are detailed in Section \myhyperref{section: eoms}. These equations are simplified in Section \myhyperref{section: linearization} by linearizing around a stable working point, which is maintained by the controllers detailed in Section \myhyperref{section: dfacs}. Numerical solving methods used for the linear system approximation, as well as the full non-linear dynamical system, are detailed in Section \myhyperref{section: solver}. Results of the dynamics simulations are shown in Section \myhyperref{section: experiments}, and in particular a demonstration of jitter suppression in Section \myhyperref{section: tdi}.

\section{Dynamical system and Reference Frames}
\label{section: dynamics}

This section is dedicated to the full derivation of \gls{lisa} \glspl{eom} and their insertion into the closed-loop system, including sensors and actuators. The \gls{eom} are indeed a core piece of the simulation and deserve thorough attention. A preliminary step is the definition of the dynamical coordinates and reference frames necessary to describe fully the dynamical state of the system. We based our mathematical modeling on a rigid-body approximation in which the dynamical state of a body is described by the motion of its \gls{com} and its angular velocity w.r.t. an inertial frame: $\left\{\myvec{r}{\text{body}}, \myvec{\omega}{\text{body}/\text{gal.}}\right\}$. 
Each \gls{lisa} spacecraft is a 20 \gls{dof} system, 
six for each of the spacecraft and test-mass translational and rotational dynamics, and an additional two for the \glspl{mosa} that will rotate along with the constellation orbital breathing. The remaining \gls{mosa} \glspl{dof} are assumed rigidly fixed to the spacecraft.
We note that in nominal operations, \gls{mosa} angle actuation is designed to be symmetric, hence suppressing one \gls{dof}. However, other modes of operation where this rotation is asymmetric are possible, so it is preferable at this stage to maintain generality.

The LISA dynamics will be operated in closed-loop control by the onboard \gls{dfacs}, locking the test-mass and spacecraft \glspl{dof} onto specific target points as described in Section \myhyperref{section: dfacs}.
To simulate the system, it is therefore required to express the dynamics of the system in the frame of reference from which they are observed. For example, the test-mass dynamics need to be expressed in the frame attached to its housing they are lodged in, since it is, to first approximation, the reference for the \gls{grs} and \gls{ifo} sensing. 
This requirement results in the presence of several imbricated, non-inertial reference frames, which, when coupled with the multidimensionality of the system increases the complexity of its description.
As a first step, we list and define the set of reference frames we use in the simulation, after which we will select a state representation of the dynamics which facilitates mapping of sensing and actuation.

\subsection{Frames of reference}
\label{subsection: Frames of reference}

\begin{figure}[h!]
\frame{\centerline{\includegraphics[width=1.0\linewidth]{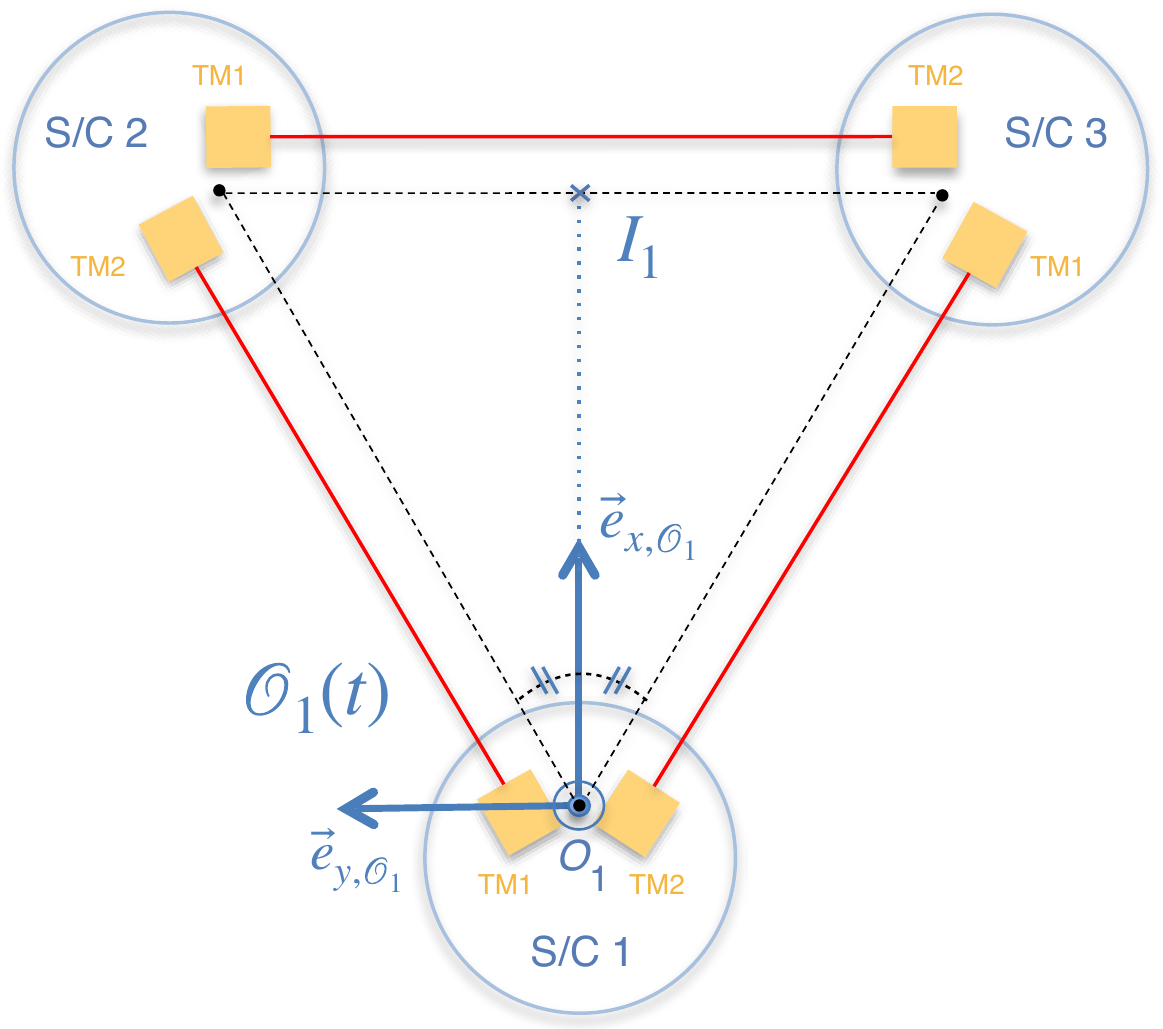}}}
\caption{Figure cut of constellation geometry in the triangle plane. The geometrical construction of the $\mathcal{O}_1$ is illustrated. The satellite positions considered in order to build the $\mathcal{O}$-frames are computed from the macroscopic orbital motions only, that is, neglecting the (dynamical state-dependent) stray forces inducing spacecraft jitter. However, state-independent {\it dc} forces deviating the spacecraft from pure free-fall can be accounted for in the definition of $\mathcal{O}$-frames.}
\label{figure: constellation}
\end{figure}

The dynamical model invokes six different types of reference frame---each with its own system of coordinates.
One reference frame per rotating, rigid body (spacecraft, test-mass, \gls{mosa}) will be used to describe relative orientation between bodies. Two frames associated with the fiducial rotation of the spacecraft and \gls{mosa} about their operation point, that facilitate linearization (as discussed in Section \myhyperref{section: linearization}) and one inertial reference frame. These frames are defined as follows:

\begin{itemize}

    \item Galilean (inertial) $\mathcal{J}$-frame, fixed w.r.t. distant stars, and with orientation defined according to the \gls{icrs} convention \cite{kaplan_iau_2006}.
    
    \item The $\mathcal{O}$-frames (or $\mathcal{B^*}$-frames) which set the target attitude for the spacecraft. It is built following the diagram in Figure \myhyperref{figure: constellation}:
    \begin{itemize}
        \item $x$-axis $\framevu{O}{x}[i]$ is constructed as the bisector of constellation, through local summit spacecraft.
        \item $z$-axis $\framevu{O}{z}[i]$ is the unit vector normal to the constellation plane.
        \item $y$-axis $\framevu{O}{y}[i]$ is built from the cross product of the two above.
        \item The origin $O_i$ of the frame $\mathcal{O}_i$ follows the ideal orbit of the spacecraft.
    \end{itemize}
    \item The $\mathcal{B}$-frames rigidly attached to the spacecraft and describing its attitude, whether w.r.t to $\mathcal{O}$-frame or $\mathcal{J}$-frame:
    \begin{itemize}
        \item $x$-axis $\vu{e}_{x,\mathcal{B}_i}$ is constructed as the bisector to the angle between the two \glspl{mosa} axes of spacecraft $i$.
        \item $z$-axis $\vu{e}_{z,\mathcal{B}_i}$ is the unit vector normal to the solar panel plane (defining the $xOy$ plane).
        \item $y$-axis $\vu{e}_{y,\mathcal{B}_i}$ is deduced from the two above.
        \item The origin $B_i$ of the frame $\mathcal{B}_i$ are the centers of mass of the spacecraft.
    \end{itemize}
    
   \item The $\mathcal{H^*}$-frames define the {\bfseries target} attitude of the \gls{mosa}s w.r.t to $\mathcal{J}$-frame.
    \begin{itemize}
        \item $x$-axis $\vu{e}_{x,\mathcal{H}^*_j}$ is the unit vector aligned to the axis normal to the incoming wavefront as intercepted by the telescope $j=(1,2)$ of the local spacecraft $i$. It can be deduced from a rotation of the unit vector $\framevu{O}{x}[i]$ around $\framevu{O}{z}[i]$ of an angle $[-1]^\exposant{j+1}\ \tfrac{\phi_{m}^{*}}{2}$ equal to half the {\it constellation}'s opening angle at local spacecraft $i$ location.
        \item $z$-axis $\vu{e}_{z,\mathcal{H}^*_j}$ is the unit vector normal to LISA constellation plane. It is equal to $\framevu{O}{z}[i]$.
        \item $y$-axis $\vu{e}_{y,\mathcal{H}^*_j}$ is deduced from the two above.
        \item The origin $H_j$ of the frame $\mathcal{H}_j$ is the geometrical center of the housing $j$ of the spacecraft $i$.
    \end{itemize}
    
    \item The $\mathcal{H}$-frames are rigidly attached to their respective \gls{mosa} and define their actual attitude, whether w.r.t to $\mathcal{H}^*$-frame, $\mathcal{B}$-frame, $\mathcal{O}$-frame or $\mathcal{J}$-frame.
    \begin{itemize}
        \item $x$-axis $\vu{e}_{x,\mathcal{H}_j}$ is the axis along which local OMS measurement of spacecraft to test-mass $j=(1,2)$ is performed. It is a drag-free axis. It can be deduced from a rotation of the unit vector $\framevu{B}{x}[i]$ around $\framevu{B}{z}[i]$ of an angle $[-1]^\exposant{j+1}\ \tfrac{\phi_{m}}{2}$ equal to half the \gls{mosa}'s opening angle of the spacecraft $i$.
        \item $z$-axis $\vu{e}_{z,\mathcal{H}_j}$ is the unit vector normal to the solar panel plane (defining the $(xOy)$ plane). It is equal to $\vu{e}_{z,\mathcal{B}_i}$ in the simulator.
        \item $y$-axis $\vu{e}_{y,\mathcal{H}_j}$ is deduced from the two above.
        \item The origin $H_j$ of the frame $\mathcal{H}_j$  is the geometrical center of the housing $j$ of the spacecraft $i$.
        \item The pivot points $P_j$ denote the center of rotation of the \gls{mosa} $j$ in the satellite $i$, and coincide with $H_j$ in the nominal, geometrical configuration.
    \end{itemize}
    
    \item The $\mathcal{T}$-frames are rigidly attached to the corresponding test-masses and describe their attitude, whether w.r.t to $\mathcal{H}$-frame, $\mathcal{B}$-frame, $\mathcal{O}$-frame or $\mathcal{J}$-frame:
    \begin{itemize}
        \item $x$-axis $\vu{e}_{x,\mathcal{T}_j}$ is constructed as the unit vector normal the $x$-faces of the \gls{tm} $j$ and aligned with $\vu{e}_{x,\mathcal{H}_j}$ when nominally oriented.
        \item $z$-axis $\vu{e}_{z,\mathcal{T}_j}$ is constructed as a unit vector normal to the $z$-faces of the \gls{tm}$j$ and aligned with $\vu{e}_{z,\mathcal{H}_j}$ when nominally oriented.
        \item $y$-axis $\vu{e}_{y,\mathcal{T}_j}$ is deduced from the two above.
        \item The origin $T_j$ of the frame $\mathcal{T}_j$ is the center of mass of the test-mass $j$.
    \end{itemize}
\end{itemize}

\begin{figure}[h!]
\frame{\centerline{\includegraphics[width=1.0\linewidth]{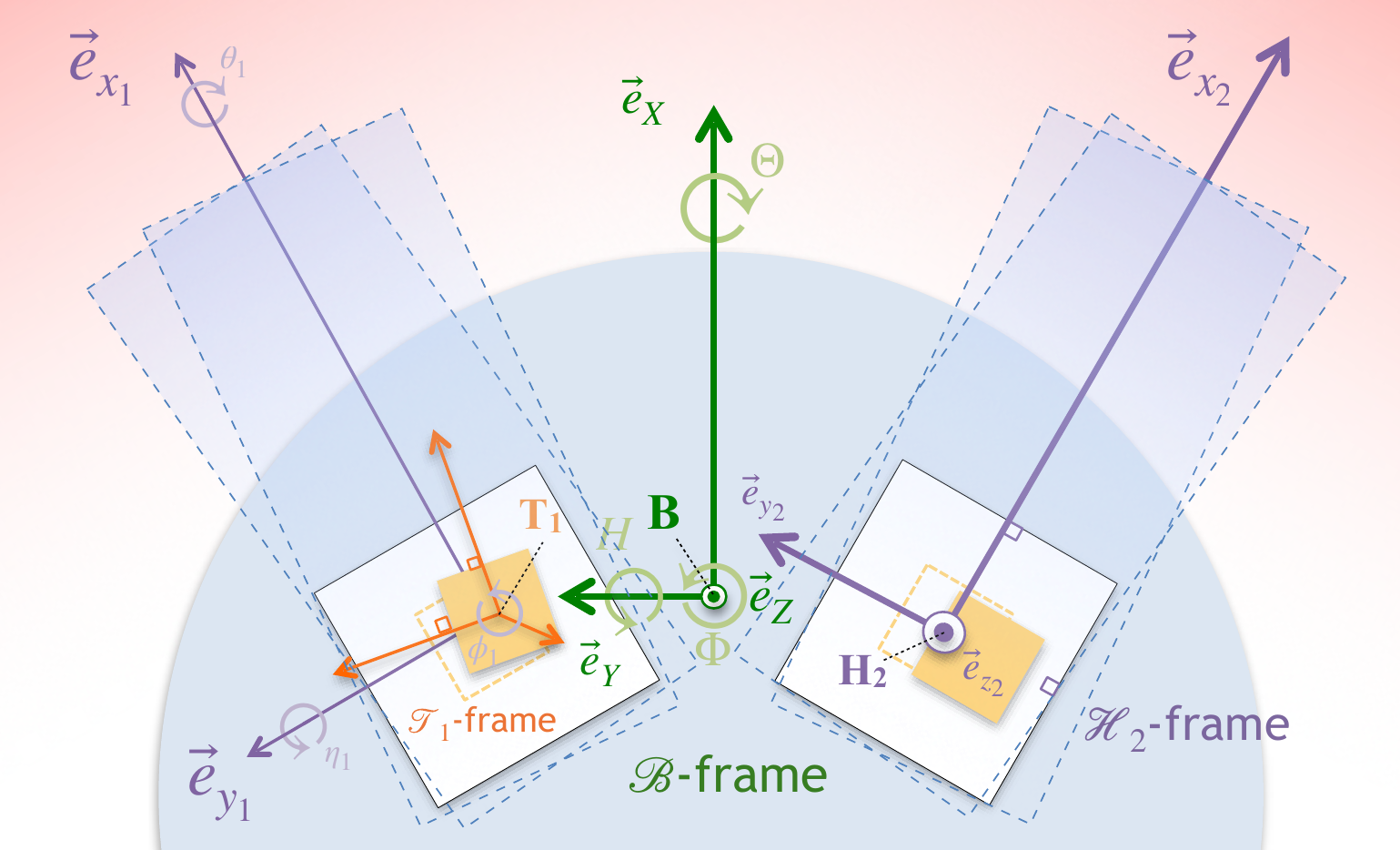}}}
\caption{Figure cut of spacecraft geometry on the $xOy$ plane of the spacecraft body frame $\mathcal{B}$. The reference frames attached to the spacecraft, the housings and the test-masses, resp. $\mathcal{B}$, $\mathcal{H}$ and $\mathcal{T}$, are drawn. Short notation has been preferred for readability to label the axes and coordinates associated to these frames, as explained in section \myhyperref{subsection: Dynamical state vector}, Equations (\myhyperref{eq: tm_posvecexp}), (\myhyperref{eq: tm_attvecexp}), (\myhyperref{eq: sc_attvecexp}) and (\myhyperref{eq: sc_angvelvecexp}).}
\label{figure: spacecraft}
\end{figure}

We finally mention the important mathematical relationship in Equation (\myhyperref{eq: Transport Theorem}), which will be used throughout the document: the so-called transport equation, sometimes called the Varignon formula, named after the late 17th century french mathematician Pierre Varignon. It relates time-derivatives of a given vector w.r.t. to different reference frames (here $\mathcal{B}$ and $\mathcal{J}$)

\begin{equation}
    \ddt{\ -\ }{\mathcal{J}} = \ddt{\ -\ }{\mathcal{B}} + \vangvel{B}{J} \times \left[\ -\ \right] \ .
\label{eq: Transport Theorem}
\end{equation}

\subsection{Geometrical construction of target frames}
\label{subsection: target frames}

While $\mathcal{J}$, $\mathcal{B}$, $\mathcal{H}$ and $\mathcal{T}$ reference frames are self-defining as each of them rely on existing bodies (distant stars, spacecraft, \gls{mosa}s and test-masses), the targeted frames $\mathcal{O}$ and $\mathcal{H}^{*}$ required a physical definition to be utilized as coordinate systems in LISA \gls{eom}. Their objective is to encode the nominal orientation of the spacecraft and the \gls{mosa}s w.r.t. the Galilean frame $\mathcal{J}$. They are formed using {\it idealized bodies}, that is, from the trajectory the spacecraft would follow if they were subject to solar system gravitational field only, hence freely following their respective geodesics. Indeed, given the dimension of the \gls{lisa} constellation compared to spacecraft dimension and noisy deviation from geodesics, target frames can reliably be constructed from orbits in a very solid approximation (hence neglecting in their definition the microscopic local jittering from geodesics).

As stated in section \myhyperref{subsection: Frames of reference}, $\mathcal{O}$ frame $X$-axis is defined as the bisector of the constellation from the spacecraft $i$ at study. The bisector vector, $\mylongvec{O}[i]{I}[i]$, for spacecraft $i$ is then
\begin{equation}
    \begin{split}
        \mylongvec{O}[i]{I}[i] = \myvec{r}{ij} + \left[ \frac{r_{ij}}{r_{ij}+r_{ik}} \right] \myvec{r}{jk} = \myvec{r}{ij} + u_i \myvec{r}{jk}\ ,
    \end{split}
\label{eq: exBasis}
\end{equation}
using the angle bisector theorem, and where $\myvec{r}{ij}$ is the relative position between two of the three spacecraft (see Equation (\myhyperref{eq: r-distance})), which are labeled with $\{i, j, k\}$ letters triplet, $i$ denoting the local spacecraft, $j$ the distant spacecraft facing \gls{tm}$1$ of spacecraft $i$, and $k$ facing its \gls{tm}$2$ (cf. Figure \myhyperref{figure: constellation}).

\begin{equation}
    \myvec{r}{ij} = \myvec{r}{j} - \myvec{r}{i} = \myvec{r}{O_\indice{j}/J} - \myvec{r}{O_\indice{i}/J}\ .
\label{eq: r-distance}
\end{equation}
Therefore, the $\framevu{O}{x}[i]$ basis vector is defined as 
\begin{equation}
    \framevu{O}{x}[i] = \frac{\mylongvec{O}[i]{I}[i]}{\norm{\mylongvec{O}[i]{I}[i]}} \ .
\label{eq: exvec}
\end{equation}
Finally, the $\framevu{O}{y}[i]$ and $\framevu{O}{z}[i]$ basis vectors are determined by
\begin{align}
    &\framevu{O}{z}[i] = \frac{\mylongvec{O}[i]{I}[i] \times \myvec{r}{jk}}{\norm{\mylongvec{O}[i]{I}[i] \times \myvec{r}{jk}}}\ , \label{eq: ezvec} \\
    &\framevu{O}{y}[i] = \framevu{O}{z}[i] \times \framevu{O}{x}[i]\ . \label{eq: eyvec}
\end{align}

The basis vectors of the $\mathcal{O}$ frame are now completely defined from orbital position data $\myvec{r}{i}$, $\myvec{r}{j}$, and $\myvec{r}{k}$. The dynamics of this frame---that is, angular velocity and acceleration---relatively to the Galilean frame $\mathcal{J}$ will come into play when deriving \gls{lisa} \gls{eom}. To avoid numerical difficulties arising from the differentiation of orbits necessary to compute these angular quantities, we should derive them purely symbolically from orbital positions, velocities and accelerations w.r.t. $\mathcal{J}$-frame.

Indeed, angular velocity can be related to the rate of changes of basis vector (mathematical proof in appendix \myhyperref{appendix: angvel}) from the following expression
\begin{align}
    \begin{split}
    \vangvel{O}[i]{J} =\ & \frac{1}{2} \ \framevu{O}{x}[i] \times \ddt{\framevu{O}{x}[i]}{J} \\
    &+ \frac{1}{2} \ \framevu{O}{y}[i] \times \ddt{\framevu{O}{y}[i]}{J}
    \\
    &+ \frac{1}{2} \ \framevu{O}{z}[i] \times \ddt{\framevu{O}{z}[i]}{J}\ ,
    \end{split}
    \label{eq: AngularVelAnalytical}
\end{align}
which then implies an analytical differentiation of expressions (\myhyperref{eq: exvec}), (\myhyperref{eq: eyvec}) and (\myhyperref{eq: ezvec}) w.r.t. time. These basis vectors being functions $\va{f}$ of orbital elements only---that is $\framevu{O}{x}[i] = \va{f}_{e_x} (\myvec{r}{i}, \myvec{r}{j}, \myvec{r}{k})$ taking the $X$-axis of $\mathcal{O}$-frame as example---it is clear that the angular velocity and acceleration can also be written as mere functions $\va{f}$ of orbital positions, velocities and accelerations, respectively $\vangvel{O}[i]{J} = \va{f}_{\omega}(\myvec{r}{i,j,k}, \myvec{v}{i,j,k})$ and $\ddt{\vangvel{O}[i]{J}}{J} = \va{f}_{\dot{\omega}} (\myvec{r}{i,j,k}, \myvec{v}{i,j,k}, \myvec{w}{i,j,k})$. Starting with differentiation of Equation (\myhyperref{eq: exvec}), we have
\begin{equation}
    \begin{split}
        \ddt{\framevu{O}{x}[i]}{J} = \frac{\text{O}_\indice{i} \text{I}_\indice{i}\dv{\mylongvec{O}[i]{I}[i]{}}{t}-\dv{\text{O}_\indice{i} \text{I}_\indice{i}}{t}\mylongvec{O}[i]{I}[i]}{\text{O}_\indice{i} \text{I}_\indice{i}^{2}}
    \end{split}
\label{eq: ExDeriv}
\end{equation}
from which one can find its contribution to  $\va{f}_{\omega}(\myvec{r}{i,j,k}, \myvec{v}{i,j,k})$ in expanding Equation (\myhyperref{eq: ExDeriv}) using
\begin{align}
       & \text{O}_\indice{i} \text{I}_\indice{i} = \sqrt{(\myvec{r}{ij}+u_{i}\myvec{r}{jk})\cdot(\myvec{r}{ij}+u_{i}\myvec{r}{jk})}\ , \\
       & \dv{\mylongvec{O}[i]{I}[i]}{t} = \myvec{v}{ij}+{\dot{u}_{i}}\myvec{r}{jk}+u_{i}\myvec{v}{jk}\ , \\
       & \dv{\text{O}_\indice{i} \text{I}_\indice{i}}{t} = \dv{\mylongvec{O}[i]{I}[i]}{t} \cdot \framevu{O}{x}[i] \ , \\
       & {\dot{u}_{i}} = \frac{(r_{ik}+r_{ij})\dot{r}_{ij}-r_{ij}(\dot{r}_{ij}+\dot{r}_{ik})}{(r_{ij}+r_{ik})^{2}}\ , \\
       & r_{ij} = \sqrt{\myvec{r}{ij}\cdot\myvec{r}{ij}} \ , \hspace{1cm} \dot{r}_{ij} = \frac{\myvec{v}{ij}\cdot\myvec{r}{ij}}{r_{ij}} \ .
\end{align}

Now treating the $Z$-axis, we define the following intermediate vectorial and scalar terms
\begin{align}
        & \!\!\!\!\myvec{w}{i} = \mylongvec{O}[i]{I}[i] \times \myvec{r}{jk}\ , && w_{i} = \norm{\myvec{w}{i}} \\
        & \!\!\!\!\dv{\myvec{w}{i}}{t} = \dv{\mylongvec{O}[i]{I}[i]}{t}\times\myvec{r}{jk}+\mylongvec{O}[i]{I}[i]\times\myvec{v}{jk}\ , && \dot{w}_{i} = \frac{\dv{\myvec{w}{i}}{t} \cdot \myvec{w}{i}}{w_{i}}
\label{eq: aEzVec}
\end{align}
and the time derivative of the basis vector $\framevu{O}{z}[i]$ is simply written as
\begin{equation}
    \ddt{\framevu{O}{z}[i]}{J} = \frac{ w_{i} \dv{\myvec{w}{i}}{t} - \dot{w}_{i} \myvec{w}{i} }{w_i^{2}}\ ,
\label{eq: ezO_dot}
\end{equation}
from which the deduction of the derivative of the third axis is straightforward
\begin{equation}
    \begin{split}
        \ddt{\framevu{O}{y}[i]}{J} = \, &\ddt{\framevu{O}{z}[i]}{J} \times \framevu{O}{x}[i] \\
        &+ \framevu{O}{z}[i] \times \ddt{\framevu{O}{x}[i]}{J}\, .
    \end{split}
\end{equation}

We now have derived all the terms necessary to expand the $\mathcal{O}$-frame angular velocity in Equation (\myhyperref{eq: AngularVelAnalytical}) as an analytical function $\va{f}_{\omega}(\myvec{r}{i,j,k}, \myvec{v}{i,j,k})$ of the constellation's orbital positions and velocities only.

The angular accelerations require further expansion of Equation (\myhyperref{eq: AngularVelAnalytical}). Without giving full details, one can find an expression of the angular acceleration vector as a function of the second order time derivatives of the basis vectors
\begin{align}
    \!\!\ddt{\vangvel{O}[i]{J}}{\mathcal{J}} =\, & \frac{1}{2} \ \framevu{O}{x}[i] \times \ddtddt{\framevu{O}{x}[i]}{J} \nonumber
    \\
    &+ \frac{1}{2} \ \framevu{O}{y}[i] \times \ddtddt{\framevu{O}{y}[i]}{J}
    \\
    &+ \frac{1}{2} \ \framevu{O}{z}[i] \times \ddtddt{\framevu{O}{z}[i]}{J}\ . \nonumber
\label{eq: AngularVelAnalyticalExpand}
\end{align}
Treating the $X$-axis first, Equation (\myhyperref{eq: ExDeriv}) is differentiated again, which gives
\begin{align}
    \!\!\ddtddt{\framevu{O}{x}[i]}{J} =\, & \frac{\dv[2]{\mylongvec{O}[i]{I}[i]{}}{t}}{\text{O}_\indice{i} \text{I}_\indice{i}}\, -\, \frac{\left[ 2\dv{\text{O}_\indice{i} \text{I}_\indice{i}}{t}\dv{\mylongvec{O}[i]{I}[i]{}}{t} + \dv[2]{\text{O}_\indice{i} \text{I}_\indice{i}}{t}\mylongvec{O}[i]{I}[i]{} \right]}{\text{O}_\indice{i} \text{I}_\indice{i}^{2}} \notag
    \\
    & \ +\  \frac{2\left(\dv{\text{O}_\indice{i} \text{I}_\indice{i}}{t}\right)^{2}\mylongvec{O}[i]{I}[i]{}}{\text{O}_\indice{i} \text{I}_\indice{i}^{3}} \ ,
\end{align}
where
\begin{align}
    &\dv[2]{\mylongvec{O}[i]{I}[i]{}}{t} = \myvec{a}{ij}+\ddot{u}_i\myvec{r}{jk}+2\dot{u}_i\myvec{v}{jk}+u_{i}\myvec{a}{jk} \ ,\\
    &\dv[2]{\text{O}_\indice{i} \text{I}_\indice{i}}{t} = \dv{\mylongvec{O}[i]{I}[i]{}}{t} \cdot \ddt{\framevu{O}{x}[i]}{J} + \dv[2]{\mylongvec{O}[i]{I}[i]{}}{t} \cdot \framevu{O}{x}[i]\ , \\
    &{\ddot{u}_{i}} = \frac{\ddot{r}_{ij}r_{ik}-\ddot{r}_{ik}r_{ij}}{(r_{ij}+r_{ik})^{2}} \nonumber \\
    &\qquad - \frac{2(\dot{r}_{ij}r_{ik}-\dot{r}_{ik}r_{ij})(\dot{r}_{ij}+\dot{r}_{ik})}{(r_{ij}+r_{ik})^{3}}\ , \\
    &\ddot{r}_{ij} = \frac{\left[(\myvec{a}{ij}\cdot\myvec{r}{ij})+(\myvec{v}{ij}\cdot\myvec{v}{ij})\right]r_{ij}-(\myvec{v}{ij}\cdot\myvec{r}{ij}) \ \dot{r}_{ij}}{r_{ij}^{2}}\ .
\end{align}

The derivatives of the variables in Equation (\myhyperref{eq: aEzVec}) are evaluated as
\begin{align}
    & \!\!\dv[2]{\myvec{w}{i}}{t} = \dv[2]{\mylongvec{O}[i]{I}[i]}{t} \times \myvec{r}{jk} + 2\dv{\mylongvec{O}[i]{I}[i]}{t} \times \myvec{v}{jk} + \mylongvec{O}[i]{I}[i] \times \myvec{a}{jk}\, , \\
    & \!\!\ddot{w}_{i} = \frac{ \left( \dv[2]{\myvec{w}{i}}{t} \cdot \myvec{w}{i} + \dv{\myvec{w}{i}}{t} \cdot \dv{\myvec{w}{i}}{t}\right)w_{i}-\left(\dv{\myvec{w}{i}}{t}\cdot\myvec{w}{i}\right)\dot{w}_{i}}{w_{i}^{2}} .
\end{align}

The second-order rate of change of the $\framevu{O}{z}[i]$-basis can now be calculated with
\begin{equation}
    \begin{split}
    \!\!\!\ddtddt{\framevu{O}{z}[i]}{J}\! = \!\frac{\dv[2]{\myvec{w}{i}}{t}}{w_{i}}-\frac{\myvec{w}{i}\ddot{w_{i}}}{w_{i}^{2}}-\frac{2\dv{\myvec{w}{i}}{t}\dot{w_{i}}}{w_{i}^{2}}+\frac{2\myvec{w}{i}\dot{w_{i}}^{2}}{w_{i}^{3}}\,.
    \end{split}
\end{equation}
Additionally, the second-order rate of change of the $\framevu{O}{y}[i]$-basis is given by
\begin{equation}
    \begin{split}
        \ddtddt{\framevu{O}{y}[i]}{J} =\,
        & \ddtddt{\framevu{O}{z}[i]}{J} \times \framevu{O}{x}[i] \\
        &+ 2 \ \ddt{\framevu{O}{z}[i]}{J} \times \ddt{\framevu{O}{x}[i]}{J} \\
        &+ \framevu{O}{z}[i] \times \ddtddt{\framevu{O}{x}[i]}{J}\ .
    \end{split}
\end{equation}

Finally, we have all the materials needed to build the function $\va{f}_{\dot{\omega}}(\myvec{r}{i,j,k}, \myvec{v}{i,j,k}, \myvec{a}{i,j,k})$ giving the target angular acceleration of the spacecraft as a function of orbital positions, velocities, and accelerations. This is an interesting result, as now, target attitude, angular velocity and acceleration are derivable purely analytically from the orbit information. It fully solves the question of numerical treatment of orbits regarding attitude in LISA dynamics simulations.

We will not treat the case of the target \gls{mosa} frames $\mathcal{H}_\indice{1}^{*}$ and $\mathcal{H}_\indice{2}^{*}$ in this article, as the procedure would be mathematically identical. Starting from the $\mathcal{O}$-frame, the derivation would consist in adding another rotation to the $\mathcal{O}$-frame corresponding to $\pm \sfrac{1}{2}$ the constellation opening angle $\phi^{*}_{m}$.

\subsection{Dynamical state vector}
\label{subsection: Dynamical state vector}
In the previous sections, we have listed and discussed all the reference frames used in the dynamics modeling, which were introduced to account for any actual or targets entities being non-rigid w.r.t. each other. It also defines the set of coordinates with which we describe longitudinal or angular displacement of the bodies involved. For each quantity, a system of coordinates will be preferred, mainly driven by the reference frame in which those dynamical \glspl{dof} are observed.

\begin{itemize}

    \item Position and orientation of test-masses are sensed either through \gls{ifo} or \gls{grs} which can be to first approximation assumed to be attached rigidly to the housing frames $\mathcal{H}_1$ and $\mathcal{H}_2$. The nominal position for the test-masses are the centers of the housings $H_1$ and $H_2$, and their nominal attitude is to be aligned with the respective $\mathcal{H}$-frame. Consequently, test-mass $1$ dynamics are described by the vector quantity $[\myvec{r}{T_\indice{1}/H_\indice{1}}, \cardanvec{T}[1]{H}[1]]$ (respectively for test mass $2$), where $\myvec{r}{T_\indice{1}/H_\indice{1}}$ denotes a position vector from point $H_\indice{1}$ to $T_\indice{1}$ ($\overrightarrow{H_\indice{1} T_\indice{1}}$) and $\cardanvec{T}[1]{H}[1]$ is an attitude pseudo-vector complying with the Cardan representation for a rotation (ZYX convention) \cite{diebel_representing_2006}. The position vectors are to be expressed in the corresponding housing frames, which gives six scalar quantities per test-mass
    \begin{align}
        \posvecexp{T}[1]{H}[1]{H}[1] =&
        \begin{bmatrix}
            \myvec{r}{T_\indice{1}/H_\indice{1}} \cdot \framevu{H}{x}[1] \\ \myvec{r}{T_\indice{1}/H_\indice{1}} \cdot \framevu{H}{y}[1] \\ \myvec{r}{T_\indice{1}/H_\indice{1}} \cdot \framevu{H}{z}[1]
        \end{bmatrix}
        =
        \begin{bmatrix}
            \myexpr{x}{T_\indice{1}/H_\indice{1}}{\mathcal{H}_\indice{1}} \\ \myexpr{y}{T_\indice{1}/H_\indice{1}}{\mathcal{H}_\indice{1}} \\ \myexpr{z}{T_\indice{1}/H_\indice{1}}{\mathcal{H}_\indice{1}},
        \end{bmatrix}
        =
        \begin{bmatrix}
            x_\indice{1} \\
            y_\indice{1} \\
            z_\indice{1}
        \end{bmatrix}\, ,
    \label{eq: tm_posvecexp}
    \end{align}
    
    \begin{align}
    \cardanvec{T}[1]{H}[1] =&
    \begin{bmatrix}
    \theta_\indice{{\mathcal{T}_\indice{1}/\mathcal{H}_\indice{1}}} \\ \eta_\indice{{\mathcal{T}_\indice{1}/\mathcal{H}_\indice{1}}} \\ \phi_\indice{{\mathcal{T}_\indice{1}/\mathcal{H}_\indice{1}}}
    \end{bmatrix}
    =
    \begin{bmatrix}
        \theta_\indice{1} \\
        \eta_\indice{1} \\
        \phi_\indice{1}
    \end{bmatrix}\, ,
\label{eq: tm_attvecexp}
\end{align}
identically for test-mass $2$ as well. Our convention uses $\theta$, $\eta$, and $\phi$ as rotations around $x$, $y$, and $z$ axes of a given frame. Test-mass velocities are represented as time-derivatives of the position vector w.r.t. their expression (observation) frame, that is, w.r.t. the housing frames
\begin{align}
    \myvec{v}{T_\indice{1}/H_\indice{1}} &= \ddt{\myvec{r}{T_\indice{1}/H_\indice{1}}}{H}[1]\, , \\
    \myexpr{v}{T_\indice{1}/H_\indice{1}}{\mathcal{H}_\indice{1}} &= \velvecexp{T}[1]{H}[1]{H}[1]
    =
    \begin{bmatrix}
        \myexpr{\dot{x}}{T_\indice{1}/H_\indice{1}}{\mathcal{H}_\indice{1}} \\
        \myexpr{\dot{y}}{T_\indice{1}/H_\indice{1}}{\mathcal{H}_\indice{1}} \\
        \myexpr{\dot{z}}{T_\indice{1}/H_\indice{1}}{\mathcal{H}_\indice{1}}
    \end{bmatrix}
    =
    \begin{bmatrix}
        \dot{x}_\indice{1} \\
        \dot{y}_\indice{1} \\
        \dot{z}_\indice{1}
    \end{bmatrix} \, ,
\end{align}
and respectively for test-mass 2. We stress again that, throughout this article, the {\it dot} symbols over a quantity refer only to time derivative w.r.t. its reference frame of expression.

    \item Spacecraft attitude will refer to rotation w.r.t. the $\mathcal{O}$-frame in the final form of the \gls{eom}, that is, to the quantity $\myvec{\alpha}{\mathcal{B}/\mathcal{O}}$, since its working point is zero. However, intermediate steps will involve angular velocity w.r.t. $\mathcal{J}$-frame. It follows the Cardan angles of the spacecraft, written as
    \begin{align}
        \myvec{\alpha}{\mathcal{B}/\mathcal{O}} =
        \begin{bmatrix}
            \theta_\indice{\mathcal{B}/\mathcal{O}} \\
            \eta_\indice{\mathcal{B}/\mathcal{O}} \\
            \phi_\indice{\mathcal{B}/\mathcal{O}}
        \end{bmatrix}
        =
        \begin{bmatrix}
            \Theta \\
            H \\
            \Phi
        \end{bmatrix}\, .
    \label{eq: sc_attvecexp}
    \end{align}

    \item Angular velocities of all the objects---either fiducial or actual---can appear in the \gls{eom} either w.r.t. the Galilean frame, target or actual other body frames. They must ultimately be expressed in the reference frame attached to the rotating body itself, since it will greatly simplify the rotational equation of motion, as inertial tensors are static in such frames. They can be expressed as
    \begin{align}
    \myexpr{\omega}{\mathcal{B}/\mathcal{O}}{\mathcal{B}} &=
    \begin{bmatrix}
        \myvec{\omega}{\mathcal{B}/\mathcal{O}} \cdot \framevu{B}{x} \\ \myvec{\omega}{\mathcal{B}/\mathcal{O}} \cdot \framevu{B}{y} \\ \myvec{\omega}{\mathcal{B}/\mathcal{O}} \cdot \framevu{B}{z}
    \end{bmatrix}
    =
    \begin{bmatrix}
        \myexpr{\omega}{x, \mathcal{B}/\mathcal{O}}{\mathcal{B}} \\
        \myexpr{\omega}{y, \mathcal{B}/\mathcal{O}}{\mathcal{B}} \\
        \myexpr{\omega}{z, \mathcal{B}/\mathcal{O}}{\mathcal{B}}
    \end{bmatrix}
    =
    \begin{bmatrix}
        \omega_{X} \\
        \omega_{Y} \\
        \omega_{Z}
    \end{bmatrix} \, ,
    \label{eq: sc_angvelvecexp} \\
    \myexpr{\omega}{\mathcal{T}_\indice{1}/\mathcal{H}_\indice{1}}{\mathcal{T}_\indice{1}} \!\!&=
    \begin{bmatrix}
        \vangvel{T}[1]{H}[1] \cdot \framevu{T}{x}[1] \\ \vangvel{T}[1]{H}[1] \cdot \framevu{T}{y}[1]
        \\ \vangvel{T}[1]{H}[1] \cdot \framevu{T}{z}[1]
    \end{bmatrix}
    =
    \begin{bmatrix}
        \myexpr{\omega}{x, \mathcal{T}_\indice{1}/\mathcal{H}_\indice{1}}{\mathcal{T}_\indice{1}} \\
        \myexpr{\omega}{y, \mathcal{T}_\indice{1}/\mathcal{H}_\indice{1}}{\mathcal{T}_\indice{1}} \\
        \myexpr{\omega}{z, \mathcal{T}_\indice{1}/\mathcal{H}_\indice{1}}{\mathcal{T}_\indice{1}}
    \end{bmatrix}\, .
    \end{align}

    \item Finally, the attitude and angular velocity of the \glspl{mosa} (labeled as {\it telescopes} in the equations) deserve a dedicated attention. They can be defined w.r.t their target reference frames $\mathcal{H}^{*}$ similarly to the spacecraft attitude---i.e. the frames w.r.t. which \glspl{mosa} orientation are observed based on the incoming wave fronts (see section \myhyperref{subsection: metrology}). However, it is also relevant to use angular coordinates in the spacecraft body frame $\mathcal{B}$, since this is the natural frame of the actuation mechanism, which only leaves a single degree of freedom per \gls{mosa} (rotations around $\framevu{H}{z}[1]$ and $\framevu{H}{z}[2]$) while approximating the other four fixed. Following this \gls{mosa} mechanism property in the simulation, we are using the following angular coordinates for the \glspl{mosa} in the \gls{eom}
    \begin{align}
        \myvec{\alpha}{\mathcal{H}_\indice{1}/\mathcal{B}} &=
        \begin{bmatrix}
            \theta_\indice{\mathcal{H}_\indice{1}/\mathcal{B}} \\
            \eta_\indice{\mathcal{H}_\indice{1}/\mathcal{B}} \\
            \phi_\indice{\mathcal{H}_\indice{1}/\mathcal{B}}
        \end{bmatrix}
        =
        \begin{bmatrix}
            0.0 \\
            0.0 \\
            \sfrac{\pi}{6} + \delta \phi_{tel, 1}
        \end{bmatrix}\, , \\
        \myexpr{\omega}{\mathcal{H}_\indice{1}/\mathcal{B}}{\mathcal{H}_\indice{1}} &=
        \begin{bmatrix}
            \vangvel{H}[1]{B} \cdot \framevu{H}{x}[1] \\ \vangvel{H}[1]{B} \cdot \framevu{H}{y}[1]
            \\ \vangvel{H}[1]{B} \cdot \framevu{H}{z}[1]
        \end{bmatrix}
        =
        \begin{bmatrix}
            \myexpr{\omega}{x, \mathcal{H}_\indice{1}/\mathcal{B}}{\mathcal{H}_\indice{1}} \\
            \myexpr{\omega}{y, \mathcal{H}_\indice{1}/\mathcal{B}}{\mathcal{H}_\indice{1}} \\
            \myexpr{\omega}{z, \mathcal{H}_\indice{1}/\mathcal{B}}{\mathcal{H}_\indice{1}}
        \end{bmatrix}
        =
        \begin{bmatrix}
            0.0 \\
            0.0 \\
            \delta \dot{\phi}_{tel, 1}
        \end{bmatrix}\, ,
    \end{align}
    where it is made clear that only $1$ \gls{dof} per \gls{mosa} remains dynamical. On the other hand, \gls{dws} will project \gls{mosa} dynamics into a different system of coordinates---implicitly containing the spacecraft angular motion---which can be denoted by 
    \begin{align}
        \myvec{\alpha}{\mathcal{H}_\indice{1}/\mathcal{H}_\indice{1}^\exposant{*}} &=
        \begin{bmatrix}
            \theta_\indice{\mathcal{H}_\indice{1}/\mathcal{H}_\indice{1}^\exposant{*}} \\
            \eta_\indice{\mathcal{H}_\indice{1}/\mathcal{H}_\indice{1}^\exposant{*}} \\
            \phi_\indice{\mathcal{H}_\indice{1}/\mathcal{H}_\indice{1}^\exposant{*}}
        \end{bmatrix} \, ,\\
        \myexpr{\omega}{\mathcal{H}_\indice{1}/\mathcal{H}_\indice{1}^\exposant{*}}{\mathcal{H}_\indice{1}} &=
        \begin{bmatrix}
            \vangvel{H}[1]{H}[1][*] \cdot \framevu{H}{x}[1]
            \\
            \vangvel{H}[1]{H}[1][*] \cdot \framevu{H}{y}[1]
            \\
            \vangvel{H}[1]{H}[1][*] \cdot \framevu{H}{z}[1]
        \end{bmatrix}
        =
        \begin{bmatrix}
            \myexpr{\omega}{x, \mathcal{H}_\indice{1}/\mathcal{H}_\indice{1}^\exposant{*}}{\mathcal{H}_\indice{1}} \\
            \myexpr{\omega}{y, \mathcal{H}_\indice{1}/\mathcal{H}_\indice{1}^\exposant{*}}{\mathcal{H}_\indice{1}} \\
            \myexpr{\omega}{z, \mathcal{H}_\indice{1}/\mathcal{H}_\indice{1}^\exposant{*}}{\mathcal{H}_\indice{1}}
        \end{bmatrix}\, ,
    \end{align}
\end{itemize}
and again identically for test-mass 2. This is an important clarification as the reference frames in which the \gls{mosa} orientation is considered can be the source of confusion, when actual ($\mathcal{H}$, $\mathcal{B}$) or fiducial bodies ($\mathcal{H}^\exposant{*}$, $\mathcal{O}$) are mistaken. We opt here for the convention of referring to actual bodies for writing the dynamics \gls{eom}s, and referring to targeted frames when considering \gls{dws} sensing. 

Finally, we note that the absolute position of the spacecraft in the solar system $\myvec{r}{B/J}$ is not part of the state vector, since this quantity is completely decoupled from the system dynamics, aside from setting the level of solar system gravity gradient on-board the spacecraft, which one can nevertheless approximate precisely enough using orbital, fiducial positions only ($\myvec{r}{O/J}$).

Gathering all these dynamical \glspl{dof} within a single state vector, one can fully represent the dynamical state of the system with:

\begin{align}
    \va{X}=
    \begin{matrix}
        \Big[
        & \myvec{\alpha}{\mathcal{B}/\mathcal{O}} & \myexpr{\omega}{\mathcal{B}/\mathcal{O}}{\mathcal{B}} & \posvecexp{T}[1]{H}[1]{H}[1] & \cardanvec{T}[1]{H}[1] & \posvecexp{T}[2]{H}[2]{H}[2]
         \\
        & \ \ \cardanvec{T}[2]{H}[2] &
        \velvecexp{T}[1]{H}[1]{H}[1] &
        \myexpr{\omega}{\mathcal{T}_\indice{1}/\mathcal{H}_\indice{1}}{\mathcal{T}_\indice{1}} & \velvecexp{T}[2]{H}[2]{H}[2] & \vangvelexp{T}[2]{H}[2]{T}[2]
         \\
        & \delta \phi_{tel, 1} & \delta \dot{\phi}_{tel, 1} & \delta \phi_{tel, 2} & \delta \dot{\phi}_{tel, 2}
        & \Big] \hspace*{1.3em}
    \end{matrix}
\label{eq: state vector}
\end{align}

\section{LISA Equations of Motion}
\label{section: eoms}
Equation (\myhyperref{eq: state vector}) shows a state vector which fully describes the dynamical state of the spacecraft-telescopes-test-masses system of a given LISA satellite. This state vector is the solution of a second order differential system---the \acrlong{eom}---relating the longitudinal and angular displacement of the spacecraft, telescopes and test-masses to the environmental and command forces and torques they are exposed to. It is important to stress that the dynamics of the three LISA satellites are treated independently, being $2.5$ million kilometers apart. The incoming spherical wavefront are indeed quasi-independent of distant spacecraft rotation, and the dynamics of the three spacecraft can interact with each other through wavefront defects only. Such defects could in principle impact attitude control locked on the \gls{dws} channels, hence contribute to spacecraft attitude jitter. Their effect is however, considerably outweighed by other contributors, especially micro-propulsion noise.

We can distinguish four types of \gls{eom} in LISA Dynamics, among which one requires scrutiny and consequently a detailed derivation: the longitudinal dynamics of the test-masses.
This \gls{eom} regarding the \gls{tm} motion is particularly complex as it introduces several nested frames and has the most stringent performance requirements in terms of residual motion.
The three other types will be angular \gls{eom}, to which angular velocities of the spacecraft, the \gls{mosa}s, and the test-masses will be solutions, provide critical information in the scope of tilt-to-length effects analysis.

\subsection{Test-mass longitudinal motion}
\label{subsection: Test mass longitudinal motion}
Starting with test-mass longitudinal motion---a treatment applicable to both test masses by symmetry---the first equation of motion is derived from Newton's third law
\begin{equation}
    \ddtddt{\myvec{r}{T/J}}{\mathcal{J}} = \sum \frac{\myvec{f}{T}}{m_{T}}\, ,
\label{eq: Newton - Test Mass}
\end{equation}
which is valid only w.r.t. a Galilean frame, here chosen to be $\mathcal{J}$, and where $\myvec{f}{T}$ is the net force vector applied to the test-mass and $m_{T}$ is its mass. The next steps will then merely consist in expanding Equation (\myhyperref{eq: Newton - Test Mass}) in order to express quantities in the correct frames of observation, or in other words, involving elements of the state vector (\myhyperref{eq: state vector}) only. The vector $\myvec{r}{T/J}$ is first decomposed as
\begin{equation}
    \myvec{r}{T/J} = \myvec{r}{T/H} + \myvec{r}{H/B} + \myvec{r}{B/J}\, .
\label{eq: TM pos vector}
\end{equation}
We recognize the double time derivative of $\myvec{r}{B/J}$ to be related to the spacecraft dynamics, and again Newton's third law gives
\begin{equation}
    \ddtddt{\myvec{r}{B/J}}{\mathcal{J}} = \sum \frac{\myvec{f}{B}}{m_{B}} \, , 
\label{eq: Newton - Spacecraft}
\end{equation}
where $m_{B}$ is the mass of the spacecraft. While the norm of the vectors $\myvec{r}{H/B}$ is constant---the \gls{mosa} pivot points $P$ are assumed to be coincident with the center of the housing $H$---their orientation is dynamical and will impact the frame of observations of test-mass displacement. Using the transport theorem from Equation (\myhyperref{eq: Transport Theorem}), we find
\begin{align}
\begin{split}
    \ddtddt{\myvec{r}{H/B}}{J} =
    & \ddtddt{\myvec{r}{H/B}}{B} + 2 \vangvel{B}{J} \times \ddt{\myvec{r}{H/B}}{B} \\ 
    & + \ddt{\vangvel{B}{J}}{J} \times \myvec{r}{H/B} \\
    & + \vangvel{B}{J} \times \left( \vangvel{B}{J} \times \myvec{r}{H/B} \right)\, .
\end{split}
\label{eq: ddtddt_J r_HB}
\end{align}
This can be simplified by remarking that for a static spacecraft \gls{com} $B$, the norm of the position vector $\norm{\myvec{r}{P/B}}$ is static in the body frame $\mathcal{B}$.
Only a non-nominal offset of the point $H$ w.r.t to the pivot point $P$ may render $\myvec{r}{H/B}$ dynamical in the spacecraft body frame. It is however important to enable such imperfection in the model and to introduce the position offset $\myvec{r}{H/P}$ between the pivot point $P$ and the housing center $H$ of a given \gls{mosa} in Equation (\myhyperref{eq: housing to pivot offset}). Lever arm effects coupling the noisy \gls{mosa} attitude to the highly stable test-mass longitudinal \gls{dof} are critical dynamical features to account for \cite{lpf_platform_stability}. Thus,
\begin{align}
    \begin{split}
    \ddt{\myvec{r}{H/B}}{B} & = \ddt{\myvec{r}{H/P}}{B} + \ddt{\myvec{r}{P/B}}{B}  \\
    & = \ddt{\myvec{r}{H/P}}{B} = \vangvel{H}{B} \times \myvec{r}{H/P}\, .
    \end{split}
\label{eq: housing to pivot offset}
\end{align}

At this stage, the model still neglects the time variations of the distribution of mass in the satellite due to \gls{mosa} rotations, as well as, for instance, due to gas depletion for $\si{\micro\newton}$ thruster system. Examining the contribution of \gls{mosa} dynamics, one can argue that the \gls{mosa} represents a substantial fraction of the mass of the spacecraft, and that a $\pm 1 \si{\degree}$ yearly modulation of the opening angle changes the \gls{com} and the inertia tensor of the spacecraft enough to impact significantly the longitudinal dynamics of the test-mass (through levers), as well as, the rotational dynamics of the spacecraft.

However, we expect such contributions to test-mass dynamics to be second order terms, and therefore they are not currently included in the simulation. We decided not to introduce an additional layer of complexity in the body of this document, although the full \gls{eom} are provided in appendix \myhyperref{appendix: eom} and future works will investigate the order of magnitude of such contributions, which are, to our knowledge, still untreated in the literature.

The last position term of Equation (\myhyperref{eq: TM pos vector}) to be treated is $\myvec{r}{T/H}$ which has to be differentiated w.r.t. the housing frame $\mathcal{H}$. From the transport equation, we simply have
\begin{equation}
\begin{split}
    \ddtddt{\myvec{r}{T/H}}{J} =
    & \ddtddt{\myvec{r}{T/H}}{H} + 2 \myvec{\omega}{\mathcal{H}/\mathcal{J}} \times \ddt{\myvec{r}{T/H}}{H} \\
    & + \ddt{\vangvel{H}{J}}{J} \times \myvec{r}{T/H} \\
    & + \myvec{\omega}{\mathcal{H}/\mathcal{J}} \times \left( \myvec{\omega}{\mathcal{H}/\mathcal{J}} \times \myvec{r}{T/H} \right)\, ,
\end{split}
\end{equation}

where the classical inertial terms, namely Coriolis, Euler, and centrifugal forces, show up. Decomposing the housing angular velocity as $\myvec{\omega}{\mathcal{H}/\mathcal{J}} = \vangvel{H}{B} + \vangvel{B}{J}$, and simplifying cross-product terms thanks to the Jacobi identity,
\begin{equation}
    \vb{a} \times \left( \vb{b} \times \vb{c} \right) + \vb{b} \times \left( \vb{c} \times \vb{a} \right) + \vb{c} \times \left( \vb{a} \times \vb{b} \right) = \vb{0}\, ,
\label{eq: Jacobi Identity}
\end{equation}
we get
\begin{align}
     \ddtddt{\myvec{r}{T/H}}{J} & =  \ddtddt{\myvec{r}{T/H}}{H} + 2 \vangvel{H}{B} \times \ddt{\myvec{r}{T/H}}{H} 
    \\
    + \ & 2 \vangvel{B}{J} \times \ddt{\myvec{r}{T/H}}{H} + \ddt{\vangvel{H}{B}}{B} \times \myvec{r}{T/H} \nonumber
    \\
    + \ & \ddt{\vangvel{B}{J}}{J} \times \myvec{r}{T/H} + \vangvel{H}{B} \times \left( \vangvel{H}{B} \times \myvec{r}{T/H} \right) \nonumber
    \\
    + \ & \ \vangvel{B}{J} \times \left( \vangvel{B}{J} \times \myvec{r}{T/H} \right) \nonumber
    \\
    + \ & \ 2 \vangvel{B}{J} \times \left( \vangvel{H}{B} \times \myvec{r}{T/H} \right)\, . \nonumber
\end{align}
Hence, putting back all the terms of \myhyperref{eq: Newton - Test Mass} and \myhyperref{eq: Newton - Spacecraft} together, one writes finally the vectorial test-mass equation of motion \myhyperref{eq: TM Long EOM - Vector}, where dynamics of bodies are considered w.r.t. their frame of observations, hence progressing towards a constraint of the state vector in Equation (\myhyperref{eq: state vector}). The missing steps, that is, the introduction of fiducial frames and expression in specific coordinate systems, will be addressed in section \myhyperref{section: linearization}. This gives

\begin{widetext}
\begin{equation}
    \begin{aligned}
        & \ddtddt{\myvec{r}{T/H}}{H} + 2 \vangvel{H}{B} \times \ddt{\myvec{r}{T/H}}{H} + 2 \vangvel{B}{J} \times \ddt{\myvec{r}{T/H}}{H} + \ddt{\vangvel{H}{B}}{B} \times \myvec{r}{T/H} + \ddt{\vangvel{B}{J}}{J} \times \myvec{r}{T/H} \\
        & + 2 \vangvel{B}{J} \times \left( \vangvel{H}{B} \times \myvec{r}{T/H} \right) + \vangvel{H}{B} \times \left( \vangvel{H}{B} \times \myvec{r}{T/H} \right) + \vangvel{B}{J} \times \left( \vangvel{B}{J} \times \myvec{r}{T/H} \right) + \ddt{\vangvel{H}{B}}{\mathcal{B}} \times \myvec{r}{H/P} \\
        & + \vangvel{H}{B} \times (\vangvel{H}{B} \times \myvec{r}{H/P}) + 2 \vangvel{B}{J} \times \left( \vangvel{H}{B} \times \myvec{r}{H/P} \right) + \ddt{\vangvel{B}{J}}{J} \times \myvec{r}{H/B} + \vangvel{B}{J} \times \left( \vangvel{B}{J} \times \myvec{r}{H/B} \right) \\
        & = \sum \frac{\myvec{f}{T}}{m_{T}} - \sum \frac{\myvec{f}{B}}{m_{B}}\ .
    \end{aligned}
\label{eq: TM Long EOM - Vector}
\end{equation}
\end{widetext}

\subsection{Angular equations of motion of spacecraft and test-masses}
\label{subsection: angular eom sc and tm}
Analogously to Section \myhyperref{subsection: Test mass longitudinal motion}, we start from Euler's equation, derivable from Newtonian mechanics of a rigid body and related to the conservation of angular momentum $\va{h}$ of an isolated body w.r.t. a Galilean frame. Treating the spacecraft dynamics first, we have
\begin{equation}
    \ddt{\myvec{h}{\text{sc}/\mathcal{J}}}{\mathcal{J}} = \ddt{\itensor{sc}{B} \ \vangvel{B}{J}}{\mathcal{J}} = \sum \myvec{t}{B}\, ,
\label{eq: Euler equation - SC}
\end{equation}
where $\itensor{sc}{B}$ is the inertia tensor matrix of the spacecraft computed at its \gls{com}, $B$, and $\myvec{t}{B}$ are the external torques applied to the spacecraft body. Again, Equation (\myhyperref{eq: Euler equation - SC}) will be expanded so that quantities appear from the viewpoint of the frames in which they are observed. Firstly, angular dynamics is most conveniently treated from the body frame, where the inertia tensor is by definition static. Using the transport theorem (\myhyperref{eq: Transport Theorem}) leads to
\begin{equation}
    \ddt{\itensor{sc}{B} \ \vangvel{B}{J}}{\mathcal{B}} + \vangvel{B}{J} \times \left( \itensor{sc}{B} \ \vangvel{B}{J} \right) = \sum \myvec{t}{B}\, .
\label{eq: Euler equation - SC - B frame}
\end{equation}

Here we encounter a similar difficulty as in section \myhyperref{subsection: Test mass longitudinal motion}, that is the mass distribution in the spacecraft is not strictly static, for the \gls{mosa} rotating along with constellation orbital breathing. However, accounting for this non-static mass distribution will introduce a large amount of second order terms, expected to be of little impact on spacecraft jitter dynamics. As before, we derive the model as currently implemented, and consequently, we will ignore this additional layer of complexity in this section. The interested reader will find more information about a proposed treatment in the appendix \myhyperref{appendix: sc ang dyn}.

Hence, assuming $\itensor{sc}{B}$ is time-invariant in the $\mathcal{B}$ frame, Equation (\myhyperref{eq: Euler equation - SC - B frame}) leads to the vectorial equation of motion constraining the spacecraft attitude
\begin{equation}
    \itensor{sc}{B} \ddt{\vangvel{B}{J}}{\mathcal{B}} + \vangvel{B}{J} \times \left( \itensor{sc}{B} \ \vangvel{B}{J} \right) = \sum \myvec{t}{B}\, .
\label{eq: S/C Angular EOM - Vector}
\end{equation}

Examining the test-mass case, we derive an identical relationship from similar arguments
\begin{equation}
\begin{split}
    \itensor{tm}{T} \ddt{\myvec{\omega}{\mathcal{T}/\mathcal{J}}}{H} - \itensor{tm}{T} \left( \vangvel{T}{H} \times \myvec{\omega}{\mathcal{T}/\mathcal{J}} \right) \\
    + \myvec{\omega}{\mathcal{T}/\mathcal{J}} \times \left( \itensor{tm}{T} \myvec{\omega}{\mathcal{T}/\mathcal{J}} \right) = \sum \myvec{t}{T}\, ,
\end{split}
\end{equation}
where we have used additional geometrical properties of the cubic test-masses, which have scalar inertia matrices $\itensorexp{tm}{T}{T} = \lambda\, \identite{3}$ in their body frames $\mathcal{T}$. However, further decomposition is needed since $\myvec{\omega}{\mathcal{T}/\mathcal{J}}$ still contains the spacecraft and \gls{mosa} rotational \gls{dof} as $\myvec{\omega}{\mathcal{T}/\mathcal{J}} = \vangvel{T}{H} + \myvec{\omega}{\mathcal{H}/\mathcal{B}} + \myvec{\omega}{\mathcal{B}/\mathcal{J}}$, which leads to
\begin{equation}
\begin{split}
    & \itensor{tm}{T} \ddt{\vangvel{T}{H}}{H} + \itensor{tm}{T} \ddt{\vangvel{H}{B}}{B} \\
    & + \itensor{tm}{T} \ddt{\vangvel{B}{J}}{J} - \itensor{tm}{T} \left( \vangvel{H}{B} \times \vangvel{B}{J} \right) \\
    & - \itensor{tm}{T} \left( \vangvel{T}{H} \times \vangvel{H}{B} \right) - \itensor{tm}{T} \left( \vangvel{T}{H} \times \vangvel{B}{J} \right) \\
    & = \sum \myvec{t}{T}\ .
\end{split}
\label{eq: TM Angular EOM - Vector}
\end{equation}

\subsection{Angular equations of motion of MOSAs}
\label{subsection: angular eom mosa}
Finally, addressing the \gls{mosa} angular dynamics, we again start from the Euler equation
\begin{equation}
    \ddt{\itensor{mo}{Q} \  \vangvel{H}{J}}{J} = \sum \myvec{t}{H}\, ,
\label{eq: Euler equation - MOSA}
\end{equation}
which we expand to consider the angular acceleration w.r.t. the \gls{mosa} frame---in which its inertia tensor is static---and in breaking down the angular velocity $\vangvel{H}{J} = \vangvel{H}{B} + \vangvel{B}{J}$ to bring out the dynamics components that belong to the spacecraft and the \gls{mosa} independently
\begin{align}
    & \itensor{mo}{Q} \ddt{\vangvel{H}{B}}{B} + \itensor{mo}{Q} \ddt{\vangvel{B}{J}}{J}
    \\
    & - \itensor{mo}{Q} \left( \vangvel{H}{B} \times \vangvel{B}{J} \right) + \vangvel{H}{B} \times \left( \itensor{mo}{Q} \  \vangvel{H}{B} \right) \notag\\
    & + \vangvel{H}{B} \times \left( \itensor{mo}{Q} \  \vangvel{B}{J} \right) + \vangvel{B}{J} \times \left( \itensor{mo}{Q} \  \vangvel{H}{B} \right)
    \notag\\
    & + \vangvel{B}{J} \times \left( \itensor{mo}{Q} \  \vangvel{B}{J} \right) = \sum \myvec{t}{H}\ .\notag
\end{align}

At this point, we have to examine the sum of applied torques, as for this \gls{eom} we want to focus on the telescope dynamics induced by the pointing mechanism. Indeed, if the spacecraft rotates through the thrust system, structure torque will be applied on the telescope such that they follow the spacecraft motion. The pointing mechanism only concerns relative motion between the \gls{mosa} and the spacecraft. We have accordingly
\begin{equation}
    \sum \myvec{t}{H} = \sum \myvec{t}{H}^{\,\text{rel}} + \myvec{t}{H}^{\,\text{struct}}\ .
\end{equation}
We can define $\myvec{t}{H}^{\,\text{struct}}$ as the torque necessary for the \gls{mosa} to rigidly follow the spacecraft in its rotation w.r.t. $\mathcal{J}$-frame
\begin{equation}
\begin{split}
    \myvec{t}{H}^{\,\text{struct}}\!
     =& \ddt{\itensor{mo}{Q} \  \vangvel{B}{J}}{J}
    \\
     =&\, \itensor{mo}{Q} \ddt{\vangvel{B}{J}}{J} - \itensor{mo}{Q}\! \left( \vangvel{H}{B} \times \vangvel{B}{J} \right) \\
    &+ \vangvel{H}{B} \times \left( \itensor{mo}{Q} \  \vangvel{B}{J} \right) \\
    &+ \vangvel{B}{J} \times \left( \itensor{mo}{Q} \  \vangvel{B}{J} \right)\, .
\end{split}
\end{equation}
Then it follows that
\begin{equation}
\begin{split}
    \itensor{mo}{Q} \ddt{\vangvel{H}{B}}{B} + \vangvel{H}{B} \times \left( \itensor{mo}{Q} \  \vangvel{H}{B} \right) \\
    + \vangvel{B}{J} \times \left( \itensor{mo}{Q} \  \vangvel{H}{B} \right) = \sum \myvec{t}{H}^{\,\text{rel}}\, .
\end{split}
\label{eq: MOSA Angular EOM - Vector}
\end{equation}

Alternatively, we could have found an identical expression in observing that
\begin{equation}
    \ddt{\itensor{mo}{Q} \  \vangvel{H}{B}}{J} = \sum \myvec{t}{H}^{\,\text{rel}}\, ,
\label{eq: Euler equation rel - MOSA}
\end{equation}
where only torques that cause relative rotation between the \gls{mosa} and the spacecraft are considered.

\section{Working point, Linearization and State Equations of Dynamics}
\label{section: linearization}

\subsection{Introduction of target attitude frames}
\label{subsection: EOM target frames}

The equations of motion \myhyperref{eq: TM Long EOM - Vector}, \myhyperref{eq: S/C Angular EOM - Vector}, \myhyperref{eq: TM Angular EOM - Vector} and \myhyperref{eq: MOSA Angular EOM - Vector} are obviously non-linear. They involve several quadratic terms caused by various inertial forces / torques and projections to the reference frame of observations. As previously discussed in section \myhyperref{subsection: Dynamical state vector} a convenient and yet reliable way to simplify system dynamics is to choose a state-space representation where dynamical quantities at play are deviations from their target values---or {\it working point}. It is particularly relevant in the presence of feedback systems, whose objective precisely is to lock the observables to target values. In \gls{lisa}, the \gls{dfacs} action will guarantee that the position and orientation of the four dynamical bodies stay locked to their working point, meaning that the nominal value for the deviation from the working point will be $0.0$.

Such parameterization eases and justifies the linearization of the system, while preventing possible numerical issues arising when different scales of motion are to be compared---since it decouples small-scale and large-scale motion by construction. It requires the introduction of the three target frames described in sections \myhyperref{subsection: Frames of reference} and \myhyperref{subsection: target frames} which allow us to break down the spacecraft and \gls{mosa} rotations into low-frequency angular motion and in-band jittering
\begin{align}
    & \vangvel{B}{J} = \vangvel{B}{O} + \vangvel{O}{J}\, , \label{eq: omega_B breakdown} \\
    & \vangvel{H}{B} = \vangvel{H}{H^{*}} + \vangvel{H^{*}}{O} + \vangvel{O}{J} \, .\label{eq: omega_H breakdown}
\end{align}

The decomposition of \gls{mosa} rotation is not used in the \gls{eom}, as for angular dynamical state the relative velocity between spacecraft and telescope is used (see subsection \myhyperref{subsection: angular eom mosa}). However, the decomposition of spacecraft angular velocity (Equation (\myhyperref{eq: omega_B breakdown})) is more critical for this section. In particular, derivatives such as equations (\myhyperref{eq: omega_B_O derivative breakdown}) and (\myhyperref{eq: omega_B_J derivative breakdown}) have to be broken down in order for the dynamical state of interest to show up---a general rule being that the angular velocities and accelerations must be computed relative to the same frame. Thus,
\begin{align}
    & \ddt{\vangvel{B}{O}}{J} = \ddt{\vangvel{B}{O}}{O} + \vangvel{O}{J} \times \vangvel{B}{O}\, , \label{eq: omega_B_O derivative breakdown} \\
    & \ddt{\vangvel{O}{J}}{B} = \ddt{\vangvel{O}{J}}{J} + \vangvel{O}{J} \times \vangvel{B}{O}\, . \label{eq: omega_B_J derivative breakdown}
\end{align}
This decomposition of rotating frames increases the number of inertial force and torque terms to address, although several of them compensate for each other and can be simplified through the Jacobi identity (Equation (\myhyperref{eq: Jacobi Identity})) such as for instance, the following inertial terms
\begin{align}
\label{eq: Jacobi O/J - O/B - T/H}
    & \vangvel{O}{J} \times \left( \vangvel{B}{O} \times \myvec{r}{T/H} \right)  = \\
    & \left( \vangvel{O}{J} \times \vangvel{B}{O} \right) \times \myvec{r}{T/H} +  \vangvel{B}{O} \times \left( \vangvel{O}{J} \times \myvec{r}{T/H} \right)\, . \nonumber
\end{align}

\vspace{0.5cm}

We apply these composition rules to all the equations of motion \myhyperref{eq: TM Long EOM - Vector}, \myhyperref{eq: TM Angular EOM - Vector}, \myhyperref{eq: S/C Angular EOM - Vector} and \myhyperref{eq: MOSA Angular EOM - Vector} derived in section \myhyperref{section: eoms}. These are tedious but straightforward expansion steps, and we will not detail them in this document for readability. They have been, however, documented and cross-checked with \texttt{Mathematica}. A dedicated \href{git@gitlab.in2p3.fr:hinchaus/lisaeom.git}{\texttt{gitlab project}} containing these \texttt{Mathematica} validation notebooks is accessible to LISA consortium members on-demand.

\subsection{Expression in a common system of coordinates}
\label{subsection: EOM expression}

In addition to these further expansions, it is now necessary to express the vectorial quantities into a specific system of coordinates. As discussed and motivated in details in section \myhyperref{subsection: Dynamical state vector}, we are using the following rules:

\begin{itemize}
    \item Translational dynamics of the test-masses are expressed in the housing frame coordinates systems, resp. $\mathcal{H}_\indice{1}$ and $\mathcal{H}_\indice{2}$.
    \item Spacecraft angular \gls{eom}s are expressed in the spacecraft body frame $\mathcal{B}$.
    \item Test-mass angular \gls{eom}s are expressed in their respective body frames $\mathcal{T}_\indice{1}$ and $\mathcal{T}_\indice{2}$
    \item Telescope opening motion is expressed in the \gls{mosa} body frames $\mathcal{H}_\indice{1}$ and $\mathcal{H}_\indice{2}$.
\end{itemize}
It means consequently that rotation matrices need to be introduced whenever an \gls{eom} expressed in a specific frame involves physical quantities that are preferably expressed in a different system of coordinates. For example, spacecraft angular velocities in Equation (\myhyperref{eq: TM Long EOM - Vector}) require rotations into $\mathcal{H}$ frames
\begin{align}
    & \vangvel{B}{O} \overset{\mathcal{H}}{=} \vangvelexp{B}{O}{H} = \rotmat{H}{B} \ \vangvelexp{B}{O}{B}\, \label{eq: omega_B_O expressed} ,
    \\
    & \vangvel{O}{J} \overset{\mathcal{H}}{=} \vangvelexp{O}{J}{H} = \rotmat{H}{B} \ \rotmat{B}{O} \ \rotmat{O}{J} \ \vangvelexp{O}{J}{J} \, \label{eq: omega_O_J expressed} ,
\end{align}
where rotation matrices are functions of the relative orientation of reference frames $\rotmat{B}{O} = \rotmat{B}{O} (\cardanvec{B}{O})$ as following
\begin{align}
    \!\!T \left( \va{\alpha} \right)\! &=\!
    \begin{bmatrix}
        & 1 & 0 & 0 \\
        & 0 & \cost{\theta} & \sint{\theta} \\
        & 0 & -\sint{\theta} & \cost{\theta} \\
    \end{bmatrix}\!\!
    \begin{bmatrix}
        & \cost{\eta} & 0 & -\sint{\eta} \\
        & 0 & 1 & 0 \\
        & \sint{\eta} & 0 & \cost{\eta}
    \end{bmatrix}\!\!
    \begin{bmatrix}
        & \cost{\phi} & \sint{\phi} & 0 \\
        & -\sint{\phi} & \cost{\phi} & 0 \\
        & 0 & 0 & 1
    \end{bmatrix}\notag
\end{align}
\begin{align}
    & =\!
    \begin{bmatrix}
        & \cost{\eta} \cost{\phi} & \cost{\eta} \sint{\phi} & -\sint{\eta} \\
        & \sint{\theta} \sint{\eta} \cost{\phi} - \cost{\theta} \sint{\phi} & \sint{\theta} \sint{\eta} \sint{\phi} + \cost{\theta} \cost{\phi} & \sint{\theta} \cost{\eta} \\
        & \cost{\theta} \sint{\eta} \cost{\phi} + \sint{\theta} \sint{\phi} & \cost{\theta} \sint{\eta} \sint{\phi} - \sint{\theta} \cost{\phi} & \cost{\theta} \cost{\eta}  \\
    \end{bmatrix}\, ,
\end{align}
adopting a short notation for sine and cosine functions (e.g. $-\sin{\theta} = -s_\theta$), and again, comply with the ZYX Cardan sequence convention \cite{diebel_representing_2006}. The matrices $T(\myvec{\alpha}{})$ and their respective rotation angles $\myvec{\alpha}{} = [\theta, \eta, \phi]$ encode for a transformation of the system of coordinates w.r.t. which the vectors are expressed: they correspond to {\it passive rotations} of vectors. We will be using the convention throughout this work.

Expressing vectors w.r.t. a specific basis now transforms tensors to matrices of $3^K$ components where $K$ is the tensor order. Consequently, and as already introduced in paragraphs above, vectors become triplets of scalars encoding coordinates in a given frame. Similarly, cross products $\vb{a} \times \vb{b}$ can be written in a specific coordinate system as a matrix product between a skew-symmetric matrix $\myskew{a}$ made out of the vector $\vb{a}$ and the target vector $\vb{b}$
\begin{align}
    \!\!\vb{a} \times \vb{b} \overset{\mathcal{H}}{=} \myskew{a^\exposant{\mathcal{H}}} b^\exposant{\mathcal{H}} =
    \begin{bmatrix}
        & 0 & -\myexpr{a}{z}{\mathcal{H}} & \myexpr{a}{y}{\mathcal{H}} \\
        & \myexpr{a}{z}{\mathcal{H}} & 0 & -\myexpr{a}{x}{\mathcal{H}} \\
        & - \myexpr{a}{y}{\mathcal{H}} & \myexpr{a}{x}{\mathcal{H}} & 0
    \end{bmatrix}
    \begin{bmatrix}
        & \myexpr{b}{x}{\mathcal{H}} \\
        & \myexpr{b}{y}{\mathcal{H}} \\
        & \myexpr{b}{z}{\mathcal{H}}
    \end{bmatrix}\,\! .
\end{align}
\vspace*{1pt}

We have now all the required material to assemble the test-mass equation of motion expressed in their reference frames of observation. Equation (\myhyperref{eq: TM Long EOM - Vector}) shows such \gls{eom}, generically written so that it holds for both test-mass $1$ and $2$---indexing $T$, $H$ bodies and indexing $\mathcal{T}$, $\mathcal{H}$ frames with a specific test-mass label. The vectorial test-mass equation is then
\begin{widetext}
\vspace*{5pt}
\begin{equation}
    \begin{aligned}
        & \accvecexp{T}{H}{H} + 2 \myskew{\vangvelexp{H}{B}{H}} \velvecexp{T}{H}{H} + 2 \myskew{\rotmat{H}{B}\vangvelexp{B}{O}{B}} \velvecexp{T}{H}{H} + 2 \myskew{\rotmat{H}{B} \rotmat{B}{O} \rotmat{O}{J} \vangvelexp{O}{J}{J}} \velvecexp{T}{H}{H}
        + \myskew{\vangaccexp{H}{B}{H}} \posvecexp{T}{H}{H} \\
        & + \myskew{ \rotmat{H}{B} \vangaccexp{B}{O}{B} } \posvecexp{T}{H}{H} + \myskew{ \rotmat{H}{B} \rotmat{B}{O} \rotmat{O}{J} \vangaccexp{O}{J}{J} } \posvecexp{T}{H}{H} + 2 \myskew{ \rotmat{H}{B} \vangvelexp{B}{O}{B} } \myskew{ \vangvelexp{H}{B}{H} } \posvecexp{T}{H}{H} \\
        & + 2 \myskew{ \rotmat{H}{B} \rotmat{B}{O} \rotmat{O}{J} \vangvelexp{O}{J}{J} } \myskew{ \vangvelexp{H}{B}{H} } \posvecexp{T}{H}{H} + \myskew{ \vangvelexp{H}{B}{H} } \myskew{ \vangvelexp{H}{B}{H} } \posvecexp{T}{H}{H} \\
        & + \myskew{ \rotmat{H}{B} \vangvelexp{B}{O}{B} } \myskew{ \rotmat{H}{B} \vangvelexp{B}{O}{B} } \posvecexp{T}{H}{H} + 2 \myskew{ \rotmat{H}{B} \rotmat{B}{O} \rotmat{O}{J} \vangvelexp{O}{J}{J} } \myskew{ \rotmat{H}{B} \vangvelexp{B}{O}{B} } \posvecexp{T}{H}{H} + \myskew{ \rotmat{H}{B} \rotmat{B}{O} \rotmat{O}{J} \vangvelexp{O}{J}{J} } \myskew{ \rotmat{H}{B} \rotmat{B}{O} \rotmat{O}{J} \vangvelexp{O}{J}{J} } \posvecexp{T}{H}{H} \\
        & - \myskew{ \posvecexp{H}{P}{H}} \vangaccexp{H}{B}{H} + \myskew{\vangvelexp{H}{B}{H}}\myskew{\vangvelexp{H}{B}{H}}\posvecexp{H}{P}{H} + 2 \myskew{ \myskew{ \posvecexp{H}{P}{H} } \vangvelexp{H}{B}{H} } \rotmat{H}{B} \vangvelexp{B}{O}{B} - 2 \myskew{ \rotmat{H}{B} \rotmat{B}{O} \rotmat{O}{J} \vangvelexp{O}{J}{J} } \myskew{ \posvecexp{H}{P}{H} } \vangvelexp{H}{B}{H} \\
        & - \myskew{ \rotmat{H}{B} \posvecexp{H}{B}{B} } \rotmat{H}{B} \vangaccexp{B}{O}{B} + \myskew{ \myskew{ \rotmat{H}{B} \posvecexp{H}{B}{B} } \rotmat{H}{B} \vangvelexp{B}{O}{B} } \rotmat{H}{B} \vangvelexp{B}{O}{B} - 2 \myskew{ \rotmat{H}{B} \rotmat{B}{O} \rotmat{O}{J} \vangvelexp{O}{J}{J} } \myskew{ \rotmat{H}{B} \posvecexp{H}{B}{B} } \rotmat{H}{B} \vangvelexp{B}{O}{B} \\
        & = \sum \frac{\force{T}{H}}{m_{T}} - \sum \frac{\rotmat{H}{B} \force{B}{B}}{m_{B}} - \myskew{ \rotmat{H}{B} \rotmat{B}{O} \rotmat{O}{J} \vangaccexp{O}{J}{J} } \rotmat{H}{B}  \posvecexp{H}{B}{B} - \myskew{ \rotmat{H}{B} \rotmat{B}{O} \rotmat{O}{J} \vangvelexp{O}{J}{J} } \myskew{ \rotmat{H}{B} \rotmat{B}{O} \rotmat{O}{J} \vangvelexp{O}{J}{J} } \rotmat{H}{B}  \posvecexp{H}{B}{B}\ .
    \end{aligned}
\label{eq: TM Long EOM - Projected}
\end{equation}
\end{widetext}

We apply the same procedure to the angular \gls{eom} of the spacecraft, the test-mass, and the \gls{mosa}, and we get the following projected \gls{eom}:
\clearpage
\begin{itemize}
    \item For the spacecraft attitude
    \begin{equation}
        \begin{split}
            & \itensorexp{sc}{B}{B} \vangaccexp{B}{O}{B} + \itensorexp{sc}{B}{B} \myskew{ \rotmat{B}{O} \rotmat{O}{J} \vangvelexp{O}{J}{J} } \vangvelexp{B}{O}{B} \\
            & + \myskew{ \vangvelexp{B}{O}{B} } \itensorexp{sc}{B}{B} \vangvelexp{B}{O}{B} \\
            &- \myskew{ \itensorexp{sc}{B}{B} \rotmat{B}{O} \rotmat{O}{J} \vangvelexp{O}{J}{J} } \vangvelexp{B}{O}{B} \\
            & + \myskew{ \rotmat{B}{O} \rotmat{O}{J} \vangvelexp{O}{J}{J} } \itensorexp{sc}{B}{B} \vangvelexp{B}{O}{B}
            \\
            & =  \sum \torque{B}{B} - \itensorexp{sc}{B}{B} \rotmat{B}{O} \rotmat{O}{J} \vangaccexp{O}{J}{J} \\
            &\ \ \ \ - \myskew{ \rotmat{B}{O} \rotmat{O}{J} \vangvelexp{O}{J}{J} } \itensorexp{sc}{B}{B} \rotmat{B}{O} \rotmat{O}{J} \vangvelexp{O}{J}{J}\, .
        \end{split}
    \label{eq: S/C Angular EOM - Projected}
    \end{equation}
    
    \item For the test-mass attitude
    \begin{equation}
        \begin{split}
            & \itensorexp{tm}{T}{T} \vangaccexp{T}{H}{T} + \itensorexp{tm}{T}{T} \rotmat{T}{H} \vangaccexp{H}{B}{H} + \itensorexp{tm}{T}{T} \rotmat{T}{H} \rotmat{H}{B} \vangaccexp{B}{O}{B} \\
            &+ \itensorexp{tm}{T}{T} \rotmat{T}{H} \rotmat{H}{B} \myskew{ \rotmat{B}{O} \rotmat{O}{J} \vangvelexp{O}{J}{J} } \vangvelexp{B}{O}{B} \\
            &+ \itensorexp{tm}{T}{T} \rotmat{T}{H} \rotmat{H}{B} \rotmat{B}{O} \rotmat{O}{J} \vangaccexp{O}{J}{J}\\
            &- \itensorexp{tm}{T}{T} \rotmat{T}{H} \myskew{ \vangvelexp{H}{B}{H} } \rotmat{H}{B} \vangvelexp{B}{O}{B} \\
            &- \itensorexp{tm}{T}{T} \rotmat{T}{H} \myskew{ \vangvelexp{H}{B}{H} } \rotmat{H}{B} \rotmat{B}{O} \rotmat{O}{J} \vangvelexp{O}{J}{J} \\
            &- \itensorexp{tm}{T}{T} \myskew{ \vangvelexp{T}{H}{T} } \rotmat{T}{H} \vangvelexp{H}{B}{H} \\
            &- \itensorexp{tm}{T}{T} \myskew{ \vangvelexp{T}{H}{T} } \rotmat{T}{H} \rotmat{H}{B} \vangvelexp{B}{O}{B} \\
            &- \itensorexp{tm}{T}{T} \myskew{ \vangvelexp{T}{H}{T} } \rotmat{T}{H} \rotmat{H}{B} \rotmat{B}{O} \rotmat{O}{J} \vangvelexp{O}{J}{J} = \sum \torque{T}{T}\, .
        \end{split}
    \label{eq: TM Angular EOM - Projected}
    \end{equation}
    
    \item For the \gls{mosa} attitude
    \begin{equation}
        \begin{split}
            & \itensorexp{mo}{Q}{H} \vangaccexp{H}{B}{H} + \myskew{ \vangvelexp{H}{B}{H} } \itensorexp{mo}{Q}{H} \vangvelexp{H}{B}{H} \\
            &+ \myskew{ \rotmat{H}{B} \vangvelexp{B}{O}{B} } \itensorexp{mo}{Q}{H} \vangvelexp{H}{B}{H} \\
            &+ \myskew{ \rotmat{H}{B} \rotmat{B}{O} \rotmat{O}{J} \vangvelexp{O}{J}{J} } \itensorexp{mo}{Q}{H} \vangvelexp{H}{B}{H} \\
            &= \sum \torque{H}{\text{rel,} H}\ .
        \end{split}
    \label{eq: MOSA Angular EOM - Projected}
    \end{equation}
\end{itemize}

\subsection{State-space representation}
\label{subsection: state space representation}
Equations \myhyperref{eq: TM Long EOM - Projected}, \myhyperref{eq: S/C Angular EOM - Projected}, \myhyperref{eq: TM Angular EOM - Projected} and \myhyperref{eq: MOSA Angular EOM - Projected} brought together provide a second order $17$ differential system fully describing the dynamics of a single \gls{lisa} spacecraft. To facilitate its implementation and integration, it is convenient to split these $17$ second order equations into $34$ first order equations using a state-space representation, introducing explicitly velocity terms in the differential equation as an intermediate step, as can be illustrated by
\begin{align}
    & a = \dv[2]{x}{t} & \iff & &
    \begin{cases}
        v = \dv{x}{t} \\
        a = \dv{v}{t}
    \end{cases}\ .
\label{eq: 2nd to 1st order diff}
\end{align}
The velocity part of the state equations acts as identification relationships or mapping system between the second components of the state vector $\va{X} = [x, v]$ and the first components of its derivative $\dot{\va{X}} = [v, a]$. In our case, most of these relationships are straightforward---filled with $1$s and $0$s, aside from the mapping between angular velocity and Cardan angle rates (reckoning that our angular state representation is made of $[\va{\alpha}, \va{\omega}]$ pairs) which relate each other non-trivially
\begin{align}
    \begin{split}
    \vangvelexp{B}{O}{B} &= E \left( \cardanvec{B}{O} \right) \cardanvecdot{B}{O} \\
    &=
    \begin{pmatrix}
        & 1 & 0 & -\sint{\eta}\\
        & 0 & \cost{\theta} & \cost{\eta} \sint{\theta} \\
        & 0 & -\sint{\theta} & \cost{\eta} \cost{\theta}
    \end{pmatrix}
    \begin{bmatrix}
        & \thetacardandot{B}{O} \\
        & \etacardandot{B}{O} \\
        & \phicardandot{B}{O}
    \end{bmatrix}\ ,
    \end{split}
\label{eq: angle rates to angvel}
\end{align}
(for a Cardan ZYX rotation convention \cite{diebel_representing_2006}). The matrix $E$ in the Equation (\myhyperref{eq: angle rates to angvel}) provides the mapping between the angular state and its derivative, and is not merely the identity matrix $\unitmat_{3}$ in the general case.

In this representation, the dynamical state is fully represented by the state vector $\va{X}$ in Equation (\myhyperref{eq: state vector}) gathering all the dynamical longitudinal and angular states and their respective derivatives---hence 34 elements in total. The equations of dynamics can now be written
\begin{equation}
    \dv{\va{X}}{t} = \va{f}\!\left( \va{X}, \va{u}, t \right) = A( \va{X}, t ) \: \va{X}(t) + B( \va{X}, t ) \: \va{u}(t)
\end{equation}
in the general case, and where we have introduced the so-called state matrices $A$ and $B$ and the source terms vector $\va{u}(t)$---here forces and torques, that is, right-hand terms of \gls{eom} (\myhyperref{eq: TM Long EOM - Projected}), (\myhyperref{eq: S/C Angular EOM - Projected}), (\myhyperref{eq: TM Angular EOM - Projected}) and (\myhyperref{eq: MOSA Angular EOM - Projected}). This matrix form is particularly useful and powerful when the system is linear and time-invariant, as in such case $A$ and $B$ are constant matrices and most questions are then reducible to problems of matrix algebra (solving, controllability and observability, stabilization, and control design...).

At this stage, the \gls{eom} are not yet linearized. Although the introduction of the target frames at section \myhyperref{subsection: EOM target frames} has greatly facilitated the process, as large-scale motion is already well separate from small-scale jittering of bodies.
We treat the terms of the \gls{eom} that are quadratic in the dynamical state---that is, involving products of elements of $X$---writing the element with the largest fluctuation as the rightmost factor in each term. The variations of the other elements of $X$ are averaged around their target point and enter into the constant A matrix.
Because most elements $x_{i}$ of the state vector $\va{X}$ have as target point $0.0$, these terms either vanish or have a trivial treatment, for instance: $\left[ \rotmat{B}{O} ( \cardanvec{B}{O} ) \right]_\indice{\text{target}} = \rotmat{O}{O} = \mathbb{1}_{3}$. State-independent, time-varying terms are treated as source terms and incorporated in $\va{u}(t)$ and have been moved consequently on right-hand side of the \glspl{eom}. The only dynamical state which is non-vanishing when set at target are the orientation and angular velocities of the \gls{mosa} $\left[ \cardanvec{H}{B} \: \vangvelexp{H}{B}{H} \right]$, since its dynamics have been described w.r.t. spacecraft body frame and not its target frames. Nevertheless, these angular parameters are being treated similarly during the linearization process
\begin{align}
    \begin{bmatrix}
        \cardanvec{H}{B} & \vangvelexp{H}{B}{H}
    \end{bmatrix}_{\text{target}}
    =
    \begin{bmatrix}
        \cardanvec{H^*}{O} & \vangvelexp{H^*}{O}{H^*}
    \end{bmatrix}\, .
\end{align}

In addition, to get a linear, time-invariant system that has constant $A$ and $B$ matrices with minimal approximation, time-varying multipliers of the states have to be treated and approximated. Thanks to our expansion using target frames discussed in section \myhyperref{subsection: EOM target frames}, these time-varying factors are well out-of-band. They involve slow varying terms such as $\vangvelexp{O}{J}{J}$ or $\vangvelexp{H^*}{O}{H^*}$ driven by orbital motion. A fair treatment consists in averaging those terms out and considering them constant over the course of a simulation run.

The final system of differential equations of motion of LISA can then be approximated by the following linear, time-invariant differential system
\begin{equation}
    \dv{\va{X}}{t} = \expval{ A \big( \va{X}^{\text{target}} \big) }_{t} \: \va{X}(t) + \expval{ B \big( \va{X}^{\text{target}} \big) }_{t} \: \va{u}(t)\, ,
\end{equation}
where matrices $A$ and $B$ are evaluated at the target state vector $\va{X}^{\text{target}}$ and averaged over simulation time
\begin{align}
    \va{X}^{\text{target}}=
    \begin{matrix}
        \Big[
        & \vb{0} & \vb{0} & \vb{0} & \vb{0} & \vb{0}
        \\
        & \vb{0} & \vb{0} & \vb{0} & \vb{0} & \vb{0}
        \\
        & \delta \phi_{tel, 1}^* & \quad \delta \dot{\phi}_{tel, 1}^* & \quad \delta \phi_{tel, 2}^* & \quad \delta \dot{\phi}_{tel, 2}^*
        & \Big] \hspace*{1.3em}
    \end{matrix}\, .
\label{eq: target state vector}
\end{align}

Note that in the simulation we use an implicit formulation of the state-space model, involving the {\it mass matrix} 
\begin{equation}
    M_{\text{im}} \dv{\va{X}}{t} = A_{\text{im}} \va{X}(t) + B_{\text{im}} \va{X}(t) \va{u}(t)\, ,
\label{eq: state space equation implicit}
\end{equation}
but this does not impact the generality of the discussion in this section, as we can---and eventually do---map the implicit and explicit matrix formulations
\begin{align}
    & A_{\text{ex}} = M_{\text{im}}^{-1} A_{\text{im}}\ , & B_{\text{ex}} = M_{\text{im}}^{-1} B_{\text{im}}\ .
\label{eq: implicit to explicit}
\end{align}

In addition to the linearized system, we also implement and solve the full non-linear system of equations for comparison, and will test and discuss their differences with simulation experiments in section \myhyperref{section: experiments}. The two models are important and will have their own scope, as the linearized version is faster, more flexible, and required for control design, whereas the non-linear simulation provides more realistic simulated data, and can resolve non-linearities and time-variability, which is especially interesting for system identification, diagnostics, and data analysis.

\section{Closing the loop: DFACS, Sensors and Actuators}
\label{section: dfacs}

The dynamics of \gls{lisa} is a closed-loop control system to ensure that all dynamical \glspl{dof} stay as close as possible to working points. To realize this, the control loop will cancel any stray forces and torques deviating the bodies from their set points, hence, commanding forces and torques that are exactly opposed to the disturbances---in the limit case where control authority is infinite. The strategy of control---the \gls{dfacs}---is designed to fulfil the three following, main objectives:

\begin{itemize}
    \item The satellite motion must be locked on the test-mass average trajectories, as monitored by the local, test-mass interferometer, to compensate for the otherwise noisy spacecraft jitter, mainly driven by its own thrust noise \cite{lisa_pathfinder_collaboration_lisa_2019}. This is called {\it Drag-Free} control. At frequencies where control gain is high, in the lower part of the \gls{lisa} band, it can successfully force the spacecraft to be nearly as quiet as the test-masses.
    \item Any relative displacement drift between the two test-masses inside the spacecraft must---and can only be---corrected by applying direct, electrostatic forces and torques on the test-masses ($x_1$ and $x_2$ axes excluded). These are inevitable actuations for the test-masses to stay within their housings over the course of the mission, mainly compensating for spacecraft self-gravity gradients. This is called {\it Suspension} control, and its use must stay minimal.
    \item The \glspl{mosa} must point constantly towards their respective distant spacecraft, and so it must be ensured that the incoming wave fronts are normal to the lines of sight of the telescopes at any time. To that end, both spacecraft rotation and \gls{mosa} opening angle actuation will be commanded. In a nominal science mode, we expect the spacecraft attitude control to ensure that the $\vu{e}_{X}$ is aligned along the constellation bisector (common-mode angle), whereas the \gls{mosa} angle mechanism control will actuate the opening angle $\phi_{m}$ (differential-mode angle).
\end{itemize}

A dedicated publication \cite{lisa_dfacs_scheme} has already thoroughly addressed the question of \gls{lisa} \gls{dfacs} and its optimization regarding the isolation of test-mass actuation from spacecraft jitter. Hence, here we will discuss more succinctly the question of \gls{dfacs} modeling, and we will refer the reader to this past publication for further details.

\begin{figure}[h!]
\centerline{\includegraphics[width=\linewidth]{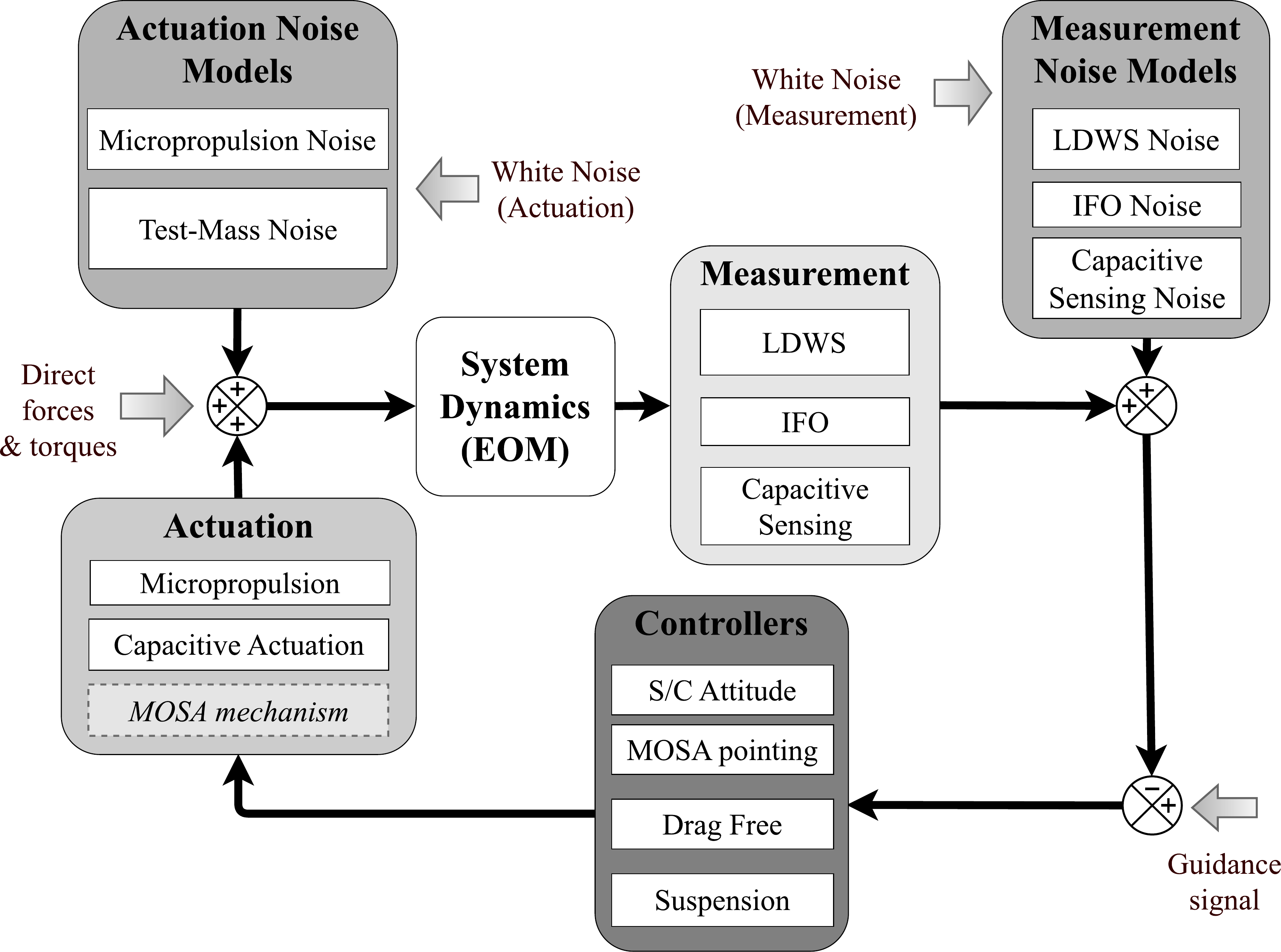}}
\caption{Block diagram of LISA closed-loop dynamics. The system \gls{eom}, intrinsically unstable,  get stabilized by the application of compensation forces and torques, based on the information from in-loop sensors, and exerted by the actuation system on-board. External disturbances and noise perturb the system, and require the continuous compensation of the control to maintain the \gls{dof} at target points. While the current \texttt{LISANode} dynamics simulation includes white noise only, noise models will be inserted, as depicted in this diagram.}
\label{figure: closed-loop diagram}
\end{figure}

The simulation we present in this article involves a detailed modeling of the closed-loop system, including models for the \gls{dfacs}, the sensor and actuator systems, noise models for these systems as well as for the test-mass acceleration noise based on LISA Pathfinder output \cite{armano_sub-femto-g_2016, armano_beyond_2018}. Figure \myhyperref{figure: closed-loop diagram} shows a diagram of such a loop and of the way it is modeled in the simulation, emphasizing in-loop and out-of-loop physical quantities. It breaks down as:

\begin{itemize}

    \item The \gls{eom} block, core of the loop and described thoroughly in section \myhyperref{section: eoms}, is the passive dynamical system to be controlled and stabilized. As previously discussed, it includes longitudinal and rotational dynamics of spacecraft and test-masses. It yields the time-series of $17$ dynamical states per spacecraft which feed in the measurement system.
    
    \item The {\it Measurement} block then models the observations of the dynamical states from optical interferometer read-out \cite{armano_sensor_2021, armano_sensor_2022} (longitudinal: $x_1$, $x_2$, rotational: $\eta_1$, $\phi_1$, $\eta_2$, $\phi_2$), electrostatic capacitive sensing \cite{lisa_pathfinder_collaboration_capacitive_2017} (all test-masses longitudinal and rotational \glspl{dof}), and \gls{ldws} informing about spacecraft and \gls{mosa} rotational \glspl{dof}. Noise is added to each of these measurement outputs, and its spectral characteristics are detailed in Table \myhyperref{table: NoiseLevels}. Geometrical imperfections are introduced with sensing cross-talk matrices for \gls{ifo} and \gls{grs} sensors.
    The \gls{dws} measurement geometry is modeled in more details. At each instant, $t$ the incident angles ($\eta_1^\sensor{ldws}$, $\phi_1^\sensor{ldws}$, $\eta_2^\sensor{ldws}$, $\phi_2^\sensor{ldws}$) of the incoming beam are computed w.r.t. the telescope axes of the local spacecraft. The Cardan angles describing the spacecraft attitude are then recovered from an attitude determination matrix (cf. Equation (\myhyperref{eq: AttDetMatrix})).
    \begin{align}
    \label{eq: AttDetMatrix}
        \!\!\begin{bmatrix}
        & \Theta^\sensor{ldws} \\
        & H^\sensor{ldws} \\
        & \Phi^\sensor{ldws}
        \end{bmatrix}\!
        =\!
        \begin{pmatrix}
            & 0.0 & -1.0 & 0.0 & 1.0 \\
            & 0.0 & -\tfrac{1}{\sqrt{3}} & 0.0 & -\tfrac{1}{\sqrt{3}} \\
            & -0.5 & 0.0 & -0.5 & 0.0
        \end{pmatrix}\!\!
        \begin{bmatrix}
        & \phi_1^\sensor{ldws} \\
        & \eta_1^\sensor{ldws} \\
        & \phi_2^\sensor{ldws} \\
        & \eta_2^\sensor{ldws}
        \end{bmatrix}
    \end{align}
    
    \item The Control block is fed in with $16$ error signals, made from the difference between sensor outputs and reference values to be tracked (all $0.0$ due to our representation of the dynamical states, as deviation from working points). It includes the transfer functions corresponding to the four control strategies: {\it drag-free}, {\it suspension}, {\it attitude} and {\it telescope pointing} control. We refer the reader to \cite{lisa_dfacs_scheme} for a detailed discussion of these control strategies, and especially to Table III. of \cite{lisa_dfacs_scheme}, which summarizes the control mapping and bandwidth.
    
    \item The {\it Actuation} block receives commands from the {\it Control} block and delivers compensation forces and torques to the \gls{eom} block. Noise time series with spectral characteristics listed in Table \myhyperref{table: NoiseLevels} are added to these actuation outputs. Presently, the actuation models consist merely in gains, cross-talks or time constants transfer functions.
\end{itemize}

\begin{table*}[t]
\centering
\caption{Table of sensing and actuation noise settings in the simulations. All noises are still assumed to be white in the simulation (apart from test-mass acceleration noise along $x_1$ and $x_2$), but will account for low-frequency roll-up in upcoming upgrades, as already considered and detailed in \cite{lisa_dfacs_scheme}.}
\label{table: NoiseLevels}
\setlength\tabcolsep{0.3cm}
\begin{tabular}{|c||c|c||c|c|}
\hline
\# & \bf{Sensing Channel} & \bf{\hspace{1 mm} Noise Floor \hspace{1 mm}} & \bf{Actuation Channel} & \bf{\hspace{1 mm} Noise Floor \hspace{1 mm}} \\ \hline

1 & $x_1^{\text{ifo}} \quad / \quad x_2^{\text{ifo}}$ & $1.0\times 10^{-12} \,\si{\metre\persqrthz}$ & Thrust $X$     &                                      $2.2\times 10^{-7}  \,\si{\newton\persqrthz}$ \\ \hline

2 & $\eta_1^{\text{ifo}} \quad / \quad \eta_2^{\text{ifo}}$ & $2.0\times 10^{-9} \,\si{\radian\persqrthz}$ & Thrust $Y$ & $1.3\times 10^{-7}  \,\si{\newton\persqrthz}$ \\ \hline

3 & $\phi_1^{\text{ifo}} \quad / \quad \phi_2^{\text{ifo}}$ & $2.0\times 10^{-9} \,\si{\radian\persqrthz}$ & Thrust $Z$ & $3.6\times 10^{-7}  \,\si{\newton\persqrthz}$ \\ \hline

4 & $x_1^{\text{grs}} \quad / \quad x_2^{\text{grs}}$ & $1.8\times 10^{-9} \,\si{\metre\persqrthz}$ &  Thrust $\Theta$ & $7.7\times 10^{-8} \,\si{\newton\metre\persqrthz}$ \\ \hline

5 & $y_1^{\text{grs}} \quad / \quad y_2^{\text{grs}}$ & $1.8\times 10^{-9} \,\si{\metre\persqrthz}$ & Thrust $H$      & $6.9\times 10^{-8} \,\si{\newton\metre\persqrthz}$ \\ \hline

6 & $z_1^{\text{grs}} \quad / \quad z_2^{\text{grs}}$ & $3.0\times 10^{-9} \,\si{\metre\persqrthz}$ & Thrust $\Phi$ & $1.3\times 10^{-7} \,\si{\newton\metre\persqrthz}$ \\ \hline

7 & $\theta_1^{\text{grs}} \quad / \quad \theta_2^{\text{grs}}$ & $120.0\times 10^{-9} \,\si{\radian\persqrthz}$ &  $f_{y}^{\text{grs}}$      &  $6.0\times 10^{-15} \,\si{\newton\persqrthz}$ \\ \hline

8 & $\eta_1^\sensor{ldws} \quad / \quad \eta_2^\sensor{ldws}$ & $0.2\times 10^{-9} \,\si{\radian\persqrthz}$ &  $f_{z}^{\text{grs}}$      &  $10.0\times 10^{-15} \,\si{\newton\persqrthz}$ \\ \hline

9 & $\phi_1^\sensor{ldws} \quad / \quad \phi_2^\sensor{ldws}$ & $0.2\times 10^{-9} \,\si{\radian\persqrthz}$ &  $t_{x}^{\text{grs}}$      &  $1.0\times 10^{-15} \,\si{\newton\metre\persqrthz}$ \\ \hline

10 &           &            &  $t_{y}^{\text{grs}}$      &  $1.0\times 10^{-15} \,\si{\newton\metre\persqrthz}$ \\ \hline

11 &          &            &  $t_{z}^{\text{grs}}$      & $1.0\times 10^{-15} \,\si{\newton\metre\persqrthz}$ \\ \hline
\end{tabular}
\end{table*}

\section{Numerical solving and LISANode simulations}
\label{section: solver}

\subsection{LISANode software}
\texttt{LISANode} is a graph-based prototype simulator created and used by the LISA collaboration \cite{bayle_simulation_2019, bayle_lisanode_2022}. The graph encoding computations is assembled in Python either from existing graphs or from atomic nodes implemented in C++. The provided Python transcriber is then able to translate the graph into a C++ file encoding the simulation. This ensures the ease-of-use of Python while keeping the performance of C++.

In this work, the library of existing nodes was extended consistently to allow not only scalars to be passed between nodes, but also vectors and $3 \times 3$ matrices. This allows for a much more convenient implementation of the dynamics described here, as well as providing faster execution speeds. Additionally, this enables an easy implementation of differential equation solvers.

\subsection{Linearized Simulation}
\label{section: solver subsection: linear}
The linear differential system to be solved can be represented in the continuous state-space formalism, cf. Equation (\myhyperref{eq: state space equation implicit}).
Solving the system in Equation (\myhyperref{eq: state space equation implicit}) amounts to translating it into a discrete state-space, i.e., the continuous equation of the form
\begin{align}
    \dv{\va{X}}{t} = A\, \va{X}(t) + B\, \va{u}(t)
\end{align}
with constant $A$ and $B$ into the discrete
\begin{align}
    \va{X}\!\left(t^{(n+1)}\right) = A_\text{disc.} \va{X}\!\left(t^{(n)}\right) + B_\text{disc.} \va{u}\!\left(t^{(n)}\right)\ .
\end{align}
For completeness, a derivation of the solution can be found in the appendix \myhyperref{appendix: to discrete ssm}. The relevant relations are given by
\begin{align}
    A_{\text{disc.}} &= \euler^{A\, dt}\ , \label{eq: ssm discrete - zoh 1}\\
    B_{\text{disc.}} &= A^{-1} \left(\euler^{A\, dt} - \unitmat\right)B\ ,
\label{eq: ssm discrete - zoh 2}
\end{align}
where $dt$ gives the discretization of the time domain, which in this case is given as the inverse of the LISA simulation sampling frequency. With the equations in this form, getting the state vector $\va{X}$ at the next time point amounts to a simple matrix multiplication. As the matrices $A$ and $B$ are known beforehand, the transition to the discrete system can be handled in Python without \texttt{LISANode}. We use an off-the-shelf algorithm for the transformation \cite{2020SciPy-NMeth}. In \texttt{LISANode}, the matrix multiplication in each time step is handled within one atomic node to make it more efficient.
Finally, discretization of the input vector $\va{u}(t)$ is of importance. The simplest version would be of zero-order hold, i.e., approximating the function with step-functions. A second option would be a first-order hold, i.e., a linear interpolation. We have verified empirically that both methods yield results of equal accuracy, which is likely explained by the high sampling rate used in our simulations ($f_s = 16\,\si{\hertz}$) relatively to the typical timescale of our experiments ($< 0.1\,\si{\hertz}$). Hence, we have opted for the simple zero-order interpolation scheme, as given in Equations \myhyperref{eq: ssm discrete - zoh 1} and \myhyperref{eq: ssm discrete - zoh 2}.

\subsection{Non-linear Simulation}
In the non-linear case, the ordinary differential equation system must be solved for each time-step. There are many algorithms available, which usually require the problem to be stated in the form  
\begin{align}
    \dv{\va{X}}{t} = \va{f}(t; \va{X})\ .
\end{align}
The \gls{eom} derived here can easily be brought into this form, and the full set can be found in the appendix \myhyperref{appendix: all eom for solver}. For discretization, we use $dt$ for the step-size again. We chose to implement three algorithms into \texttt{LISANode}, as the graph-based computations made it almost impossible to use an off-the-shelf code. There are different ways to characterize these algorithms, most important is the order of the method in terms of the \gls{lte}, i.e., the error done in one time-step, or the \gls{gte}, i.e., the error accumulated over multiple time-steps until a final time. Another important aspect is the computational complexity, which here amounts to how many function evaluations of $\va{f}$ are needed to predict the next time-step. We will use the shorthand 
\begin{align}
        \va{X}_n = \va{X}\!\left(t^{(n)}\right)
\end{align}
for the discretized state vector. The three algorithms considered are:

\begin{itemize}
    \item \emph{Euler Method:} The simplest method, which approximates the derivative as a difference quotient
    \begin{align}
        \frac{\va{X}_{n+1} - \va{X}_{n}}{dt} &= \va{f}(t; \va{X}_n)\ , \\
        \va{X}_{n+1} &= dt \cdot \va{f}(t; \va{X}_n) + \va{X}_{n}\ .
    \end{align}
    The method is nevertheless interesting because it is quick (only requires 1 evaluation of $\va{f}$) and gives good results for a small step-size. The \gls{lte} is of order $dt^2$, the \gls{gte} of order $dt$.
    \item \gls{rk4}: A commonly used algorithm when it comes to ordinary differential equations. It can be formulated as 
    \begin{align}
        \ \ \va{X}_{n+1}= \va{X}_{n}\! +\! \frac{dt}{6} \!\left( \va{k}_1 + 2 \va{k}_2 + 2 \va{k}_3 + \va{k}_4 \right)
    \end{align}
    where the $\va{k}_i$ depend successively on each other and are given by
    \begin{align}
        \begin{split}
        \va{k}_1 &= \va{f}\!\left(t^{(n)}; \va{X}_n\right) \ , \\
        \va{k}_2 &= \va{f}\!\left(t^{(n)}+\frac{dt}{2}; \va{X}_n + \frac{dt}{2} \va{k}_1\right) \ , \\
        \va{k}_3 &= \va{f}\!\left(t^{(n)}+\frac{dt}{2}; \va{X}_n + \frac{dt}{2} \va{k}_2\right) \ , \\
        \va{k}_4 &= \va{f}\!\left(t^{(n)}+dt; \va{X}_n + dt \cdot \va{k}_3\right) \ .
        \end{split}
    \end{align}
    It requires 4 function evaluations of $\va{f}$, the \gls{lte} is of order $dt^5$, and the \gls{gte} of order $dt^4$.
    \item \gls{rkf45}: This is an interesting extension to RK4. The \gls{lte} and \gls{gte} are the same for \gls{rk4} and \gls{rkf45}, but by using 6 function evaluations of $\va{f}$ the algorithm can also estimate the local error itself. The estimator is of order $dt^5$, hence the name. The constants of the algorithm were chosen in accordance with \cite{galassi_gnu_2009}, the scheme is conceptually the same as for RK4, the interdependence of the $\va{k}_i$ is more complicated.
\end{itemize}
Note that due to the \texttt{LISANode} framework, the functions $\va{f}$ are represented by graphs. In the \gls{rk4} and \gls{rkf45} algorithms, this leads to 4 and 6 copies of this ``function graph'' due to the interdependence of the $\va{k}_i$.

The estimated truncation error of the \gls{rkf45} algorithm can be useful to provide an upper bound for the local error, providing an indication of a diverging solution. Its reliability is reduced in the presence of strongly non-linear solutions.

\subsection{Efficiency Analysis}
In Table \myhyperref{table_ode_solve_time} there is a comparison of run times for the linear and non-linear models with different solving algorithms. We chose a simulated time of $10^5$ seconds, but each time-step should take the same amount of time as they each have the same complexity. Thus, the results are presented as speed-ups of simulated time over computation time.

The code has not been fully optimized for run-time yet, and we expect further improvements. Several performance optimizations are possible, e.g. using multiple threads during computation. Thus, the non-linear Euler solver is currently slightly faster than the linear one. We understand this as due to their different implementation: the linear simulation uses a matrix formulation, which requires numerous matrix-vector multiplications. The matrices are quite sparse ($\sim\! 5\%$ non-zero elements), which results in some overhead compared to the non-linear Euler implementation.

\begin{table}[h]
    \centering
    \caption{Table comparing computation time of the linear and non-linear implementations with different solving algorithms. The results are given as dimensionless ratios of simulated time and computation time, i.e., how much speed-up is gained in the simulation. For a factor $100$ this means that it takes $0.01$ seconds on a computer to simulate $1$ second in the model.
    All numbers are generated using an Apple M1 Pro processor (single-thread execution) with the native clang compiler with optimization level 1.}
    \label{table_ode_solve_time}
    \begin{tabular}{|c || c| c| c| c|} 
        \hline
        Simulation & Linear & \, NL (Euler) \, & NL (RK4) & NL (RKF45) \\ [0.5ex] 
        \hline
        Single S/C & 513 & 635 & 373 & 296 \\ 
        Full LISA & 65.6 & 84.4 & 64.8 & 52.3 \\ [1ex] 
        \hline
    \end{tabular}
\end{table}

\section{Simulation experiments}
\label{section: experiments}

Solving the 17-dimensional differential systems, the simulation yields the time evolution of all the system dynamical \glspl{dof}, as well as the in-loop sensor outputs, the commanded and applied forces on each dynamical body. In addition, the possibility of injecting (sinusoidal) excitation signals is implemented, such as {\it disturbance signals} $\va{d}$ ({\it “direct forces \& torques”} in Figure \myhyperref{figure: closed-loop diagram}) or {\it guidance signals} $\va{g}$ (biasing the in-loop sensors, see Figure \myhyperref{figure: closed-loop diagram}). These input ports are useful to probe the closed-loop transfer functions of the system and perform experiments to check the dynamical model. This is the purpose of this section.

We present a series of $8$ simulated experiments realized on-board a single spacecraft, where excitation signals are injected in the closed-loop to stimulate the dynamics and probe its response to disturbances. We inject $3$ guidance signal injections simultaneously at $3$ distinct frequencies, and for $8$ different injection amplitudes:

\begin{itemize}
    \item On spacecraft control angle $H$: \\at $f_{g_H} = 1\,\si{\milli\hertz}$, amplitude:\\$A_{g_H} = \big[10^{5}, 10^{4}, 10^{3}, 100, 10, 1, 0.1, 10^{-2}\big]\,\si{\micro\radian}$
    \item On test-mass longitudinal $x$-displacement $x_1$:\\ at $f_{g_{x_1}} = 5\,\si{\milli\hertz}$, amplitude:\\ $A_{g_{x_1}} = \big[10^{5}, 10^{4}, 10^{3}, 100, 10, 1, 0.1, 10^{-2}\big]\,\si{\micro\metre}$
    \item On test-mass longitudinal $x$-displacement $x_2$:\\ at $f_{g_{x_2}} = 10\,\si{\milli\hertz}$, amplitude:\\$A_{g_{x_2}} = \big[10^{5}, 10^{4}, 10^{3}, 100, 10, 1, 0.1, 10^{-2}\big]\,\si{\micro\metre}$
\end{itemize}

We refer the reader to the Figure \myhyperref{figure: spacecraft} helping the visualization of the \glspl{dof} and of the geometry of these experiments. These five injection experiments are realized with both linear and non-linear models for comparison. The simulations have a duration of $3 \times 10^4\,\si{\second}$ and are sampled to $4\,\si{\hertz}$ each, from which the first $1 \times 10^4\,\si{\second}$ have been truncated, ensuring the slow \gls{mosa} control has stabilized fully and the steady-state regime has been reached. The residual time-series (between linear and non-linear simulations) are computed for each experiment. Figures \myhyperref{figure: sc dyn test} and \myhyperref{figure: tm dyn test} show the simulation results for the largest amplitude experiments ($A_{g_H} = 10.0 \,\si{\micro\radian}$, $A_{g_{x_1}} = A_{g_{x_2}} = 10.0 \,\si{\micro\metre}$).

\subsection{Testing spacecraft dynamics}
\label{subsection: sc dyn test}

\begin{figure*}[htb]
    \includegraphics[width=\linewidth, trim={1.25cm 0.0cm 1.35cm 0.1cm}, clip]{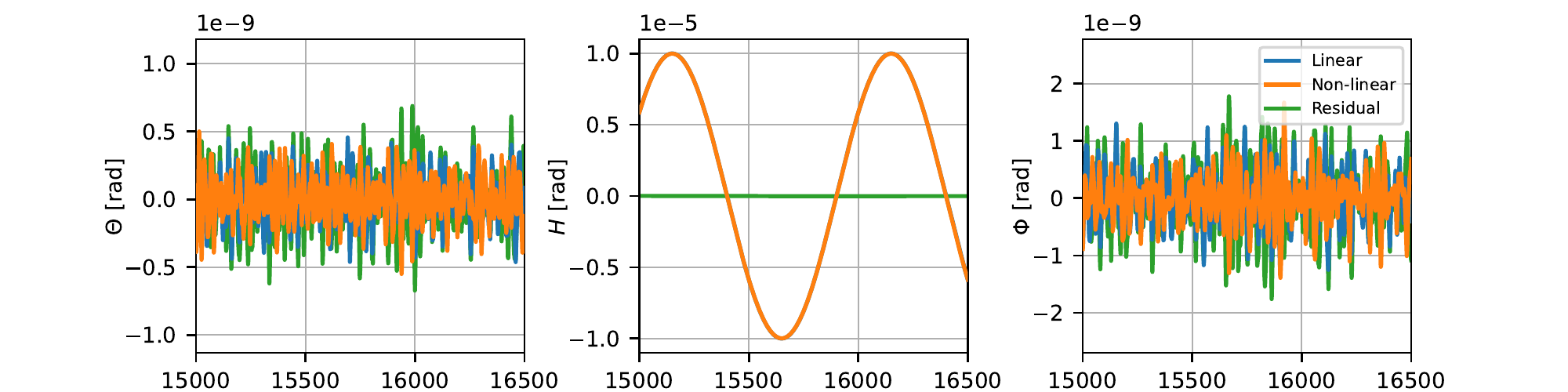}
    \caption{Time evolution of the spacecraft $3$ \glspl{dof} (angular $[\Theta, H, \Phi]$) during the largest amplitude injection experiment. Guidance signals are injected on the spacecraft angle $H$, and test-masses $x_1$ and $x_2$ controlled \glspl{dof}, respectively of amplitude $10\,\si{\micro\radian}$, $10\,\si{\micro\metre}$ and $10\,\si{\micro\metre}$ at $f = 1\,\si{\milli\hertz}$, $5\,\si{\milli\hertz}$, and $10\,\si{\milli\hertz}$. Simulations using the linear (in blue) and non-linear (orange) closed-loop system model are shown, together with the modeling residual (in green).}
    \centering
    \label{figure: sc dyn test}
\end{figure*}

Figure \myhyperref{figure: sc dyn test} shows the spacecraft angular \glspl{dof} time series $\left[ \Theta(t), H(t), \Phi(t) \right]$ for the linear (blue) and the non-linear (orange) simulations. The green traces give the residuals. We observe the $10.0 \,\si{\micro\radian}$ $H$ injections around the $Y$-axis as expected. $\Theta$ and $\Phi$ are compatible with simulated noise (sensing and actuation noise). The linear and non-linear simulations use different realizations of the noise, explaining the shape of the observed residual, which yields the sum of the uncorrelated noise. The $H$ time series at the center shows an excellent agreement between linear and non-linear simulations. This will be investigated more quantitatively with the test-mass longitudinal motion.

\subsection{Testing test-mass dynamics}
\label{subsection: tm dyn test}

\begin{figure*}[htb]
    \includegraphics[width=\linewidth, trim={1.5cm 0.0cm 1.35cm 0.5cm}, clip]{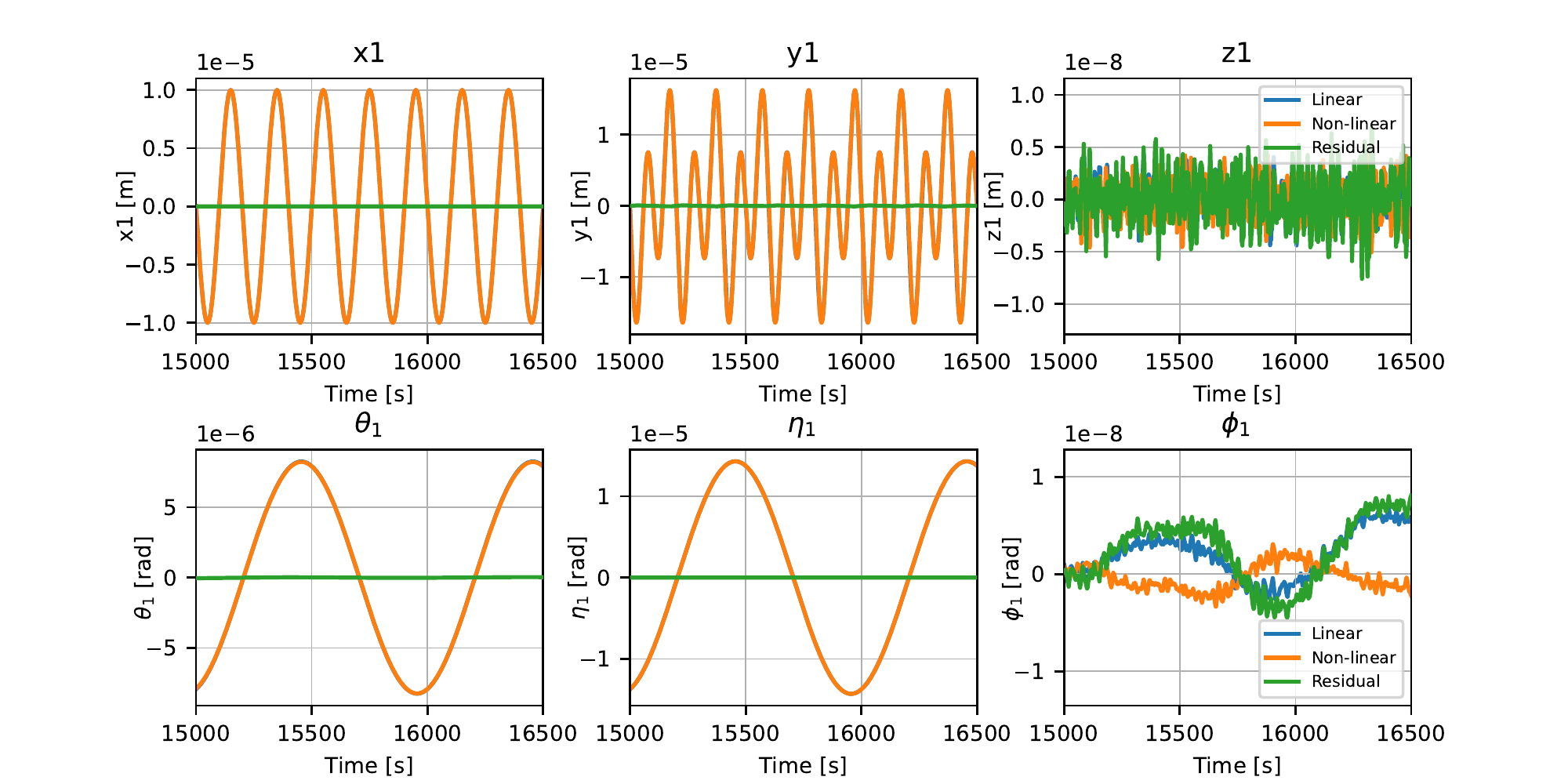}
    \caption{Time evolution of the test-mass $6$ \glspl{dof} (longitudinal $[x_1, y_1, z_1]$ on the top subplots, angular $[\theta_1, \eta_1, \phi_1]$ on the bottom subplots) during the largest amplitude injection experiment. Guidance signals are injected on spacecraft angle $H$, and test-masses $x_1$ and $x_2$ controlled \glspl{dof}, respectively of amplitude $10\,\si{\micro\radian}$, $10\,\si{\micro\metre}$ and $10\,\si{\micro\metre}$ at $f = 1\,\si{\milli\hertz}$, $5\,\si{\milli\hertz}$, and $10\,\si{\milli\hertz}$. Simulations using the linear (in blue) and non-linear (orange) closed-loop system model are shown, together with the modeling residual (in green).}
    \label{figure: tm dyn test}
    \centering
\end{figure*}

Test-mass motion has more complex dynamics properties to probe since it is impacted by both spacecraft (lever arms, inertial forces, rotations in housing) and test-mass disturbances. Hence, in the Figure \myhyperref{figure: tm dyn test} one sees the imprint of the three injections altogether. While {\it drag-free} control suppresses (lever-arm) spacecraft-rotation-driven longitudinal motion of the test-masses, the stimulated rotation of spacecraft is observable on both $\theta$ and $\phi$ rotational \glspl{dof} of the test-masses, since the $Y$-axis of the spacecraft $\mathcal{B}$-frame, around which the spacecraft rotation is performed, has non-zero components along the $x$ and $y$ axes of the housing frames $\mathcal{H}_1$ and $\mathcal{H}_2$. The excitation signal is amplified compared to the stimulation signal, exceeding $10\,\si{\micro\radian}$ for $\eta_1$. This is coincidentally due to the injection frequency $f_{g_{H}} = 1\,\si{\milli\hertz}$ sitting at the end of the suspension control bandwidth (see Table III. of \cite{lisa_dfacs_scheme} for more details), where the suspension struggles to compensate for external disturbances, amplifying them in a narrow bandwidth around the millihertz. We have verified that, when injecting at the lower frequency of $f_{g_{H}} = 0.1\,\si{\milli\hertz}$, where suspension control is still efficient, the imprint of the injection on $\theta_1$ and $\eta_1$ is mitigated down to below $0.1\,\si{\micro\radian}$, and the test-masses are well forced to rotate together with the spacecraft guidance at low frequency \cite{lisa_dfacs_scheme}.

The top subplots of Figure \myhyperref{figure: tm dyn test} show the longitudinal motion time series, where the $x_1$ and $x_2$ guidance injections are visible. The left-hand plot shows the $x_1$ \gls{dof} time series, which indicates that the {\it drag-free} control is doing well to force this dynamical \gls{dof} to track the guidance sinusoidal injection. The middle plot shows the $y_1$ \gls{dof} which exhibits a composition of the two injection frequencies, as an imprint of both $x_1$ and $x_2$ time-evolution. Indeed, {\it drag-free} control here has the task to force $x_1$ and $x_2$ variables to track down two different frequencies, despite $\framevus{x}{1}$ and $\framevus{x}{2}$ being not perpendicular. In practice, the command will then have to request compensation of spacecraft motion along $x_2$ in order to correct $f_{x_1}$ injected motion along $x_2$ originating from the $x_1$ control, which has necessarily leaked to the $x_2$ direction due to the non-orthogonality between $\framevus{x}{1}$ and $\framevus{x}{2}$. This explains why one sees traces of both frequencies in the $y_1$ plot of Figure \myhyperref{figure: tm dyn test}. All the dynamics take place in the $XOY$ plane here; the $z_1$ dynamical \gls{dof} time-series are compatible with noise in this case. Again, Figure \myhyperref{figure: tm dyn test} is presenting the largest amplitude injection experiment, showing excellent agreement between the linear and non-linear simulation. There are, however, observable discrepancies when the quantities---and in particular the residuals---are represented in the frequency domain. In the next section, we show that the observed discrepancies indicate the presence of dynamical time-varying and  non-linear features the \gls{lti} model is failing to capture by construction.

\subsection{Resolution of non-linearities}
\label{subsection: res nl}

Computing the frequency spectra of the time series in Figure \myhyperref{figure: tm dyn test}, discrepancies between linear and non-linear simulations become visible. This is quantified using the relative error of spectrum amplitudes at injection frequencies
\begin{equation}
    \delta_{\text{res}} = \abs{\frac{\abs{\tilde{X}^{\text{nl}}(f_{\text{inj}})} - \abs{\tilde{X}^{\text{l}}(f_{\text{inj}})}}{\abs{\tilde{X}^{\text{l}}(f_{\text{inj}})}}}\ .
\end{equation}

\begin{figure}[htb]
    \includegraphics[width=0.95\linewidth, trim={0.0cm 0.0cm 0.0cm 0.0cm}, clip]{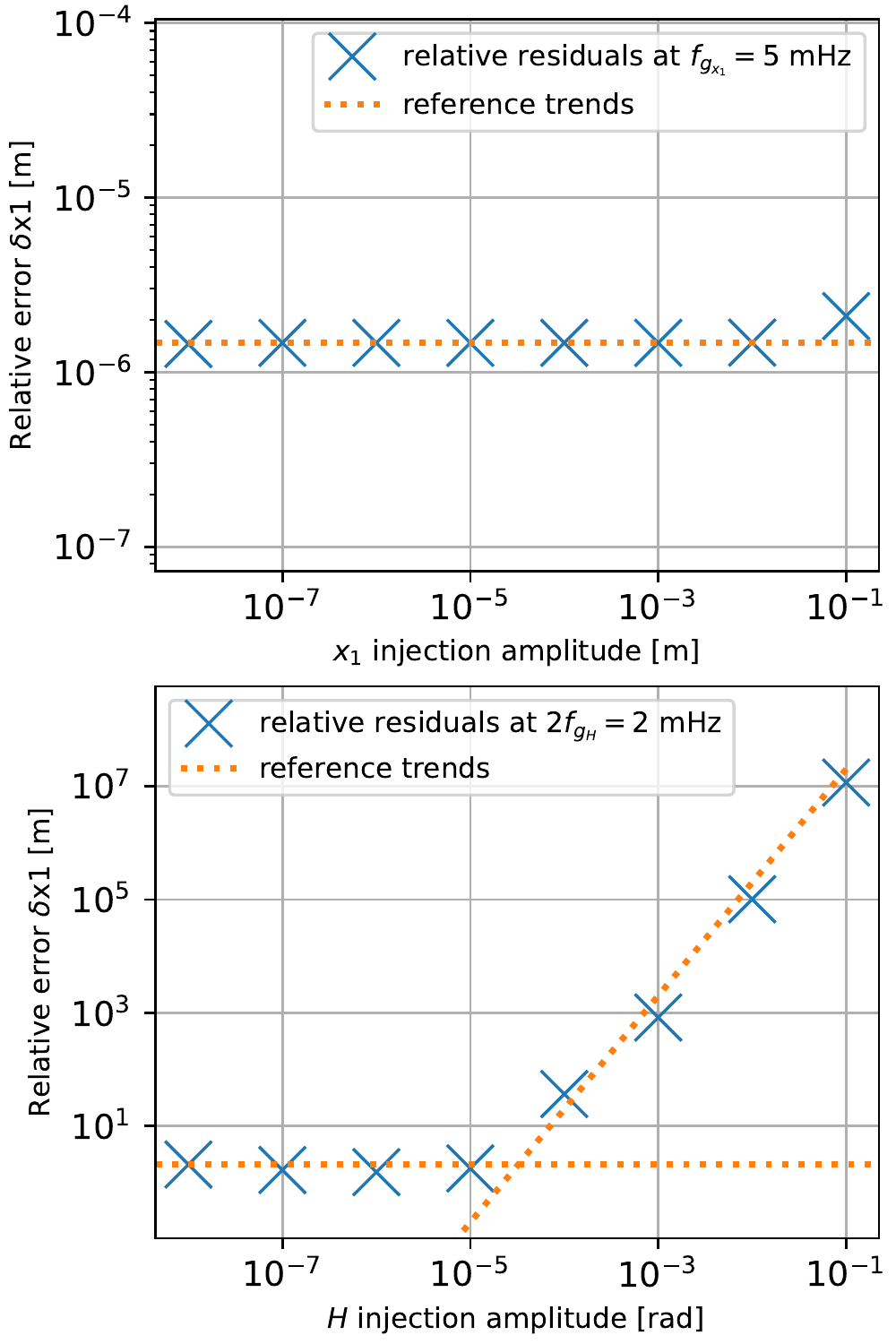}
    \caption{Measured relative errors of the $\delta x_1$ \gls{dof} between linear and non-linear simulations, for the range of eight excitation amplitudes covered. The blue crosses $\times$ show the measurements of the error at $f=10\,\si{\milli\hertz}$ corresponding to the $g_{x_1}$ longitudinal injection frequency (top subplot), and $f=2\,\si{\milli\hertz}$ corresponding to {\it twice} the $g_{H}$ angular injection frequency (bottom subplot). The orange curves represent linear reference trends in {\it log-log} scale, facilitating the reading of the order of dependence of the dynamical response to the injection amplitude scale. On the top subplot, the relative error $\delta x_1$ stays constant w.r.t. the scale of the excitation, and is the consequence of a time-varying component the \gls{lti} model cannot capture, such as the time-varying opening angle between \glspl{mosa}. On the bottom plot, we observe a relative error that scales quadratically with the excitation signal $g_H$ after $A_{g_H} = 0.1\,\si{\milli\radian}$. It reflects the linear nature of the \gls{lti} model, unable to capture non-linear dynamical contributions such as the inertial forces exerting on test mass along $x_1$ and induced by the spacecraft angular jitter.}
    \label{figure: tm1 dyn res}
    \centering
\end{figure}

Figure \myhyperref{figure: tm1 dyn res} shows the behavior of the relative error $\delta_{\text{res}}$ as a function of the injection amplitudes. Specific simulations with sensing and actuation noise turned off have been realized for this analysis, to better resolve small discrepancies between the linear and non-linear models. Figure \myhyperref{figure: tm1 dyn res} focuses on the dynamical behavior of the \gls{dof} $x_1$, inspected at two different injection frequencies: $f_{g_{x_1}}$ and $2\ f_{g_H}$. The top subplot shows agreement between the linear and non-linear model of the response to the $x_1$ guidance signals at $f_{g_{x_1}} = 5\,\si{\milli\hertz}$. The residual scales linearly w.r.t. the injection amplitude: it is the signature of the contribution of an extra linear component the linear model is missing. Indeed, the linear model discussed is also time-invariant (see Section \myhyperref{subsection: state space representation}). Hence, it cannot account for time-variability of the dynamical system such as the \gls{mosa} rotation, which can have an important impact on geometrical projections and inertial response to external input forces and torques. In the particular case of Figure \myhyperref{figure: tm1 dyn res}, the discrepancy comes from the \gls{mosa} opening angle departing from $60 \si{\degree}$ in the non-linear model, since only the latter accounts for telescope pointing constrained by the satellite's orbital motion. The linear simulation stays ignorant of such feature, hence inducing a small projection error.

On the other hand, the bottom subplot of Figure \myhyperref{figure: tm1 dyn res} presents the non-linear behavior of the $x_1$ \gls{dof} dynamics, in particular its response to spacecraft rotational excitation. It shows that when the spacecraft is forced to rotate around its $Y$-axis above an amplitude of $A_{g_H} = 0.1\,\si{\milli\radian}$, a quadratic component dominates the response to the rotational excitation, as we observe a trend in the $\delta x_1$ residual proportional to $A_{g_H}^2$. Note that, here, the frequency inspected is twice of the angular injection frequency, since the non-linear response manifests as additional signal harmonics. Since the linear response cannot create a $2f_{g_H}$ signal, the relative error at $2f_{g_H}$ is $\delta x_1 \approx 1.0$ for $A_{g_H} \leq 10\,\si{\micro\radian}$.

Figure \myhyperref{figure: tm1 dyn res} hence verifies that the \gls{lti} modeling is indeed not, by construction, capable of capturing non-linear or time-varying dynamical terms, since all quadratic terms have been truncated when the state-space matrices have been evaluated at working points (cf. section \myhyperref{subsection: state space representation}), and the time-varying components averaged out over the simulation duration. The residuals observed in Figure \myhyperref{figure: tm1 dyn res} are the residuals of this truncation, and this motivates the introduction of the non-linear modeling, for instance, in the context of system identification experiments involving large probing signals.

\section{TDI and S/C jitter suppression: a numerical demonstration}
\label{section: tdi}

\subsection{Time-Delay Interferometry and Dynamics}
\label{section: tdi and dyn}

A variety of noise sources, in particular the spacecraft thrusters, are expected to create random motion, or ``jitter'', between the test-masses and spacecraft. In addition to suppressing the laser frequency noise, one of the goals of \gls{tdi} is to suppress the influence of the spacecraft jitter, by cancelling equal and opposite effects in the long and short-arm \glspl{ifo}. With an ideal instrument, e.g., in the absence of tilt-to-length couplings, the test-mass interferometer will capture all the spacecraft noisy accelerations along the long-range interferometer. Hence, combining the signals within the TDI scheme will result in the cancellation of this contribution. Again, the situation just described is idealized, and misalignment between test-mass and long-range interferometer axes, as well as reference points mismatch (see section \myhyperref{section: spacecraft jitter}), will let jitter residuals through that one can interpret as geometrical tilt-to-length effects. In this subsection, however, we will restrict our analysis to the ideal, no-\gls{ttl} case, and utilize the jitter suppression expectation as a figure-of-merit for demonstrating the correct interfacing between the dynamical model and optical interferences on-board and across the constellation.

\subsection{Interferometer observables}
\label{section: observables}

The simulator delivers time-series of the metrology sensors on-board each spacecraft. It also yields {\it Mother-Nature} quantities such as the actual velocity $\myexpr{v}{T_\indice{1}/H_\indice{1}}{\mathcal{H}_\indice{1}}$ and $\myexpr{v}{T_\indice{2}/H_\indice{2}}{\mathcal{H}_\indice{2}}$ of the test-masses w.r.t. the housing frames $\mathcal{H}_\indice{1}$ and $\mathcal{H}_\indice{2}$, or the true acceleration $\myvec{a}{B} = \tfrac{\myvec{f}{B}}{m_{B}}$ of the spacecraft \gls{com} w.r.t. its local inertial frame, which can be projected along the interferometer long-arm axes
\begin{align}
\label{eq: sc_acc}
    & a_1 = \frac{\myvec{f}{B}}{m_{B}} \cdot \myvec{e}{\mathcal{H}_\indice{1}^{*}, x}\ ,
    & a_2 = \frac{\myvec{f}{B}}{m_{B}} \cdot \myvec{e}{\mathcal{H}_\indice{2}^{*}, x}\ .
\end{align}

These physical quantities are measured by the local interferometers (\acrshort{tmi}) and the inter-spacecraft, long-arm interferometers (\acrshort{isi}) respectively. They provide the dynamics contributions to the phase modulation of the heterodyne interferometers beat notes, once post-processed and rescaled into equivalent frequency fluctuation units w.r.t to the laser carrier frequency as 
\begin{align}
    & \nu_\indice{12}^{\text{tmi}} \approx \frac{\myexpr{v}{T_\indice{1}/H_\indice{1}}{\mathcal{H}_\indice{1}}}{c}\ ,
    && \nu_\indice{13}^{\text{tmi}} \approx \frac{\myexpr{v}{T_\indice{2}/H_\indice{2}}{\mathcal{H}_\indice{2}}}{c} \ ,\label{eq: nu_tmi}
    \\ 
    & \nu_\indice{12}^{\text{isi}} \approx \frac{1}{c} \int_{t_0}^{t}{a_1 dt'}\ ,
    && \nu_\indice{13}^{\text{isi}} \approx \frac{1}{c} \int_{t_0}^{t}{a_2 dt'} \ .\label{eq: nu_isi}
\end{align}

\subsection{What does spacecraft jitter mean?}
\label{section: spacecraft jitter}
Particular attention must be set on the definition of the spacecraft acceleration used in Equation (\myhyperref{eq: sc_acc}). Indeed, one still needs to identify the geometrical point of the spacecraft whose noisy dynamical motion contributes to the \gls{isi} beat note in Equation (\myhyperref{eq: nu_isi})---a geometrical point that we will denote with the letter $C$ from now on. That is, not only does the \gls{com} longitudinal noisy motion contribute to $\nu_{\text{isi}}$, but so does the spacecraft rotational motion around its \gls{com} which translates into a translational acceleration of the point $C$ relative to its local inertial frame. Accounting for such a lever-arm effect, Equation (\myhyperref{eq: nu_isi}) now becomes
\begin{align}
    & \nu_\indice{12}^{\text{isi}} \approx \frac{1}{c} \left[ \int_{t_0}^{t}{a_1 dt'}\ -\ \left( \vangvel{B}{O} \times \posvec{C}[\indice{1}]{B} \right) \cdot \myvec{e}{\mathcal{H}_1^{*}, x} \right]\, , \label{eq: nu_isi_21}
    \\
    & \nu_\indice{13}^{\text{isi}} \approx \frac{1}{c} \left[ \int_{t_0}^{t}{a_2 dt'}\ -\ \left( \vangvel{B}{O} \times \posvec{C}[\indice{2}]{B} \right) \cdot \myvec{e}{\mathcal{H}_2^{*}, x} \right]\, . \label{eq: nu_isi_22}
\end{align}

As confirmed by simulations, the nominal location for $C_\indice{1}$ and $C_\indice{2}$ are actually the housing centers $H_\indice{1}$ and $H_\indice{2}$, in order to suppress large geometrical tilt-to-length, deteriorating detector sensitivity significantly beyond requirements. In such a nominal case, the geometrical point of any local spacecraft imaged to the distant spacecraft corresponds to the nominal position of the test-masses, and the test-mass to test-mass optical measurement can be reconstituted independently of the spacecraft dynamics.

\subsection{Simulating spacecraft jitter suppression}
\label{section: spacecraft jitter supp}

\begin{figure}[h!]
\centerline{\includegraphics[width=\linewidth, trim={0.1cm 1.8cm 2.5cm 3.1cm}, clip]{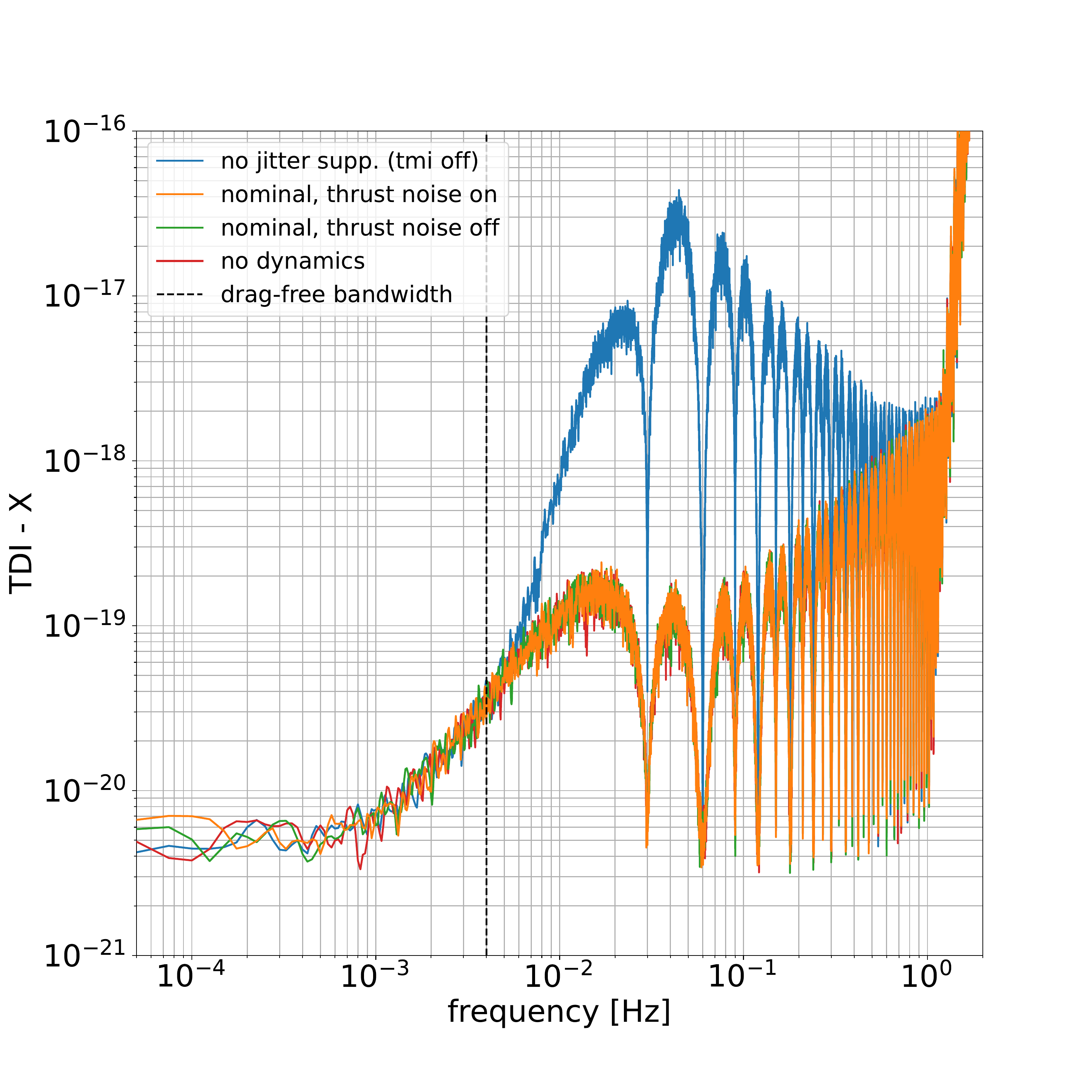}}
\caption{Spectrum of \gls{tdi}-X channel in four different analysis cases: no dynamics data (red), thrust noise on (orange) and off (green), no jitter information from local test-mass \gls{ifo} (blue). This plot demonstrates \gls{tdi} efficiency of suppressing realistic spacecraft longitudinal and angular jitters.}
\label{figure: Plots/JitterSupressionTdiX}
\end{figure}

We performed a series of simulation runs using \texttt{LISANode} to demonstrate this suppression. We use the \texttt{PyTDI} software \cite{staab_pytdi_2023} to build the \gls{tdi} data streams out of the interferometer beat notes generated by \texttt{LISANode}, while \texttt{LISAOrbits} Python software \cite{bayle_lisa_2022} is utilized to compute the light travel times between spacecraft required for \texttt{PyTDI} computations. Figure \myhyperref{figure: Plots/JitterSupressionTdiX} shows the TDI-X spectrum for 4 simulations: A run using default, expected values for all noise (orange), disabling noise on the Micro-Propulsion System (green), disabling dynamics entirely (red), and finally default noise with disabled \gls{tmi} channel (filled zith $0$'s) to emulate a failure of spacecraft and \gls{mosa} jitters correction by \gls{tdi} (blue). In all but the latter case, the summing method used by \gls{tdi} reduces the error to a comparable level. In the final case, as expected, the post-processing combination of \gls{tmi} and \gls{isi} cannot isolate the long-range measurement from spacecraft jitter, which starts to impact significantly the sensitivity around $4\,\si{\milli\hertz}$---above the {\it drag-free} frequency bandwidth. It results in greater noise leaking into the \gls{tdi} channels in this case.

Hence, the blue curve provides an estimate of how the spacecraft jitter would impact LISA sensitivity in the case where \gls{tdi} would not be efficient at isolating the detector performance from spacecraft dynamics noise or in the case of a failure in the \gls{tmi}.
The green and red traces are both giving the {\it no spacecraft jitter} reference curve: in the first case, force noise on spacecraft is essentially turned off, while in the second case, there are no spacecraft dynamics modelled in the first place, so no jitter exists by construction. Figure \myhyperref{figure: Plots/JitterSupressionTdiX} then shows that the orange curve, including both force noise on spacecraft and a correct set-up for \gls{tdi}, lines up well with the {\it no jitter} reference spectra, demonstrating that \gls{tdi} is mitigating efficiently the spacecraft jitter, and bringing it down to subdominant contributions. It is the first numerical demonstration of TDI jitter suppression within a full, time-domain LISA simulation.

\section{Conclusion}
\label{section: conclusion}

With this work, we provide a comprehensive framework for describing mathematically the closed-loop dynamics of the $3 \times 17$ \glspl{dof} at play in the LISA constellation system. We have derived the differential \gls{eom}s for the longitudinal and rotational \glspl{dof} of the moving bodies, both in a vectorial form (Equation (\myhyperref{eq: S/C Angular EOM - Vector}), (\myhyperref{eq: TM Long EOM - Vector}), (\myhyperref{eq: TM Angular EOM - Vector}), (\myhyperref{eq: MOSA Angular EOM - Vector})) and expressed in coordinate systems of interest specified in section \myhyperref{subsection: Frames of reference} (Equation (\myhyperref{eq: S/C Angular EOM - Projected}), (\myhyperref{eq: TM Long EOM - Projected}), (\myhyperref{eq: TM Angular EOM - Projected}), (\myhyperref{eq: MOSA Angular EOM - Projected})). These \gls{eom} were inserted in a \gls{mimo} feedback system, interfaced with sensors, actuators, controllers, and noise sources within a control loop, the so-called \gls{dfacs} loop (cf. section \myhyperref{section: dfacs}). The resulting \gls{mimo} differential system has been solved using several numerical schemes, depending on whether the system dynamics was linearized or not. For linear systems, semi-analytical solutions are available, while for the full non-linear simulations, we have preferred Runge-Kutta 4 numerical solving scheme (cf. section \myhyperref{section: solver}). In section \myhyperref{section: experiments}, the simulation physics has been probed and interpreted with a selection of injection experiments, and we have illustrated the excellent agreement between the linear and the non-linear simulations, the existing residual being consistent with time-varying and non-linear dynamical contributions the \gls{lti} model cannot capture (see Figure \myhyperref{figure: tm1 dyn res}). Finally, in section \myhyperref{section: tdi}, we have discussed the non-trivial interface between the simulated physical quantities (spacecraft and test-mass accelerations) with the interferometers' beat notes, and concluded with the numerical demonstration that the \gls{tdi} algorithm can suppress spacecraft and \gls{mosa} noisy motion from the final interferometer data streams, in the idealized case where no dynamical couplings (\gls{ttl}, actuation cross-talks, stiffnesses) are turned on.

This first \acrlong{e2e} simulation offers the opportunity to start quantitatively evaluating and optimizing post-processing techniques aimed at suppressing dynamical jitters' imprints on \gls{tdi} data streams (\gls{ttl} mitigation, glitch detection and suppression). More generally, it enables the study of the impact of artifacts of dynamical origin (jitter noise, micrometeoroids hitting the spacecraft \cite{thorpe_micrometeoroid_2019}) and propagating due to physical couplings such as \gls{ttl} or stiffness forces \cite{hartig_geometric_2022, paczkowski_postprocessing_2022}, through the instrument response and all the way down to the \gls{tdi} time series. It breaks down and reproduces the complex interplay between the \gls{dfacs} response, the propagation time-delays and various echo phenomena due to post-processing which transforms an impulse excitation signal into a much richer signature. Because LISA will open a novel window on the Universe, potentially capturing unexpected, transient signals, it is crucial to characterize instrumental artifacts with the best accuracy to discriminate instrumental from astrophysical events. This \gls{e2e} simulation will provide key information to that end. We finally note that all the codes and the software dependencies (\texttt{LISAOrbits}, \texttt{PyTDI}, ...) are available (on-demand) to the consortium in the \href{https://gitlab.in2p3.fr/hinchaus/my-lisanode-sim}{my-lisanode-sim} \texttt{Gitlab} repository, from which one can download the full environment encapsulated in a \texttt{docker} or a \texttt{singularity} container and from which one can run the \gls{e2e} and reproduce the results hereby presented (dedicated experiment scripts are also provided).

\section{Acknowledgement}
\label{S:ack}

Lavinia Heisenberg would like to acknowledge financial support from the European Research Council (ERC) under the European Unions Horizon 2020 research and innovation programme grant agreement No 801781 and by the Swiss National Science Foundation grant 179740. LH further acknowledges support from the Deutsche Forschungsgemeinschaft (DFG, German Research Foundation) under Germany’s Excellence Strategy EXC 2181/1 - 390900948 (the Heidelberg STRUCTURES Excellence Cluster).   Henri Inchausp\'e would like to acknowledge the Centre Nationale d'\'Etudes Spatiales (CNES) for its financial support. Peter Wass, Orion Sauter and Henri Inchausp\'e were supported by NASA LISA Preparatory Science program, grant number 80NSSC19K0324. Numerical computations were performed on the DANTE platform, APC, France.

\clearpage

\appendix
\section{Angular velocity and acceleration from basis vectors}
\label{appendix: angvel}

In this appendix, we derive an analytical expression of the angular velocity of the body as a function of the unit vectors basis of a reference frame attached rigidly to the body.

Using the transport theorem at Equation \myhyperref{eq: Transport Theorem} as well as cross-product properties, it can be found that
\begin{align}
    \begin{split}
        \ddt{\framevu{B}{x}}{J} &= \ddt{\framevu{B}{x}}{B}+\myvec{\omega}{B/J}\times\framevu{B}{x} \\
        &= \myvec{\omega}{B/J}\times\framevu{B}{x}\, ,
    \end{split}\\
    \begin{split}
        \ddt{\framevu{B}{y}}{J} &= \ddt{\framevu{B}{y}}{B}+\myvec{\omega}{B/J}\times\framevu{B}{y} \\
        &= \myvec{\omega}{B/J}\times\framevu{B}{y} \, ,
    \end{split}\\
    \begin{split}
        \ddt{\framevu{B}{z}}{J} &= \ddt{\framevu{B}{z}}{B}+\myvec{\omega}{B/J}\times\framevu{B}{z} \\
        &= \myvec{\omega}{B/J}\times\framevu{B}{z}\, ,
    \end{split}
\end{align}
which simplify since the basis vectors appear static in the body frame $\mathcal{B}$ by definition. We now cross-multiply both sides by its respective basis
\begin{align}
        & \framevu{B}{x}\times\ddt{\framevu{B}{x}}{J} = \framevu{B}{x}\times \myvec{\omega}{B/J}\times\framevu{B}{x}\, , \label{appendix eq: trpr 1}\\
        & \framevu{B}{y}\times\ddt{\framevu{B}{y}}{J} = \framevu{B}{y}\times \myvec{\omega}{B/J}\times\framevu{B}{y}\, , \label{appendix eq: trpr 2}\\
        & \framevu{B}{z}\times\ddt{\framevu{B}{z}}{J} = \framevu{B}{z}\times \myvec{\omega}{B/J}\times\framevu{B}{z}\, .
    \label{appendix eq: trpr 3}
\end{align}
From Lagrange's vector triple product formula: $\vb{a} \times (\vb{b} \times \vb{c} ) = (\vb{a} \cdot \vb{c} )\vb{b} -(\vb{a} \cdot \vb{b} )\vb{c}$, and adding up Equations \myhyperref{appendix eq: trpr 1} - \myhyperref{appendix eq: trpr 3}, we find
\begin{align}
    \begin{split}
    & \framevu{B}{x}\times\ddt{\framevu{B}{x}}{J} + \framevu{B}{y}\times\ddt{\framevu{B}{y}}{J} \\
    &+ \framevu{B}{z}\times\ddt{\framevu{B}{z}}{J}
    \\
    & = 3\, \vangvel{B}{J} - \myexpr{\omega}{x, \mathcal{B}/\mathcal{J}}{\mathcal{B}}\ \framevu{B}{x} - \myexpr{\omega}{y, \mathcal{B}/\mathcal{J}}{\mathcal{B}}\ \framevu{B}{y} \\
    &\ \ \ - \myexpr{\omega}{z, \mathcal{B}/\mathcal{J}}{\mathcal{B}}\ \framevu{B}{z}
    \\
    & = 2\, \vangvel{B}{J}\ .
    \end{split}
\end{align}
This leads to the final equation for angular velocity
\begin{align}
    \begin{split}
    \vangvel{B}{J} = &\frac{1}{2} \ \framevu{B}{x}\times\ddt{\framevu{B}{x}}{J}
    \\
    &+ \frac{1}{2} \  \framevu{B}{y}\times\ddt{\framevu{B}{y}}{J}
    \\
    &+ \frac{1}{2} \  \framevu{B}{z}\times\ddt{\framevu{B}{z}}{J}\ .
    \end{split}
\label{appendix eq: AngularVelAnalytical - Appendix}
\end{align}

\section{Full derivation of equations of motion for non-static mass distribution}
\label{appendix: eom}
In the main part of this document, we have considered a fully static mass distribution of the satellite system in its body frame. However, since the two \gls{mosa} on-board will rotate accounting for yearly breathing of the constellation, all masses within the spacecraft are not strictly static w.r.t. the $\mathcal{B}$-frame. As a result, the \gls{com} will be non-stationnary in the body frame $\mathcal{B}$. To model this extra physical feature, it is convenient to define a new geometrical point, denoted $S$, which corresponds to the \gls{com} of the spacecraft platform alone, that is, excluding the two \gls{mosa} bodies.

\subsection{Test-mass longitudinal dynamics}
\label{appendix: tm long dyn}

Equipped with this new definition, the position of the housing geometrical centers $H$ relatively to the \gls{com} $B$ writes
\begin{equation}
    \myvec{r}{H/B} = \myvec{r}{H/P} + \myvec{r}{P/S} - \myvec{r}{B/S}\, ,
\label{eq: r_HB}
\end{equation}
where again $P$ is the pivot point of the \gls{mosa}s rotation, and $S$ is the position of the \gls{com} of the spacecraft platform only (excluding the \glspl{mosa}).

The term $\myvec{r}{P/S}$ is assumed to be constant by construction, and the term $\myvec{r}{H/P}$ is the null vector since $P$ coincides with $H$ by design (and in first approximation), so that
\begin{equation}
    \ddtddt{\myvec{r}{H/B}}{B} = - \ddtddt{\myvec{r}{B/S}}{B}\, .
\label{eq: ddtddt_B r_HB}
\end{equation}

Deriving the term $\ddtddt{\myvec{r}{B/S}}{B}$ requires expressing the spacecraft's \gls{com} as a function of the telescopes orientation. The equation of the \gls{com} provides
\begin{equation}
    m_{S} \mylongvec{B}{S} + m_{H_\indice{1}} \mylongvec{B}{Q}[1] + m_{H_\indice{2}} \mylongvec{B}{Q}[2] = \vec{0}\, ,
\end{equation}
introducing the \gls{com} of the two \gls{mosa}s, respectively $Q_\indice{1}$ and $Q_\indice{2}$, the masses $m_{H_\indice{1}}$ and $m_{H_\indice{2}}$ of the two \glspl{mosa}, and the mass $m_{S}$ of the platform alone. Hence, using our notation convention
\begin{equation}
    m_{S} \myvec{r}{S/B} + m_{H_\indice{1}} \myvec{r}{Q_1/B} + m_{H_\indice{2}} \myvec{r}{Q_2/B} = \vec{0}\, ,
\end{equation}
using $\myvec{r}{B/Q} = \myvec{r}{B/S} + \myvec{r}{S/P} + \myvec{r}{P/Q}$, we get after a few steps of basic algebra
\begin{equation}
    \myvec{r}{B/S} = \epsilon_\indice{1} \left( \myvec{r}{S/P_\indice{1}} - \myvec{r}{Q_\indice{1}/P_\indice{1}} \right) + \epsilon_\indice{2} \left( \myvec{r}{S/P_\indice{2}} - \myvec{r}{Q_\indice{2}/P_\indice{2}} \right)\, ,
\end{equation}
where we have introduced mass ratio parameters (Equation \myhyperref{appendix eq: mass ratios}) between the \gls{mosa} masses and the total mass in order to lighten the writing,
\begin{align}
    & \epsilon_\indice{1} = \tfrac{m_{H_\indice{1}}}{m_{S} + m_{H_\indice{1}} + m_{H_\indice{2}}}\, ,
    & \epsilon_\indice{2} = \tfrac{m_{H_\indice{2}}}{m_{S} + m_{H_\indice{1}} + m_{H_\indice{2}}}\, .
\label{appendix eq: mass ratios}
\end{align}

\vspace{0.75cm}

The term $\myvec{r}{S/P}$ is constant so that the time derivative of $\myvec{r}{B/S}$ writes
\begin{align}
    \ddt{\myvec{r}{B/S}}{B} = \,&\epsilon_\indice{1} \left( \myvec{\omega}{H_\indice{1}/B} \times \myvec{r}{Q_\indice{1}/P_\indice{1}} \right) + \epsilon_\indice{2} \left( \myvec{\omega}{H_2/B} \times \myvec{r}{Q_\indice{2}/P_\indice{2}} \right)\, ,
\label{appendix eq: ddt rbs}
\end{align}
and its second derivative writes
\begin{align}
\label{appendix eq: ddtddt rbs}
    & \! \ddtddt{\myvec{r}{B/S}}{B} =
    \\
    & \! \epsilon_\indice{1} \Bigg[ \ddt{\myvec{\omega}{H_\indice{1}/B}}{B} \times \myvec{r}{Q_\indice{1}/P_\indice{1}} + \myvec{\omega}{H_\indice{1}/B} \times \left( \myvec{\omega}{H_\indice{1}/B} \times \myvec{r}{Q_\indice{1}/P_\indice{1}} \right) \Bigg] \nonumber
    \\
    \! + & \epsilon_\indice{2} \Bigg[ \ddt{\myvec{\omega}{H_2/B}}{B} \times \myvec{r}{Q_\indice{2}/P_\indice{2}} + \myvec{\omega}{H_2/B} \times \left( \myvec{\omega}{H_2/B} \times \myvec{r}{Q_\indice{2}/P_\indice{2}} \right) \Bigg]\, . \nonumber
\end{align}

Finally, adding the two contributions \myhyperref{appendix eq: ddt rbs} and \myhyperref{appendix eq: ddtddt rbs} from the spacecraft \gls{com} time variations to the test-mass longitudinal \gls{eom} (\myhyperref{eq: TM Long EOM - Vector}), one finds after a few straightforward, additional steps that

\small
\begin{widetext}
    \begin{equation}
        \begin{split}
            & \ddtddt{\myvec{r}{T/H}}{H} + 2 \myvec{\omega}{H/B} \times \ddt{\myvec{r}{T/H}}{H} + 2 \myvec{\omega}{B/J} \times \ddt{\myvec{r}{T/H}}{H} + \ddt{\myvec{\omega}{H/B}}{B} \times \myvec{r}{T/H} + \ddt{\myvec{\omega}{B/J}}{J} \times \myvec{r}{T/H} \\
            & + 2 \left( \myvec{\omega}{B/J} \times \myvec{\omega}{H/B} \right) \times \myvec{r}{T/H} + 2 \myvec{\omega}{H/B} \times \left( \myvec{\omega}{B/J} \times \myvec{r}{T/H} \right) + \myvec{\omega}{H/B} \times \left( \myvec{\omega}{H/B} \times \myvec{r}{T/H} \right) + \myvec{\omega}{B/J} \times \left( \myvec{\omega}{B/J} \times \myvec{r}{T/H} \right) \\
            & - \epsilon_\indice{1} \left[ \ddt{\myvec{\omega}{H_\indice{1}/B}}{B} \times \myvec{r}{Q_\indice{1}/H_\indice{1}} + \myvec{\omega}{H_\indice{1}/B} \times \left( \myvec{\omega}{H_\indice{1}/B} \times \myvec{r}{Q_\indice{1}/H_\indice{1}} \right) \right] - \epsilon_\indice{2} \left[ \ddt{\myvec{\omega}{H_2/B}}{B} \times \myvec{r}{Q_\indice{2}/H_\indice{2}} + \myvec{\omega}{H_2/B} \times \left( \myvec{\omega}{H_2/B} \times \myvec{r}{Q_\indice{2}/H_\indice{2}} \right) \right] \\
            & - 2 \myvec{\omega}{B/J} \times \Bigg[ \epsilon_\indice{1} \left( \myvec{\omega}{H_\indice{1}/B} \times \myvec{r}{Q_\indice{1}/H_\indice{1}} \right) + \epsilon_\indice{2} \left( \myvec{\omega}{H_2/B} \times \myvec{r}{Q_\indice{2}/H_\indice{2}} \right) \Bigg] + \ddt{\myvec{\omega}{B/J}}{J} \times \Big[ \myvec{r}{H/S} - \epsilon_\indice{1} \left( \myvec{r}{Q_\indice{1}/H_\indice{1}} + \myvec{r}{H_\indice{1}/S} \right) - \epsilon_\indice{2} \left( \myvec{r}{Q_\indice{2}/H_\indice{2}} + \myvec{r}{H_\indice{2}/S} \right) \Big] \\
            & + \myvec{\omega}{B/J} \times \left( \myvec{\omega}{B/J} \times \Big[ \myvec{r}{H/S} - \epsilon_\indice{1} \left( \myvec{r}{Q_\indice{1}/H_\indice{1}} + \myvec{r}{H_\indice{1}/S} \right) - \epsilon_\indice{2} \left( \myvec{r}{Q_\indice{2}/H_\indice{2}} + \myvec{r}{H_\indice{2}/S} \right) \Big] \right) \\
            & = \sum \frac{\myvec{f}{T}}{m_{T}} - \sum \frac{\myvec{f}{B}}{m_{B}}\ .
        \end{split}
    \end{equation}
\end{widetext}
\normalsize

\subsection{Spacecraft angular dynamics}
\label{appendix: sc ang dyn}

As seen in the previous section, the stationarity condition of the mass distribution within the spacecraft -- that is, the stationarity of the inertia tensor in the body frame $\mathcal{B}$ -- is not strictly met, for the \gls{mosa} rotating along with constellation orbital breathing. A full mathematical treatment for spacecraft attitude modeling would consist in breaking down the inertia tensors into non-deformable pieces, thus static in their respective frames: spacecraft body and its frame $\mathcal{B}$, \gls{mosa} bodies and their respective frames $\mathcal{H}$, as well as an additional body named {\it platform} to which is attached a frame $\mathcal{S}$ of center $S$ -- the \gls{com} of the platform considered alone, that is, \gls{mosa}s excluded. This gives a decomposition
\begin{equation}
    I_{B/B} = I_{S/B} + I_{H_\indice{1}/B} + I_{H_\indice{2}/B}\ .
\end{equation}
With this, one can treat bodies separately in applying the transport theorem suitably so that inertia tensors are constant in time in proper frames. In addition, one also needs to consider that the spacecraft \gls{com} about which the inertia tensor is considered is itself dynamical. Therefore, $S$ is preferred as the point about which to evaluate the mass distributions, and one has
\begin{equation}
\begin{split}
    \!\!I_{B/B} \ \vangvel{B}{J} = \ & I_{S/S} \ \vangvel{B}{J} \\
    & + m_{S} \myvec{r}{B/S} \times \left( \myvec{r}{B/S} \times \ \vangvel{B}{J} \right) \\
    & + I_{H_\indice{1}/Q_\indice{1}} \ \vangvel{B}{J} + I_{H_\indice{2}/Q_\indice{2}} \ \vangvel{B}{J} \\
    & - m_{H_\indice{1}} \myvec{r}{Q_\indice{1}/B} \times \left( \myvec{r}{Q_\indice{1}/B} \times \ \vangvel{B}{J} \right) \\
    & - m_{H_\indice{2}} \myvec{r}{Q_\indice{2}/B} \times \left( \myvec{r}{Q_\indice{2}/B} \times \ \vangvel{B}{J} \right),
\end{split}
\label{appendix eq: S/C AngMomentum Decomposition}
\end{equation}
where we have to introduce the \gls{com} of the \gls{mosa}s alone as well, denoted $Q$ and labeled by test-mass index. The next steps would consist in injecting \myhyperref{appendix eq: S/C AngMomentum Decomposition} into Equation \myhyperref{eq: Euler equation - SC}, although one immediately sees that the number of terms will explode while going through developments similar to section \myhyperref{subsection: Test mass longitudinal motion}. Accounting for this non-static mass distribution is likely to introduce numerous second order terms, and although full derivation of the corresponding equations of motion have been performed symbolically, a {\tt Mathematica} treatment and cross-check will be necessary in order to produce robust, cross-checked equations of motion able to evaluate the magnitude of such corrections, and their relevancy in the model.

\section{From continuous to discrete State-Space Model}\label{appendix: to discrete ssm}
We start out in the continuous state-space form
\begin{align}
    \dv{\va{X}}{t} = A\, \va{X}(t) + B \, \va{u}(t)\ .
\end{align}
For constant matrices $A$ and $B$ the solution to this is known to be represented by matrix exponentials and given by
\begin{align}
    \va{X}(t) = \euler^{A (t-t_0)} \va{X}(t_0) + \int_{t_0}^{t} \euler^{A (t-\tau)} B  \va{u}(\tau) \operatorname{d}\!\tau \ .
\end{align}
Then we discretize the solution with the grid spacing $dt$ in the time domain, using the short-hand notation 
\begin{align}
    \va{X}_{0} &:= \va{X}\!\left(t_0\right)\ , \\
    \va{X}_{n} &:= \va{X}\!\left(t^{(n)}\right)\ , \\
    \va{u}_{n} &:= \va{u}\!\left(t^{(n)}\right)\ .
\end{align}
This yields
\begin{align}
    \begin{split}
    \va{X}_{n+1} = &\euler^{A \left(t^{(n+1)}-t_0\right)} \va{X}_0 \\
        &+ \int_{t_0}^{t^{(n+1)}} \euler^{A \left(t^{(n+1)}-\tau\right)} B  \va{u}(\tau) \operatorname{d}\!\tau \ , \label{eq: app disc. sol cont. ssm 1}
    \end{split}\\
    \begin{split}
    \va{X}_n = &\euler^{A \left(t^{(n)}-t_0\right)} \va{X}_0 \\
    &+ \int_{t_0}^{t^{(n)}} \euler^{A \left(t^{(n)}-\tau\right)} B  \va{u}(\tau) \operatorname{d}\!\tau \ . \label{eq: app disc. sol cont. ssm 2}
    \end{split}
\end{align}
To combine these two equations, multiply Equation \myhyperref{eq: app disc. sol cont. ssm 2} with $\exp(A\, dt)$, use $t^{(n+1)}=t^{(n)}+dt$, and then solve for the first term
\begin{align}
    \begin{split}
    \!\!\euler^{A \left(t^{(n+1)}-t_0\right)} \va{X}_0 = &\euler^{A\, dt}\va{X}_n \\
    &- \int_{t_0}^{t^{(n)}}\!\! \euler^{A \left(t^{(n+1)}-\tau\right)} B  \va{u}(\tau) \operatorname{d}\!\tau \, . 
    \end{split}
\end{align}
This in turn can the be plugged into the previous Equation \myhyperref{eq: app disc. sol cont. ssm 1} to get
\begin{align}
    \begin{split}
    \va{X}_{n+1} = &\euler^{A\, dt}\va{X}_n \\
    &+ \int_{t^{(n)}}^{t^{(n+1)}} \euler^{A \left(t^{(n+1)}-\tau\right)} B  \va{u}(\tau) \operatorname{d}\!\tau \ . \label{eq: app disc. sol cont. ssm comb 2}
    \end{split}
\end{align}
To proceed, a choice needs to be made on how $\va{u}$ is discretized. As mentioned in subsection \myhyperref{section: solver subsection: linear}, we chose a zero-order hold here. Thus, in the relevant interval it is given by
\begin{align}
    \va{u}(\tau) = \va{u}_n \ , \quad \tau\in [t^{(n)}, t^{(n+1)}] \ .
\end{align}
Plugging this into the integral of Equation \myhyperref{eq: app disc. sol cont. ssm comb 2} and using a new variable $\lambda = \tau - t^{(n)}$ gives
\begin{align}
    \begin{split}
    &\int_{0}^{dt} \euler^{A \left(dt-\lambda\right)} \operatorname{d}\!\lambda\, B  \va{u}_n \ .
    \end{split}
\end{align}
Given that $A$ is invertible, this can be solved by the trick
\begin{align}
    \begin{split}
    \int_{0}^{dt} \euler^{A \left(dt-\lambda\right)} \operatorname{d}\!\lambda &= A^{-1} \int_{0}^{dt} A\euler^{A \left(dt-\lambda\right)} \operatorname{d}\!\lambda \\
    &=-A^{-1} \left[\euler^{A \left(dt-\lambda\right)}\right]_0^{dt} \\
    &=A^{-1} \left(\euler^{A dt} - \unitmat\right) \ .
    \end{split}
\end{align}
Together, we get the final form
\begin{align}
    \begin{split}
    \va{X}_{n+1} & = \euler^{A dt}\va{X}_n 
    + A^{-1}\left[\euler^{A dt} - \unitmat\right] B  \va{u}_n \\
    & =: A_\text{disc.}\va{X}_n + B_\text{disc.}\va{u}_n\ .
    \end{split}
\end{align}
This solution to the first-order hold is more complex. There, the external input $\va{u}$ is approximated between time-steps by
\begin{align}
    \va{u}(\tau) = \va{u}_n + \frac{\tau - t^{(n)}}{dt} \left[ \va{u}_{n+1} - \va{u}_n \right]\ .
\end{align}
More details on this can be found either in the source code of the library \cite{2020SciPy-NMeth} or in \cite{franklin1998digital}.

\section{Full set of equations of motion for DE solver}\label{appendix: all eom for solver}
As a reminder, the state vector has the following form:
\begin{align}
    \va{X}=
    \begin{matrix}
        \Big[
        & \myvec{\alpha}{\mathcal{B}/\mathcal{O}} & \myexpr{\omega}{\mathcal{B}/\mathcal{O}}{\mathcal{B}} & \posvecexp{T}[1]{H}[1]{H}[1] & \cardanvec{T}[1]{H}[1] & \posvecexp{T}[2]{H}[2]{H}[2]
         \\
        & \ \ \cardanvec{T}[2]{H}[2] &
        \velvecexp{T}[1]{H}[1]{H}[1] &
        \myexpr{\omega}{\mathcal{T}_\indice{1}/\mathcal{H}_\indice{1}}{\mathcal{T}_\indice{1}} & \velvecexp{T}[2]{H}[2]{H}[2] & \vangvelexp{T}[2]{H}[2]{T}[2]
         \\
        & \delta \phi_{tel, 1} & \delta \dot{\phi}_{tel, 1} & \delta \phi_{tel, 2} & \delta \dot{\phi}_{tel, 2}
        & \Big] \hspace*{1.3em}
    \end{matrix}
\label{appendix eq: state vector}
\end{align}
The derivative of $\va{X}$ is the interest of this section. Firstly, the derivatives of $\myvec{\alpha}{\mathcal{B}/\mathcal{O}}$, $\cardanvec{T}[1]{H}[1]$, and $\cardanvec{T}[2]{H}[2]$ can be related to $\myexpr{\omega}{\mathcal{B}/\mathcal{O}}{\mathcal{B}}$, $\vangvelexp{T}[1]{H}[1]{T}[1]$, and $\vangvelexp{T}[2]{H}[2]{T}[2]$ respectively by the following type of equation (c.f. Equation \myhyperref{eq: angle rates to angvel}) 
\begin{align}
   \vec{\omega} = E(\vec{\alpha}) \dot{\vec{\alpha}} \quad \Leftrightarrow \quad \dot{\vec{\alpha}} = E(\vec{\alpha})^{-1}\vec{\omega} \ .
\end{align}
For $\delta \phi_{tel, i}$ the relation becomes trivial ($i\in\{1,2\}$). Then, take Equation \myhyperref{eq: S/C Angular EOM - Projected} and invert the moment of inertia matrix to get 
\begin{align}
    \begin{split}
        \!\!\!\!\vangaccexp{B}{O}{B}\! = &-\myskew{ \rotmat{B}{O} \rotmat{O}{J} \vangvelexp{O}{J}{J} } \vangvelexp{B}{O}{B} \\
        &- \left(\itensorexp{sc}{B}{B}\right)^{-1}\myskew{ \vangvelexp{B}{O}{B} } \itensorexp{sc}{B}{B} \vangvelexp{B}{O}{B} \\
        &+ \left(\itensorexp{sc}{B}{B}\right)^{-1} \myskew{ \itensorexp{sc}{B}{B} \rotmat{B}{O} \rotmat{O}{J} \vangvelexp{O}{J}{J} } \vangvelexp{B}{O}{B} \\
        &- \left(\itensorexp{sc}{B}{B}\right)^{-1}\myskew{ \rotmat{B}{O} \rotmat{O}{J} \vangvelexp{O}{J}{J} } \itensorexp{sc}{B}{B} \vangvelexp{B}{O}{B} \\
        &+ \left(\itensorexp{sc}{B}{B}\right)^{-1}\sum \torque{B}{B} - \rotmat{B}{O} \rotmat{O}{J} \vangaccexp{O}{J}{J} \\
        &-\! \left(\itensorexp{sc}{B}{B}\right)^{-1}\!\myskew{ \rotmat{B}{O} \rotmat{O}{J} \vangvelexp{O}{J}{J} } \!\!\itensorexp{sc}{B}{B} \rotmat{B}{O} \rotmat{O}{J} \vangvelexp{O}{J}{J} \, .
    \end{split}
\end{align}
Note that the derivative $\vangaccexp{O}{J}{J}$ is an external input, i.e., $\vangvelexp{O}{J}{J}$ is not part of the state vector.

Afterwards, the two derivatives of the MOSA attitude (for the two MOSAs separately) of Equation \myhyperref{eq: MOSA Angular EOM - Projected} can be put as
\begin{align}
    \begin{split}
        \!\!\vangaccexp{H}{B}{H}\! = &- \left(\itensorexp{mo}{Q}{H}\right)^{-1} \myskew{ \vangvelexp{H}{B}{H} } \itensorexp{mo}{Q}{H} \vangvelexp{H}{B}{H} \\
        &- \left(\itensorexp{mo}{Q}{H}\right)^{-1} \myskew{ \rotmat{H}{B} \vangvelexp{B}{O}{B} } \itensorexp{mo}{Q}{H} \vangvelexp{H}{B}{H}  \\
        &- \left(\itensorexp{mo}{Q}{H}\right)^{-1} \myskew{ \rotmat{H}{B} \rotmat{B}{O} \rotmat{O}{J} \vangvelexp{O}{J}{J} } \!\!\itensorexp{mo}{Q}{H} \vangvelexp{H}{B}{H} \\
        &+ \left(\itensorexp{mo}{Q}{H}\right)^{-1} \sum \torque{H}{\text{rel,} H} \ .
    \end{split}
\end{align}

Then next, the two derivatives of the test-mass attitude (for the two TMs separately) of Equation \myhyperref{eq: TM Angular EOM - Projected} can be rewritten as
\begin{align}
    \vangaccexp{T}{H}{T} = &- \rotmat{T}{H} \vangaccexp{H}{B}{H} - \rotmat{T}{H} \rotmat{H}{B} \vangaccexp{B}{O}{B} \nonumber\\
    &- \rotmat{T}{H} \rotmat{H}{B} \myskew{ \rotmat{B}{O} \rotmat{O}{J} \vangvelexp{O}{J}{J} } \vangvelexp{B}{O}{B} \nonumber\\
    &- \rotmat{T}{H} \rotmat{H}{B} \rotmat{B}{O} \rotmat{O}{J} \vangaccexp{O}{J}{J} \nonumber\\
    &+ \rotmat{T}{H} \myskew{ \vangvelexp{H}{B}{H} } \rotmat{H}{B} \vangvelexp{B}{O}{B} \\
    &+ \rotmat{T}{H} \myskew{ \vangvelexp{H}{B}{H} } \rotmat{H}{B} \rotmat{B}{O} \rotmat{O}{J} \vangvelexp{O}{J}{J} \nonumber\\
    &+ \myskew{ \vangvelexp{T}{H}{T} } \rotmat{T}{H} \vangvelexp{H}{B}{H} + \myskew{ \vangvelexp{T}{H}{T} } \rotmat{T}{H} \rotmat{H}{B} \vangvelexp{B}{O}{B} \nonumber\\
    &+ \myskew{ \vangvelexp{T}{H}{T} } \rotmat{T}{H} \rotmat{H}{B} \rotmat{B}{O} \rotmat{O}{J} \vangvelexp{O}{J}{J} \nonumber\\
    &+ \left(\itensorexp{tm}{T}{T}\right)^{-1} \sum \torque{T}{T}\ .\nonumber
\end{align}
Note here that the derivatives $\vangaccexp{B}{O}{B}$ and $\vangaccexp{H}{B}{H}$ have already been calculated in the previous equations, so can be regarded as a known input here. Equivalently you could substitute the previous equation here to completely disentangle the equations, but we believe this is conceptually not necessary and would only increase computation times by redundant computations, if implemented in the completely disentangled form.

Almost the last identity to use is that the derivatives of $\posvecexp{T}[1]{H}[1]{H}[1]$ and $\posvecexp{T}[2]{H}[2]{H}[2]$ are already known values recorded in the state vector, namely $\velvecexp{T}[1]{H}[1]{H}[1]$ and $\velvecexp{T}[2]{H}[2]{H}[2]$.

As a last step, the two second derivatives of test-mass position (for the two TMs separately) of Equation \myhyperref{eq: TM Long EOM - Projected} can be reformulated as
\begin{widetext}
    \begin{equation}
        \begin{split}
            \accvecexp{T}{H}{H} = &- 2 \myskew{\vangvelexp{H}{B}{H}} \velvecexp{T}{H}{H} - 2 \myskew{\rotmat{H}{B}\vangvelexp{B}{O}{B}} \velvecexp{T}{H}{H} - 2 \myskew{\rotmat{H}{B} \rotmat{B}{O} \rotmat{O}{J} \vangvelexp{O}{J}{J}} \velvecexp{T}{H}{H}
            - \myskew{\vangaccexp{H}{B}{H}} \posvecexp{T}{H}{H} \\
            &- \myskew{ \rotmat{H}{B} \vangaccexp{B}{O}{B} } \posvecexp{T}{H}{H} - \myskew{ \rotmat{H}{B} \rotmat{B}{O} \rotmat{O}{J} \vangaccexp{O}{J}{J} } \posvecexp{T}{H}{H} - 2 \myskew{ \rotmat{H}{B} \vangvelexp{B}{O}{B} } \myskew{ \vangvelexp{H}{B}{H} } \posvecexp{T}{H}{H} \\
            &- 2 \myskew{ \rotmat{H}{B} \rotmat{B}{O} \rotmat{O}{J} \vangvelexp{O}{J}{J} } \myskew{ \vangvelexp{H}{B}{H} } \posvecexp{T}{H}{H} - \myskew{ \vangvelexp{H}{B}{H} } \myskew{ \vangvelexp{H}{B}{H} } \posvecexp{T}{H}{H} \\
            &- \myskew{ \rotmat{H}{B} \vangvelexp{B}{O}{B} } \myskew{ \rotmat{H}{B} \vangvelexp{B}{O}{B} } \posvecexp{T}{H}{H} - 2 \myskew{ \rotmat{H}{B} \rotmat{B}{O} \rotmat{O}{J} \vangvelexp{O}{J}{J} } \myskew{ \rotmat{H}{B} \vangvelexp{B}{O}{B} } \posvecexp{T}{H}{H} \\
            &- \myskew{ \rotmat{H}{B} \rotmat{B}{O} \rotmat{O}{J} \vangvelexp{O}{J}{J} } \myskew{ \rotmat{H}{B} \rotmat{B}{O} \rotmat{O}{J} \vangvelexp{O}{J}{J} } \posvecexp{T}{H}{H} + \myskew{\posvecexp{H}{P}{H}} \vangaccexp{H}{B}{H} - \myskew{\vangvelexp{H}{B}{H}}\myskew{\vangvelexp{H}{B}{H}}\posvecexp{H}{P}{H} \\
            &- 2 \myskew{ \myskew{ \posvecexp{H}{P}{H} } \vangvelexp{H}{B}{H} } \rotmat{H}{B} \vangvelexp{B}{O}{B} + 2 \myskew{ \rotmat{H}{B} \rotmat{B}{O} \rotmat{O}{J} \vangvelexp{O}{J}{J} } \myskew{ \posvecexp{H}{P}{H} } \vangvelexp{H}{B}{H} \\
            &+ \myskew{ \posvecexp{H}{B}{B} } \rotmat{H}{B} \vangaccexp{B}{O}{B} - \myskew{ \myskew{ \rotmat{H}{B} \posvecexp{H}{B}{B} } \rotmat{H}{B} \vangvelexp{B}{O}{B} } \rotmat{H}{B} \vangvelexp{B}{O}{B} + 2 \myskew{ \rotmat{H}{B} \rotmat{B}{O} \rotmat{O}{J} \vangvelexp{O}{J}{J} } \myskew{ \rotmat{H}{B} \posvecexp{H}{B}{B} } \rotmat{H}{B} \vangvelexp{B}{O}{B} \\
            &+ \sum \frac{\force{T}{H}}{m_{T}} - \sum \frac{\rotmat{H}{B} \force{B}{B}}{m_{B}} - \myskew{ \rotmat{H}{B} \rotmat{B}{O} \rotmat{O}{J} \vangaccexp{O}{J}{J} } \rotmat{H}{B}  \posvecexp{H}{B}{B} - \myskew{ \rotmat{H}{B} \rotmat{B}{O} \rotmat{O}{J} \vangvelexp{O}{J}{J} } \myskew{ \rotmat{H}{B} \rotmat{B}{O} \rotmat{O}{J} \vangvelexp{O}{J}{J} } \rotmat{H}{B}  \posvecexp{H}{B}{B}\, .
        \end{split}
    \end{equation}
\end{widetext}
Note that all quantities other than the second derivatives of test-mass positions are known either as external inputs or from previous equations in this section.

This gives all the derivatives of the state vector $\va{X}$ in a way that can be interpreted as functions $\va{f}(t;\va{X})$ and can thus directly be fed into a numerical differential equation solver, which then gives an update for $\va{X}$ at the next time-step.

\bibliographystyle{unsrt2}
\bibliography{references}
\end{document}